\documentclass[iop,twocolumn]{emulateapj}
\usepackage{amsmath, amsthm, amssymb, mathrsfs, textcomp, verbatim, graphicx, times, float}

\slugcomment{Draft of \today}
\shorttitle{Stripe 82 VLA and PTF}

\begin{document}

\title{The Caltech-NRAO Stripe 82 Survey (CNSS) Paper I: The Pilot Radio Transient Survey In 50 deg$^2$}
\author{K.~P.~Mooley\altaffilmark{1,2,9,10},
G.~Hallinan\altaffilmark{1},
S.~Bourke\altaffilmark{1},
A. Horesh\altaffilmark{3},
S.~T.~Myers\altaffilmark{2},
D.~A.~Frail\altaffilmark{2},
S.~R.~Kulkarni\altaffilmark{1},
D.~B.~Levitan\altaffilmark{4},
M.~M.~Kasliwal\altaffilmark{5},
S.~B.~Cenko\altaffilmark{6,7},
Y.~Cao\altaffilmark{1},
E.~Bellm\altaffilmark{1},
R.~R.~Laher\altaffilmark{8}}

\altaffiltext{1}{Cahill Center for Astronomy, MC 249-17,
 California Institute of Technology, Pasadena, CA 91125, USA. Email: kunal@astro.caltech.edu}
\altaffiltext{2}{National Radio Astronomy Observatory, P.O. Box O,
  Socorro, NM 87801}
\altaffiltext{3}{Benoziyo Center for Astrophysics, Faculty of Physics,
  The Weizmann Institute for Science, Rehovot 76100, Israel}
\altaffiltext{4}{Microsoft, Bellevue, WA, USA}
\altaffiltext{5}{Carnegie Observatories, 813 Santa Barbara St, Pasadena, CA 91101}
\altaffiltext{6}{Astrophysics Science Division, NASA Goddard Space Flight Center, Mail Code 661, Greenbelt, MD 20771, USA}
\altaffiltext{7}{Joint Space-Science Institute, University of Maryland, College Park, MD 20742, USA}
\altaffiltext{8}{Spitzer Science Center, California Institute of Technology, M/S 314-6, Pasadena, CA 91125, USA}
\altaffiltext{9}{NRAO Grote Reber Fellow}
\altaffiltext{10}{Current Address: Oxford Centre for Astrophysical Surveys, Denys Wilkinson Building, Keble Road, Oxford OX1 3RH}

\begin{abstract}

We have commenced a multi-year program, the Caltech-NRAO Stripe 82 Survey (CNSS), to search for radio transients with the Jansky VLA in the Sloan Digital Sky Survey Stripe 82 region. 
The CNSS will deliver five epochs over the entire $\sim$270 deg$^2$ of Stripe 82, an eventual deep combined map with a rms noise of $\sim$40 $\mu$Jy and catalogs at a frequency of 3 GHz, 
and having a spatial resolution of 3\arcsec.
This first paper presents the results from an initial pilot survey of a 50 deg$^2$ region of Stripe 82, involving four epochs spanning 
logarithmic timescales between one week and 1.5 years, with the combined map having a median rms noise of 35\,$\mu$Jy.
This pilot survey enabled the development of the hardware and software for rapid data processing, as well as transient detection and follow-up, necessary for the full 270 deg$^2$ survey. 
Data editing, calibration, imaging, source extraction, cataloging and transient identification were completed in a semi-automated fashion within six hours of completion 
of each epoch of observations, using dedicated computational hardware at the NRAO in Socorro, and custom-developed data reduction and transient detection pipelines. 
Classification of variable and transient sources relied heavily on the wealth of multi-wavelength legacy survey data in the Stripe 82 region, supplemented by repeated 
mapping of the region by the Palomar Transient Factory (PTF).
$3.9^{+0.5}_{-0.9}$\% of the few thousand detected point sources were found to vary by greater than 30\%, consistent with similar studies at 1.4 GHz and 5 GHz.
Multi-wavelength photometric data and light curves suggest that the variability is mostly due to shock-induced flaring in the jets of AGN.
Although this was only a pilot survey, we detected two bona fide transients, associated with an RS CVn binary and a dKe star.
Comparison with existing legacy survey data (FIRST, VLA-Stripe 82) revealed additional highly variable and transient sources on timescales between 5--20 years, largely associated with renewed AGN activity.
The rates of such AGN possibly imply episodes of enhanced accretion and jet activity occurring once every $\sim$40,000 years in these galaxies.
We compile the revised radio transient rates and make recommendations for future transient surveys and joint radio-optical experiments.

\end{abstract}


\section{Introduction}\label{sec:intro}


Owing to rapidly advancing detector technology and faster computing
speeds, optical and high energy astronomy have enjoyed a synoptic
survey capability\footnote{We refer to a ``synoptic survey'' here as a 
blind survey (as opposed to a targeted survey) sampling a 
large part of the sky (tens of square degrees or more) with a cadence of 
days to weeks.} for over a decade. A rich discovery phase space has been revealed through
synoptic experiments such as the Fermi Gamma-Ray Satellite \citep{michelson2010}, the 
Swift Gamma-Ray Burst Alert Mission \citep{gehrels2009}, and optical imagers such as the 
Catalina Real-Time Transient Survey (CRTS), the Panoramic Survey Telescope 
Rapid Response System (Pan-Starrs), and the Palomar Transient Factory \citep[PTF;][]{kasliwal2012}.


A similar science yield likely awaits centimeter radio surveys. In
Table~\ref{tab:transients_summary} we give a summary of known
extragalactic and Galactic slow transients (timescale $>$1s) at GHz
frequencies \citep[see also][]{murphy2013,lazio2014}. These
known sources highlight that radio observations trace high energy
particles; arise in locations of high magnetic field; and probe the
interaction of fast outflows with the surrounding medium. They are
complementary to optical and high energies in that they are not
affected by extinction and the emission from fast outflows is often
not as narrowly beamed; providing reliable calorimetry and the true
rates of cosmic explosions. 

In recognition of these discovery opportunities, a new generation of centimeter wavelength facilities
have been built or are underway. These include the Karl G. Jansky Array \citep[Jansky
VLA;][]{perley2011}, ASKAP \citep{johnston2008}, MeerKAT \citep{booth2012}, and Apertif/WSRT \citep{oosterloo2010}.
The upgraded Jansky VLA is the first operational interferometer with the survey speed to routinely 
detect the extragalactic explosive population with modest time allocation. 
Unlike the legacy VLA, the Jansky VLA possesses frequency agility essential to characterize the broadband radio spectra of detected transients.
It has a much improved snapshot multi-frequency UV coverage; imaging artifacts and contamination by sidelobes \citep{bower2007,frail2012} are no longer major impediments in transient search.
Furthermore, unlike the SKA pathfinders, the Jansky VLA has the spatial resolution to allow the accurate localization within a host 
galaxy (nuclear vs non-nuclear) necessary to rule out the AGN activity that dominates the extragalactic transient and variable population.

\begin{table*}
\scriptsize
\centering
\caption{Summary of Slow Radio Transient Phenomena at 1--6 GHz Frequencies}
\begin{tabular}{llllllllll} 
\hline \hline
No  & Object       & Variability           & Location & \multicolumn{2}{c}{Timescale}        & \multicolumn{2}{c}{Peak Luminosity}          & Rate         & Ref.  \\
\cline{5-6}\cline{7-8}
    &              & Process               &             & Radio            & Optical           & Radio                   & Optical         & ($>$0.3 mJy) & \\
    &              &                       &             &                  &                   & (erg s$^{-1}$ Hz$^{-1}$)& (erg s$^{-1}$)  & (deg$^{-2}$) & \\
\hline
\multicolumn{10}{c}{Extragalactic}\\
\hline
1   & AGN          & a) Quiescent          & Nuclear     & years -- decades & years             & 10$^{27}$--10$^{34}$& 10$^{42}$--10$^{47}$& 0.6        &  1--7   \\ 
    &              & b) Shock-in-jet       & ''          & days -- years    & ---               & 10$^{27}$--10$^{34}$& ---                 & 1.25       &  6--10  \\
    &              & c) ISS                & ''          & minutes -- days  & ---               & ---                 & ---                 & 60         &  11--13 \\
    &              & d) ESE                & ''          & months           & ---               & ---                 & ---                 & 0.001      &  14--15 \\
    &              & e) Jet-precession     & ''          & years            & ---               & ---                 & ---                 & unknown    &  16--17 \\
2   & SN-II/Ib/c   & Shocked CSM           & Traces SF   & months -- decades& days -- months    & 10$^{25}$--10$^{29}$& 10$^{41}$--10$^{43}$& 0.04 (II), &  18--23\\
    &              &                       &             &                  &                   &                     &                     & 6$\times$10$^{-6}$ (Ib/c) & \\
3   & SN-Ia        & Shocked CSM           & Non-nuclear & days -- weeks    & days -- months    & $\lesssim$10$^{23}$ & 10$^{39}$--10$^{43}$& $<$10$^{-5}$       & 23--28\\
4   & Long-GRB     & Shocked CSM / Jet     & Traces SF   & days -- years    & days -- months    & 10$^{29}$--10$^{31}$& 10$^{43}$--10$^{47}$& 6$\times$10$^{-5}$ & 29   \\
    & OA           & Shocked CSM           & Traces SF   & weeks -- years   & days?             & 10$^{29}$--10$^{30}$& 10$^{43}$--10$^{46}$& 7$\times$10$^{-4}$ & 30--32\\
5   & Short-GRB    & Shocked CSM / Jet     & Non-nuclear & days -- years?   & days              & $\sim$10$^{25}$     & $10^{39}\sim10^{41}$& $<$10$^{-6}$ & 33--35 \\
    & BNS merger   & Shocked CSM           & Non-nuclear & weeks -- years   & days              & $10^{22}\sim10^{25}$& $10^{39}\sim10^{40}$& 2$\times$10$^{-4}$ &  31,36--37 \\
6   & TDE          & Shocked CNM / Jet     & Nuclear     & years?           & months -- years?  & $10^{30}\sim10^{31}$& 10$^{42}\sim10^{44}$& 0.005      &  31,38--40 \\
    & Off-axis TDE & Shocked CNM           & Nuclear     & years?           & unknown           & $\sim$10$^{30}$     & unknown             & 0.01       &  31\\
7   & AIC          & Shocked CSM           & Non-nuclear & months           & day               & unknown             & $\sim10^{39}$       & 6$\times$10$^{-5}$ & 70--72\\
\hline
\multicolumn{10}{c}{Galactic}\\
\hline
1   & Active star  & Flaring (magnetic)    & Isotropic   & hours -- days    & seconds -- hours  & 10$^{12}$--10$^{15}$& $10^{30}\sim10^{33}$& 0.02   &  7,41--44 \\
2   & Active binary& Flaring (magnetic)    & Isotropic   & hours -- days    & hours -- days     & 10$^{14}$--10$^{22}$& $<10^{33}\sim10^{34}$&0.01   &  7,41,43,46  \\  
3   & X-ray binary & Mass accretion / flare& GP          & days -- weeks    & minutes -- hours  & 10$^{20}\sim10^{21}$& $\sim$10$^{35}$     & 0.005  &  7,44,46--50\\ 
4   & CV           & Dwarf nova / jet      & GP          & hours -- days    & days -- weeks     & 10$^{16}\sim10^{17}$& $10^{33}\sim10^{35}$& 0.001  &  44,51--54\\ 
5   & YSO          & Mass accretion / flare& GP          & hours -- weeks   & hours -- weeks    & 10$^{16}$--10$^{18}$& $10^{31}\sim10^{34}$& $<$0.05&  7,41,55      \\ 
6   & Pulsar       & Scattering            & GP          & weeks -- months  & ---               & 10$^{15}\sim10^{17}$& ---                 & 0.003  &  44,56,57      \\
7   & Magnetar     & Flaring (magnetic)    & GP          & weeks            & unknown           & $\sim$10$^{20}$?    & unknown             & $<$0.05&  7,58--60   \\
8   & Brown Dwarf  & Pulsing (magnetic)    & Isotropic   & seconds -- hours & seconds -- hours  & 10$^{13}\sim10^{14}$& $\sim$10$^{24}$     & $<$0.05&  7,61,62      \\
9   & Novae        & Ejecta / Shocked CSM  & GP          & years            & days -- months    & $\sim10^{20}$       & 10$^{38}$--10$^{40}$& $<$0.05&  23,63--65\\ 
10  & GCRT         & Unknown               & Unknown     & minutes -- years & unknown           & unknown             & unknown             & unknown&  66--69\\
\hline
\multicolumn{10}{p{7in}}{References: $^1$\cite{aller1999}, $^2$\cite{valtaoja1992}, $^3$\cite{arshakian2012}, $^4$\cite{hovatta2008}, $^5$\cite{padovani2011}, $^6$\cite{woo2002}, $^7$this work, 
  $^8$\cite{teraesranta1998}, $^9$\cite{mooley2013}, $^{10}$\cite{turler2000}, 
  $^{11}$\cite{dennett2002}, $^{12}$\cite{lovell2008}, $^{13}$\cite{ofek2011},
  $^{14}$\cite{fiedler1994}, $^{15}$\cite{murphy2013}, 
  $^{16}$\cite{lister2013}, $^{17}$\cite{chen2013}, 
  $^{18}$\cite{weiler2002}, $^{19}$\cite{stockdale2009}, $^{20}$\cite{soderberg2010}, $^{21}$\cite{gal-yam2006}, $^{22}$\cite{berger2003}, $^{23}$\cite{kasliwal2012}, 
  $^{24}$\cite{boffi1995}, $^{25}$\cite{panagia2006}, $^{26}$\cite{hancock2011}, $^{27}$\cite{chomiuk2012}, $^{28}$\cite{perez-torres2014}, 
  $^{29}$\cite{chandra2012}, 
  $^{30}$\cite{ghirlanda2014}, $^{31}$\cite{metzger2015}, $^{32}$\cite{cenko2013},
  $^{33}$\cite{hjorth2005}, $^{34}$\cite{tanvir2013}, $^{35}$\cite{fong2014}, 
  $^{36}$\cite{nakar2011}, $^{37}$\cite{piran2013}, 
  $^{38}$\cite{zauderer2011}, $^{39}$\cite{cenko2012}, $^{40}$\cite{zauderer2013}, 
  $^{41}$\cite{gudel2002}, $^{42}$\cite{kovari2007}, $^{43}$AAVSO, $^{44}$\cite{thyagarajan2011}, 
  $^{45}$\cite{henry1996}, 
  $^{46}$\cite{augusteijn1992}, $^{47}$\cite{motch1989}, $^{48}$\cite{miller-jones2004}, $^{49}$\cite{miller-jones2012}, $^{50}$\cite{williams2013}, 
  $^{51}$\cite{kording2008}, $^{52}$Mooley et al. (2015), in prep., $^{53}$Groot et al. (2015), in prep., $^{54}$\cite{patterson2011}, 
  $^{55}$\cite{findeisen2013}, 
  $^{56}$\cite{levinson2002}, $^{57}$\cite{huguenin1973}, 
  $^{58}$\cite{gaensler2005}, $^{59}$\cite{cameron2005}, $^{60}$\cite{fender2006}, 
  $^{61}$\cite{hallinan2007}, $^{62}$\cite{harding2013}, 
  $^{63}$\cite{chomiuk2012b}, $^{64}$\cite{roy2012}, $^{65}$\cite{kantharia2007}, 
  $^{66}$\cite{hyman2002}, $^{67}$\cite{hyman2005}, $^{68}$\cite{hyman2007}, $^{69}$\cite{hyman2009}
  $^{70}$\cite{piro2013}, $^{71}$\cite{metzger2009}, $^{72}$\cite{darbha2010}}\\
\multicolumn{10}{p{7in}}{Notes: (1) Location refers to the position within the host galaxy where the transient class is expected to be concentrated. 
(2) Timescale refers to the approximate time duration for which the flux density of a transient is within an order of magnitude of the flux density at the peak of the light curve.
(3) The transient rates (instantaneous snapshot rates) are compiled from a variety of sources. 
In some cases, a N$\propto$S$^{-3/2}$ scaling has been applied to find the rate above the 0.3 mJy flux density threshold. 
The expected uncertainty / scatter in the rate is between 10--50\%.
(4) Variability of the quiescent AGN emission refers to the sustained change in the quiescent flux density level, similar to that seen in VTC233002-002736.
(5) Extrinsic variability phenomena considered here, viz. interstellar scattering (ISS) and extreme scattering events (ESEs), are discussed only in 
  the context of AGN. While the rates of these events are dominated by AGN, they could very well affect any class of compact sources such as pulsars.
(6) Orphan afterglows (OAs) and binary neutron star (BNS) mergers refer to the orphan counterparts of beamed long- and short-GRBs respectively.
(7) TDE rates assume Swift J1644+57-like events. The deg$^{-2}$ rates quoted here for on- and off-axis TDEs have been calculated using the (logarithmic) mean of the Gpc$^{-3}$ yr$^{-1}$ rate 
from \cite{metzger2009} and that implied by the two on-axis TDEs discovered till date: Swift J1644+57 and Swift J2058+05 \citep{zauderer2011,cenko2012}. }
\label{tab:transients_summary}
\end{tabular}
\end{table*}


To date, there have been rather few radio surveys dedicated to slow
variables and transients\footnote{see http://www.tauceti.caltech.edu/kunal/radio-transient-surveys/index.html }, 
and they all have a number of limitations. The majority of these
surveys were single, multi-epoch interferometric pointings with a
limited field of view and as a result the number of variables and
transients is low \citep[e.g.][]{carilli2003,mooley2013}. Existing wide-area
surveys are either based on archival data or the data reduction and
candidate source identification was carried out significantly delayed
from the observing dates. This approach has drawbacks since without
near-real-time data reduction and multi-wavelength follow-up, the
candidate lists contain ambiguous transient classifications 
\citep{thyagarajan2011,bannister2011a,bannister2011b}.

We note that wide field, shallow surveys are superior to narrow field, deep
surveys, since they have the advantage of bringing the detectable
population of transients closer in distance, thus improving the
ability to find optical/infrared counterparts and to characterize host
galaxies and/or progenitors.  In appendix A 
we show mathematically that wide and shallow surveys are ideal for radio
transient searches.


In the light of these factors and to address some of the limitations of past surveys, we have commenced a multi-year program, 
the Caltech-NRAO Stripe 82 Survey (CNSS), to search for slow radio transients with the Jansky VLA at 3 GHz in the 
Sloan Digital Sky Survey Stripe 82 region. 
The CNSS is a dedicated transient survey carried out in five epochs (cadence of days, months and years) over the 
entire $\sim$270 deg$^2$ of Stripe 82, with a uniform single-epoch rms noise of $\sim$80 $\mu$Jy and a spatial 
resolution of $\sim$3\arcsec.

In this paper we present the CNSS pilot survey, a sub-mJy Jansky VLA survey at 3 GHz in 50 deg$^2$ of Stripe 82.
This is a prototype survey to demonstrate the fast imaging capabilities of the VLA, and to develop 
near real-time data processing, source identification, and transient search. 
Another unique aspect of this radio survey is that it was undertaken with a
contemporaneous high-cadence optical survey with the PTF so that a 
direct comparison could be made of the dynamic radio and optical skies.
Technical details of the radio and the optical surveys are given in \S\ref{sec:survey}.
The calibration, RFI flagging, imaging and source cataloging of the Jansky VLA data carried out in 
near-real-time as well as during the final careful processing is described in \S\ref{sec:radio_processing}.
\S\ref{sec:optical_processing} details the optical data processing for the contemporaneous survey 
with the PTF.
A description of the radio transient search on timescales less than one week, one month, 
1.5 years and longer timescales is provided in \S\ref{sec:radio_transients}.
A subset of the radio variable and transient sources representative of the full sample found in the pilot radio survey are discussed in detail in \S\ref{sec:radio_transients}.
The optical counterparts of radio sources from the Jansky VLA survey, and the optical variability of the radio 
transients from \S\ref{sec:radio_transients}, are studied in \S\ref{sec:optical_properties} using photometry 
from PTF and SDSS.
\S\ref{sec:optical_properties} also gives a brief discussion of the optical transients found in PTF 
independently of the radio survey.
A summary of the results from our joint radio-optical experiment and their implications are discussed in \S\ref{sec:summary}.

\section{The Survey}\label{sec:survey}

\subsection{Radio Observations}\label{sec:survey:radio_observations}

The Caltech-NRAO Stripe 82 Survey (CNSS) was designed to: 
1) logarithmically sample timescales roughly between one week and one year, 
2) have a high survey speed (which is a function of the bandwidth and the antenna system temperature), 
3) have a relatively high angular resolution to facilitate a precise location of transients, 
4) be carried out in a part of the sky where ample multi-wavelength archival data were available so as to enable the identification of host galaxies or progenitors of the radio transients, 
5) find non-thermal transients relatively early-on in their evolution, and 
6) be wide and sensitive enough to potentially detect or place strong constraints on binary neutron star (BNS) mergers and orphan long-duration gamma-ray burst afterglows (OAs),

For the pilot survey, we chose a $\sim$50 deg$^2$ region in SDSS Stripe 82 with similar characteristics.
The region had to satisfy other scheduling constraints in the radio and optical, so we specifically chose the region bounded in right 
ascension by 329.127\arcdeg and 353.158\arcdeg, and in declination by -1.132\arcdeg and +1.167\arcdeg, approximately.
Observations were carried out across four epochs with the Jansky VLA in B array configuration and S-band was chosen to maximize survey speed.
To maximize the continuum imaging sensitivity, the observing setup chosen was: Wideband Interferometric Digital Architecture (WIDAR) correlator with 16 spectral windows, 
64 2-MHz-wide channels each to get 2 GHz of total bandwidth centered on 3.0 GHz, and 1-sec integrations.

\begin{figure}[htp]
\centering
\includegraphics[width=3.4in]{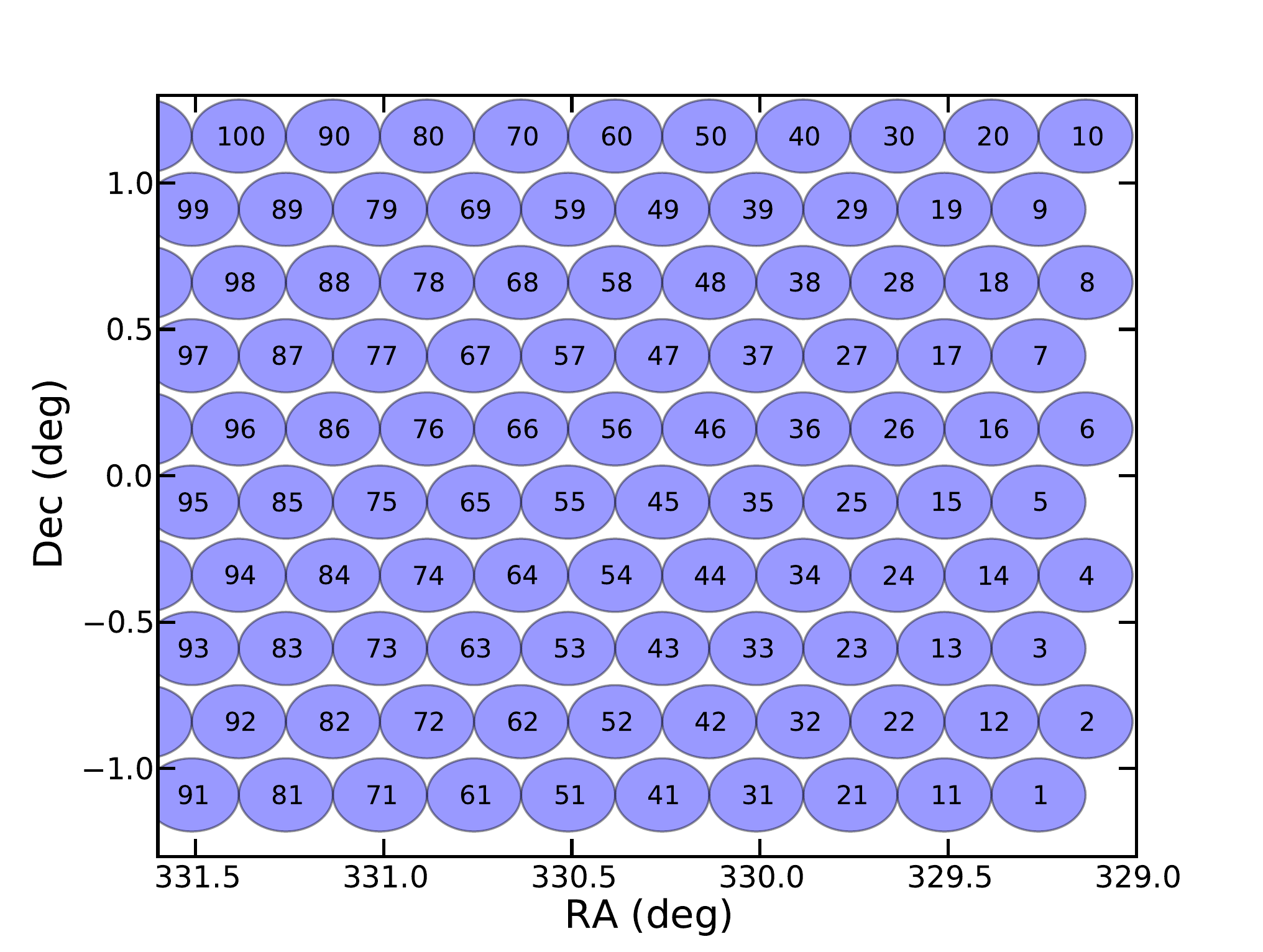}
\caption{Setup of target pointings for a part of region R1 of the Jansky VLA survey. 
The circles indicate primary beam FWHM at the mean frequency of 2.9 GHz, and the pointing numbers are labelled.
Nearest neighbors are separated by 15\arcmin.
See \S\ref{sec:survey:radio_observations} for details on the pointing setup.}
\label{fig:radio_obs_setup}
\end{figure}

The first three epochs were observed under the project code 12A-371 between Jul--Aug 2012.
Due to telescope scheduling constraints, each of these epochs was divided into two nights of observing of 7 hours each.
All the six 7-hour observing blocks were centered on 23h local sidereal time (LST).
We refer to the three epochs as E1/2/3 and the regions observed on the two nights of the first epoch as R1 and R2 respectively.
{\it Observations were carried out at the same LST in the three epochs in order to minimize systematic effects associated with sidelobes and beam squint.}
Each of the regions, R1 and R2, was divided into 485 pointings arranged in a hexagonal mosaic pattern (Figure~\ref{fig:radio_obs_setup}), 
optimized using the {\it makeschedule} task in CASA written by Andreas Brunthaler.
In order to maximize the volume probed by our survey, we compromised on uniform-sensitivity coverage, placing neighboring pointings at $\geqslant$15\arcmin.
The mosaic is therefore not precisely hexagonal close packed\footnote{Although epochs E1--E3 do not have uniform sensitivity across the survey region, the final CNSS survey is designed to have uniform RMS noise.}. 
Nearest neighbors having the same declination are separated by 15\arcmin, and those having offset in declination are $\sqrt{15\arcmin^2+7.5\arcmin^2}=16.8\arcmin$ apart.
During each night of observing, one 39-second snapshot observation was obtained at each of the 485 target pointings.
J2212+0152 and J2323-0317 were chosen as the phase calibrators for the two regions respectively, and 3C48 as the flux and bandpass calibrator.
The phase calibrator was observed every 15 minutes.
A summary of the first three observing epochs is given in Table~\ref{tab:radio_observations}.

\begin{table*}[htp]
\centering
\caption{Summary of the Jansky VLA Observations}
\begin{tabular}{lllllllll} 
\hline \hline
No.   & Date           & Epoch/ & Array   & RA range           & rms       & Obs.    & phase         & LST\\
      & (UT)           & Reg.   & config. & (deg)              & ($\mu$Jy) & mode    & calib.        & (h)\\
\hline
1     & 2012 Jul. 13  & E1R1    & B       & 329.127 -- 341.143 & 71       & STD     & J2212+0152    & 19.5 -- 02.5\\
2     & 2012 Jul. 14  & E1R2    & B       & 341.143 -- 353.158 & 61       & STD     & J2323-0317    & 19.5 -- 02.5\\
3     & 2012 Aug. 15  & E2R2    & B       & 341.143 -- 353.158 & 52       & STD     & J2323-0317    & 19.5 -- 02.5\\
4     & 2012 Aug. 18  & E2R1    & B       & 329.127 -- 341.143 & 52       & STD     & J2212+0152    & 19.5 -- 02.5\\
5     & 2012 Aug. 23  & E3R2    & B       & 341.143 -- 353.158 & 58       & STD     & J2323-0317    & 19.5 -- 02.5\\
6     & 2012 Aug. 24  & E3R1    & B       & 329.127 -- 341.143 & 55       & STD     & J2212+0152    & 19.5 -- 02.5\\
7     & 2013 Dec. 20  & E4R5    & B       & 346.000 -- 355.000 & 93       & OTF     & J2323-0317    & 23.5 -- 02.5\\ 
8     & 2013 Dec. 21  & E4R3    & B       & 328.000 -- 337.000 & 96       & OTF     & J2212+0152    & 18.6 -- 21.6\\ 
9     & 2013 Dec. 22  & E4R4    & B       & 337.000 -- 346.000 & 96       & OTF     & J2212+0152    & 19.1 -- 22.1\\ 
10    & 2014 Jan. 13  & E4R3    & B       & 328.000 -- 337.000 & 95       & OTF     & J2212+0152    & 19.8 -- 22.8\\ 
11    & 2014 Jan. 18  & E4R5    & B       & 346.000 -- 355.000 & 102      & OTF     & J0016-0015    & 23.3 -- 02.3\\ 
12    & 2014 Jan. 19  & E4R4    & B       & 337.000 -- 346.000 & 95       & OTF     & J2212+0152    & 19.8 -- 22.8\\ 
13    & 2014 Feb. 08  & E4R3    & BnA     & 328.000 -- 337.000 & 96       & OTF     & J2212+0152    & 19.3 -- 22.3\\ 
14    & 2014 Feb. 09  & E4R4    & BnA     & 337.000 -- 346.000 & 96       & OTF     & J2212+0152    & 19.3 -- 22.3\\ 
15    & 2014 Feb. 09  & E4R5    & BnA     & 346.000 -- 355.000 & 92       & OTF     & J0016-0015    & 23.3 -- 02.3\\ 
\hline
\multicolumn{9}{p{5in}}{Note: The rms noise tabulated for epochs E1, E2, and E3 are before correcting the pointings for the primary beam. 
For the flat-sky rms noise, see Figure~\ref{fig:sensitivity}. For epoch E4, the rms noise for the individual sub-epochs from the 
near-real-time processing are tabulated. See Figure~\ref{fig:sensitivity} for the rms noise of the final E4 co-add.}
\label{tab:radio_observations}
\end{tabular}
\end{table*}

The fourth epoch (E4) was observed under the project code 13B-370, and is essentially a co-addition of three epochs of observation carried out between Nov 2013 and Feb 2014.
For these observations, we took advantage of On-The-Fly mosaicing (OTF\footnote{https://science.nrao.edu/facilities/vla/docs/manuals/opt/otf}), 
available as a Resident Shared Risk Observing (RSRO) capability of the Jansky VLA.
In the OTF mode, data was taken while the telescopes were slewing at a speed of 1\arcmin~per second along right ascension, 
and the correlator phase center being stepped every four seconds.
In each observing block, 9\arcdeg~in right ascension and 2.5\arcdeg~in declination centered on 0\arcdeg~were observed to get a total areal coverage of 22.5 deg$^2$. 
Observations were carried out with the telescopes moving in a zig-zag basket-weave pattern on the sky: 
slewing 9\arcdeg~in decreasing right ascension along a constant declination strip at -1.25\arcdeg, then stepping up in 
declination by 10.6\arcmin~and slewing along increasing right ascension, and so on.
Covering 2.5\arcdeg~along declination required 15 such zig-zag strips offset by 10.6\arcmin~in declination from nearest neighbors.
This observing plan is designed to yield 2025 four-second-long scans, and a mosaic with a uniform rms noise of $\sim$80~$\mu$Jy across 22.5 deg$^2$.
Since these observations were carried out as part of a proposal to observe 270 deg$^2$ of SDSS Stripe 82, the 
details will be provided in a later publication (Mooley et al. 2015, in prep).
In this paper, we will be interested only in the region bounded in right ascension by 329.127\arcdeg~and 353.158\arcdeg, and in 
declination by -1.132\arcdeg~and +1.167\arcdeg.
A summary of the fourth epoch of observations is given in Table~\ref{tab:radio_observations}.

\subsection{Optical Observations}\label{sec:survey:optical_observations}
\begin{figure}[htp]
\centering
\includegraphics[width=3.4in]{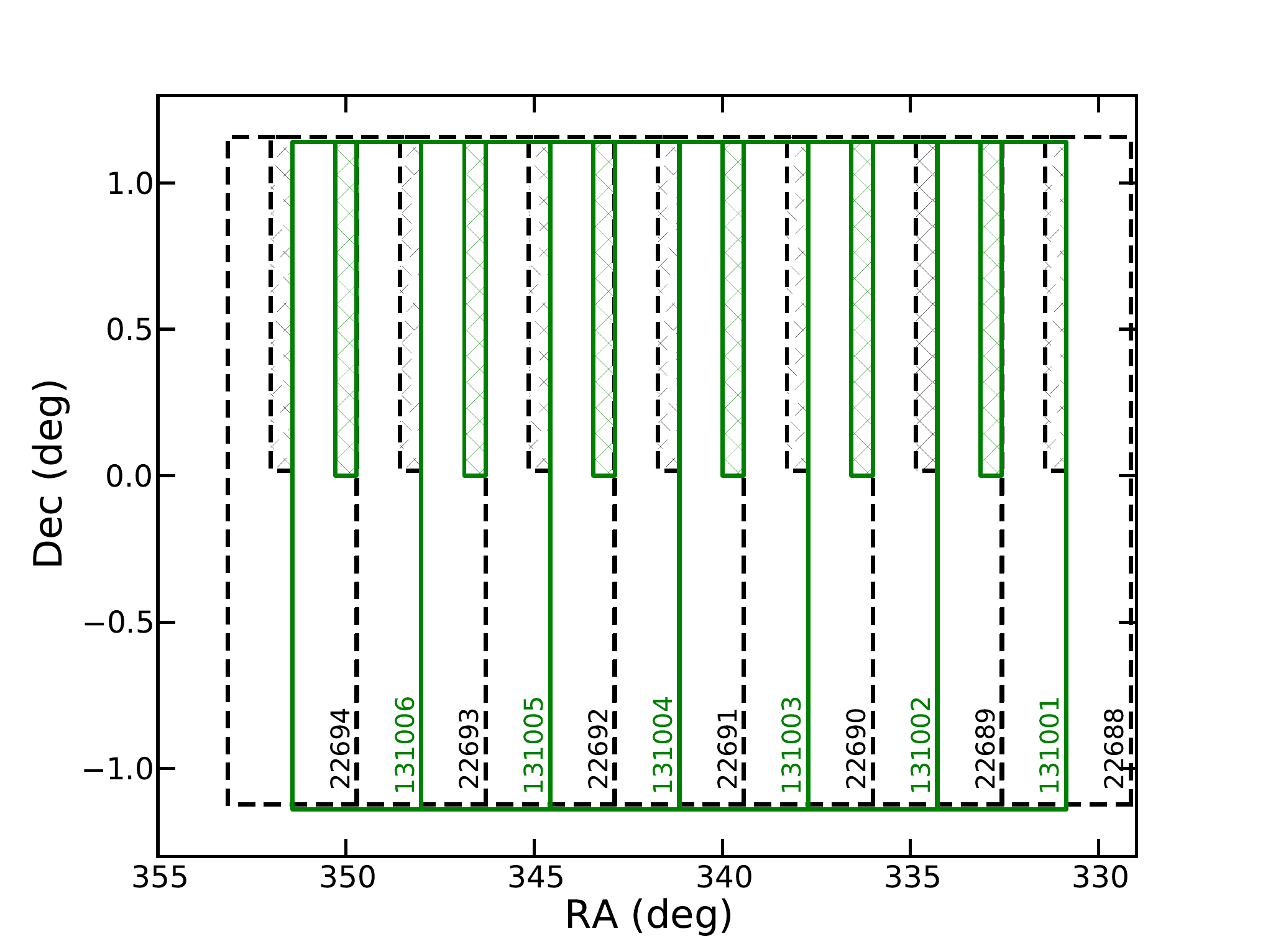}
\caption{Setup of pointings for the PTF survey. The dashed and solid lines indicate the footprint of the different pointings.
The PTF field IDs for the pointings are shown at the bottom right corner of the respective footprints.
The shaded regions indicate the missing CCD.}
\label{fig:optical_obs_setup}
\end{figure}

The optical survey, designed to be contemporaneous with the radio survey, was carried out with the 1.2-meter (48-inch) Samuel Oschin Telescope 
at the Palomar Observatory as part of the Palomar Transient Factory \citep[PTF; ][]{rahmer2008,rau2009,law2009}.
PTF uses a large field camera (CFH12k) consisting of a $6\times2$ array of $2048\times4096$ pix$^2$ CCDs out of which one CCDs are inactive. 
The camera subtends $\sim$7.2 deg$^2$ on the sky.
The 50 deg$^2$ survey region was covered in 14 pointings with each pointing overlapping by about 50\% with its adjacent one (necessitated by one inactive CCD).
The footprints of the 14 pointings (fields) are shown in Figure~\ref{fig:optical_obs_setup}.

\begin{figure}[htp]
\centering
\includegraphics[width=3.4in]{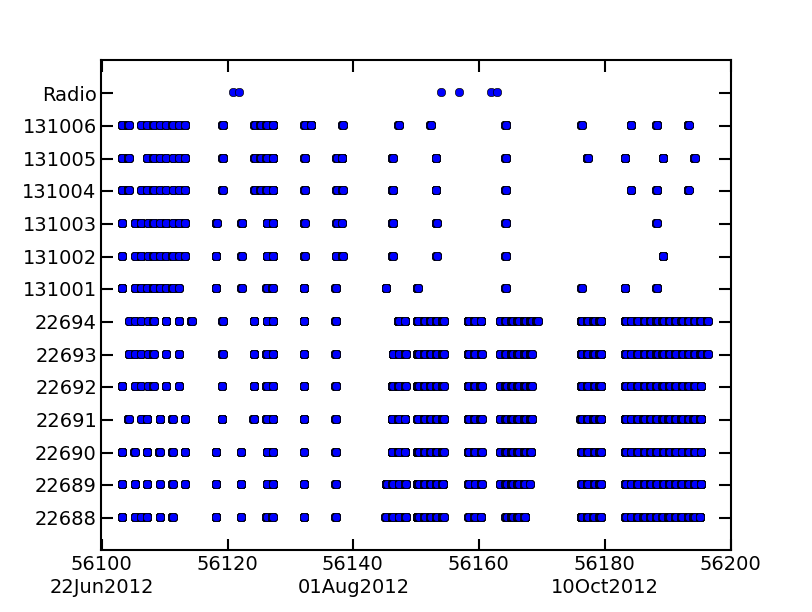}
\caption{Log of PTF observations. 
MJD / Gregorian date is on the x-axis and PTF field ID is on the y-axis.
Each point corresponds to an observation. 
The first three Jansky VLA epochs (Table~\ref{tab:radio_observations}) are shown at the top for reference.
See \S\ref{sec:survey:optical_observations} for details.}
\label{fig:optical_obs_log}
\end{figure}

The PTF observations of the 50 deg$^2$ region were carried out between 25 June 2012 and 25 September 2012.
The survey was done primarily in the R-band, but g-band observations are available for some of the nights.
The Stripe 82 fields were dynamically queued with the other PTF program fields, such that the frequency of observing was determined by the priorities of different PTF fields 
and the weather.
As a result, each PTF field was observed a maximum of five times each night.
The log of PTF observations of Stripe 82 carried out as part of our coordinated program is shown in Figure~\ref{fig:optical_obs_log}.
Standard exposure time per frame is 60 seconds, yielding 3-sigma limiting magnitudes of 20.5 and 21 in the R- and g-band respectively.

\section{Radio Data Processing}\label{sec:radio_processing}

With the new wideband system in place for the Jansky VLA, the data rates are quite large, and this poses a computational challenge in terms of rapid data processing 
needed for triggered transient follow-up.
The size of each of our observation blocks (listed in Table~\ref{tab:radio_observations}) is about 250 GB, and to enable near-real-time data processing, 
dedicated computational hardware was set in place at the NRAO in Socorro.
In general, the calibration, source cataloging and transient search was carried out over a single cluster node at Socorro, and the imaging was distributed across multiple (up to six) cluster nodes.
All this processing was done using a custom-developed, semi-automated AIPSLite/Python-based pipeline developed at Caltech \citep[e.g.][]{bourke2014}.
Details of the near-real-time data processing are given in \S\ref{sec:radio_processing:NRT}.
After the completion of all observing epochs, a more detailed calibration and imaging of the raw data was carried out using CASA 4.1\footnote{http://casa.nrao.edu/. 
Although calibration and imaging was seen to be faster in AIPSLite than in CASA, the latter was preferred due to the user-friendliness of plotting the data 
and the availability of wide-band wide-field imaging algorithms. 
Through our tests, no significant difference is seen in the quality of data calibrated using either software.} (\S\ref{sec:radio_processing:final}).

\subsection{Near-Real-Time Processing}\label{sec:radio_processing:NRT}
For each observing block, the data were downloaded onto the NRAO lustre file system (typical download time of 20 min) 
immediately after they appeared on the VLA archive (only several minutes lag after the completion of the observations).
Next, the raw data was calibrated on a single cluster node using AIPS/Obit tasks accessed via ObitTalk, ParselTongue and AIPSLite.
The calibration part of the pipeline follows the procedures outlined in Chapter E of the AIPS Cookbook.
The first three spectral windows (SPWs), spanning 2.00--2.38 GHz in frequency, and as well as the last three SPWs, spanning 3.62--4.00 GHz, were completely flagged.
Calibration as per the prescription is carried out on a per-SPW basis and almost always failed for many of these SPWs.
This was caused by strong satellite-induced RFI in these bands which was not automatically removed by RFLAG.
Therefore, to speedup the data processing and meet the goal of rapid follow-up, we flagged these SPWs.
For the fourth epoch, we recognized that the first SPW was recoverable with minor manual flagging and the entire SPW was not flagged.

Post-calibration imaging was carried out somewhat differently for the first three epochs and the fourth epoch, since different observing modes were used.
For the first three epochs, the derived calibration was applied to each pointing during imaging.
The imaging stage of the pipeline was distributed over 16 cores across two cluster nodes such that, at any point in time, 16 pointings were being imaged simultaneously using the AIPS task IMAGR.
CLEANing was done with 800 iterations per pointing and natural weighting.
For each pointing, the CLEANed $4096\times4096$ pix$^2$, 0.58 \arcsec pix$^{-1}$ images, uncorrected for the primary beam to get a uniform rms noise across the image, were 
sub-imaged to $2048\times2048$ pix$^2$ using the AIPS task SUBIM.
After the imaging of each observing block, the 485 single-pointing sub-images were copied over to a local machine at Caltech, where the source cataloging and transient detection part of the pipeline was run.
The thermal noise in each pointing is expected to be $\sim$50~$\mu$Jy, and our automated flagging produced data which is within 10\% of this value, except for E1 where larger RFI as present.
For source-finding, the use of the {\it sfind} algorithm from MIRIAD was motivated by the short runtime and reasonable completeness and reliability of catalogs \citep{huynh2012,hancock2012,mooley2013}.
We cataloged all sources above the 5$\sigma$ threshold and used these for the variability and transient search described in detail in \S\ref{sec:radio_transients:nrt}.

The imaging for the fourth epoch was carried out with the CASA task {\it clean} after the calibrated data from AIPSLite were exported into UVFITS format and then imported into CASA as measurement sets.
Before the imaging step, we clipped the visibility amplitudes of all SPWs at a threshold determined by the SPW with the least RFI.
For each SPW, a RFI-proxy value was calculated as the sum of the mean and three times the standard deviation of the scalar-averaged amplitude.
The clipping amplitude was then chosen to be the least of the RFI-proxy values.
We carried out tests on several pointings to ensure that this clipping threshold was robust and did not alter the flux densities and spectral indices of sources.
After the clipping of visibility amplitudes, all 4-second-long scans, which have unique phase centers, were imaged individually.
Imaging was distributed over 60 cores across five cluster nodes such that, at any given time, 60 pointings were being imaged simultaneously using the CASA task {\it clean}.
CLEANing was done with 500 iterations per scan and natural weighting with a {\it cyclefactor}\footnote{The {\it cyclefactor} parameter determines the 
number of minor cycles carried out between successive major clean cycles. 
A large {\it cyclefactor} value thus minimizes the confusion between sidelobes and true sources during minor cycles that may sometimes result in 
strong line or spike-shaped imaging artifacts.} of 4.5 chosen to mitigate imaging artifacts.
A common 3\arcsec~circular restoring beam was chosen to facilitate the co-addition of scans in the image plane.
$2800\times2800$ pix$^2$, 0.75\arcsec~pix$^{-1}$ images, uncorrected for the primary beam, were produced and then sub-imaged to $1700\times1700$ pix$^2$ using the image analysis toolkit in CASA.
1700 pixels corresponds to 20\% of the primary beam at 3 GHz.
The CASA images were exported to FITS format and combined into a mosaic using the AIPS task {\it FLATN} after correcting for the primary beam.
The mean rms noise is $\sim$95 $\mu$Jy while the theoretically expected value is 80 $\mu$Jy.
The increased rms noise is likely due to low-level RFI and joint-deconvolution not being incorporated during the CLEANing process.
Sources beyond 5$\sigma$ were found using RMSD and SAD tasks in AIPS to get reasonably complete and reliable catalogs \citep{mooley2013}.
Once all the E4 observations were complete, the images were combined using the AIPS task {\it COMB} to make a final E4 mosaic.
This mosaic was used to carry out a deeper variability and transient search with respect to epoch E2, the FIRST survey, and the VLA-Stripe 82 survey (\S\ref{sec:radio_transients}).

Since computing time required for the various stages of data processing and transient search are critical for any near-real-time search, we briefly describe this aspect here.
Data I/O is the rate-determining factor in our near-real-time data processing. 
As a result, the processing time is a strong function of the percentage of free disk space on the NRAO lustre file system at Socorro which hosted the data-processing for our survey.
For the first three epochs, the lustre had 75\% free disk space, and the calibration, imaging, cataloging and transient search for each observing 
block, covering 25 deg$^2$ over 485 pointings or scans and amounting to 250 GB of visibility data, required about 4.5 hours, 0.5 hours, 0.5 hours, and 0.5 hours respectively. 
Thus, we were able to detect transient candidates within six hours from the completion of observations.
This fast turnaround allowed same-night triggering of optical follow-up and also rapid broadband radio follow-up, which are crucial for 
understanding transients that evolve on relatively short timescales. 
For the fourth epoch, the lustre had 15--20\% free disk space and imaging problem was much more lengthy owing to the 2025 scans to be imaged.
The calibration and imaging for each observing block within E4, which had 250 GB of visibility data and covered a 22.5 deg$^2$ area on the sky, required 12 hours and 3 hours respectively, 
while the transient search was completed in 1 hour.

\subsection{Final Processing}\label{sec:radio_processing:final}

During the near-real-time processing, the data editing and source identification (\S\ref{sec:radio_processing:NRT}) was not done optimally in order to enable rapid follow-up of 
interesting radio transient candidates.
Also, some problems with a small subset of the data were discovered after the near-real-time processing.
Specifically, 1) the automated RFI algorithm and manual flagging likely excised both
terrestrial interference and valid astronomical data, 2) the gain
calibration for several pointings was affected by strong satellite signals,
and 3) a subset the initial calibrated data suffered from a systematic phase distortion (\S\ref{sec:radio_processing:final:imaging}).
Thus, following the completion of the observing epochs E1, E2 and E3, we carried out a more detailed processing of the raw data using CASA.
Epoch E4 was not subjected to final detailed processing; calibrated data from the near-real-time processing was directly used for analysis.

\subsubsection{Calibration}\label{sec:radio_processing:final:calibration}
The final calibration was carried out using the NRAO calibration pipeline (pipeline version 1.2.0 implemented in CASA 4.1), 
modified to accommodate manual flagging and additional calibration diagnostic plots, and to bypass Hanning smoothing.
The calibration was done with two iterative runs of the pipeline.
Diagnostic plots for the flux and gain calibrators (plots of the gain calibration tables and calibrated phases and amplitudes) from the initial run were used to determine visibilities 
with incorrect amplitude and/or phase calibration or bad calibrator data.
Antenna-, baseline-, correlation-, SPW-, and time-based manual flagging of the flux and gain calibrators was incorporated in the second run to remove these visibilities.
Especially, five SPWs having frequencies between 2.12--2.38 GHz and 3.62--4.00 GHz, for which the amplitude and phase calibration failed in most instances, were manually flagged before calibration.
After the second iteration of the NRAO calibration pipeline followed by imaging, two issues persisted: 
1) the amplitude gain solutions were not stable as a function of time likely due to RFI-induced non-linearities in the signal chain (see \S\ref{sec:radio_processing:final:rfi}), and 
2) Y-shaped imaging artifact in the first 95 scans of the 24Jul2012 epoch, indicating bad raw phase data for the gain calibrator (see \S\ref{sec:radio_processing:final:imaging}).

\subsubsection{RFI}\label{sec:radio_processing:final:rfi}
Since Stripe 82 is close to the Clarke belt, radio observations are prone to severe RFI from satellites in geostationary and geosynchronous (GSO) orbits.
The RFI in the frequency range 3.62--4.00 GHz is low-level in amplitude, but it distorts the phase information quite significantly.
Those data that are irreparably affected are flagged for our final reduction.
GSO satellites seen by the Jansky VLA have not been individually characterized in terms of downlink frequencies and polarizations, and our target data could potentially be affected. 

\begin{figure}[htp]
\centering
\includegraphics[width=3in]{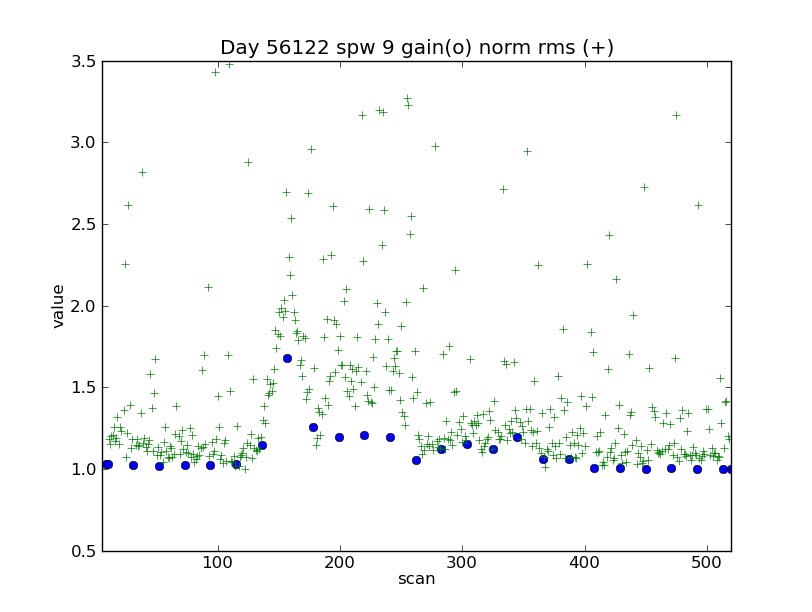}
\includegraphics[width=3in]{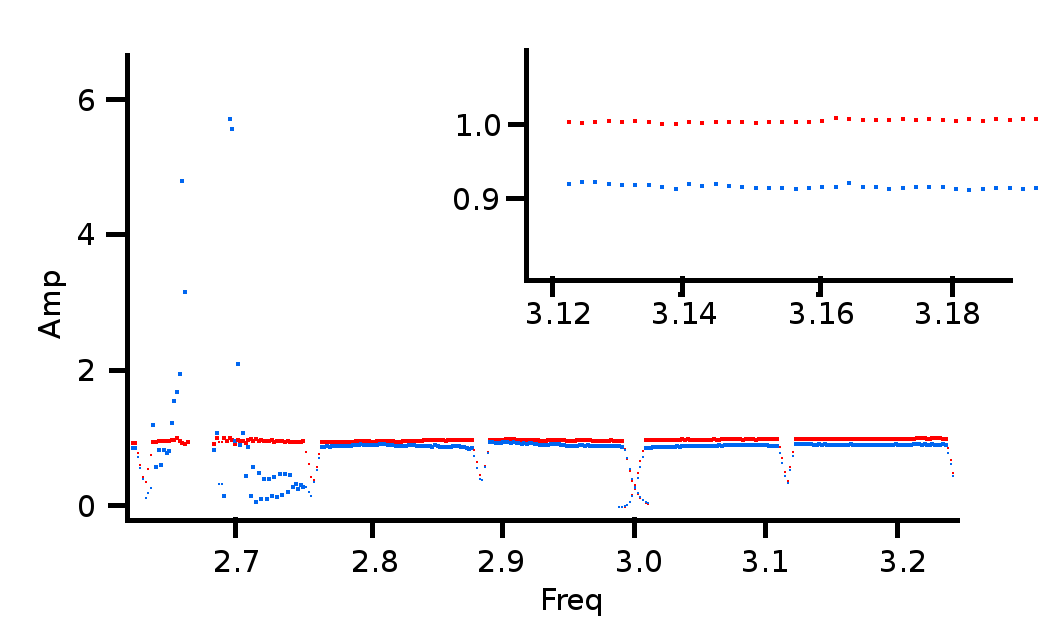}
\caption{Demonstration of gain compression in the data. 
{\it Top}: The inverse of gain amplitudes derived for the gain calibrator J2323-0317 observed during the survey epoch 14Jul2012 (E1R2; blue circles) 
correlate well with the rms noise (green 'plus' symbols) in the pointings (scans). 
The gain amplitudes and the rms have been arbitrarily normalized. 
Note that the scan number is a proxy for time. 
The large rms noise relative to the gain amplitudes between scans 200--250 is likely due to large amount of low-level RFI.
{\it Bottom}: Calibrated and normalized amplitude of J2323-0317 (scan 157) observed during the 23Jul2012 (E3R2) survey epoch. 
RR and LL correlations are shown in red and blue respectively.
There is large RFI in the LL correlation of SPW 5 (between 2.65 and 2.75 GHz), which has affected the gains in the other spectral windows as well. 
The inset shows a zoom-in for SPW 9 (known to be RFI-free), where the gains in the LL correlation are seen to differ from the RR by almost 10\% (the latter is assumed to have the true or expected gain).}
\label{fig:GainAmp}
\end{figure}

The derived calibration also shows instability in the form of sporadically varying amplitude gains from the gain calibrator.
This effect is pronounced in those gain calibrator scans that are severely affected by RFI.
In epochs E1, E2 and E3, where the observations were carried out over the same LST range (\S\ref{sec:survey:radio_observations}), the aberration in gain is reproducible within a specific LST range.
This suggests that the aberrant amplitude gains occur at a particular range of azimuth: between 124\arcdeg~and 255\arcdeg~for region R1 (J2212+0152 used as gain calibrator) and 
between 135\arcdeg~and 220\arcdeg~for region R2 (gain calibrator J2323-0317).
The scans of the southern calibrator, J2323-0317, are more severely affected than those of J2212+0152 due to the proximity of the former to the Clarke belt.
The upper panel of Figure~\ref{fig:GainAmp}~ shows the normalized gain amplitudes for J2323-0317 and the normalized rms noise in the target pointings as a function of the observing scan number (as a proxy for time).
It is evident that the amplitude gains are correlated with the rms noise of the target pointings, suggestive of anomalous gain values.
Our analysis of the calibrated data is complicated by the fact that, although the RFI is localized to typically a single SPW, amplitude 
gains in SPWs that are free from RFI are also significantly affected (irrespective of the baseband in which the RFI is present; lower panel of Figure~\ref{fig:GainAmp}).
We refer to this as ``gain compression''.
The ``compression'' signifies that the amplitude gain values are reduced with respect to their true values and result in increased rms noise in the target fields.
Initial assessment attributes the aberrant gains with non-linearities in the amplification stage in the analog signal chain somewhere upstream from the 
correlator caused by the the high power of the satellite signals.
However, this hypothesis has not been thoroughly tested.

We devised a fix for the gain compression in the gain calibrator scans by altering the amplitude gain calibration table generated by the NRAO calibration pipeline.
Towards the end of each epoch there is a time interval lasting several minutes where the gains are relatively stable and close to unity (to within a few percent).
We therefore applied constant amplitude gains of unity for the duration of each epoch, and for all epochs.
With this correction, the rms noise values for epochs E1, E2 and E3 are more stable with time and are usually within 10\% of the thermal noise.
The spread in the baseline-based amplitude gains in the stable time duration is $<$0.05, and hence we estimate that this correction will affect the true amplitudes of the target sources by $<$5\%.
{\it Given that gain calibrators in the vicinity of the Clarke belt are susceptible to gain compression, we have avoided using such gain calibrators for the final CNSS survey.}
Accordingly, there is no indication of severe gain compression in the gain calibrator scans from epoch E4.
While the gain compression in the gain calibrator scans has a relatively straightforward workaround, the effect of gain compression on the target pointings, if present, is much more challenging to deduce.
We investigate the possible issue of gain compression in target fields in \S\ref{sec:radio_transients}.

\subsubsection{Imaging}\label{sec:radio_processing:final:imaging}
\begin{figure}[htp]
\centering
\includegraphics[width=3.4in]{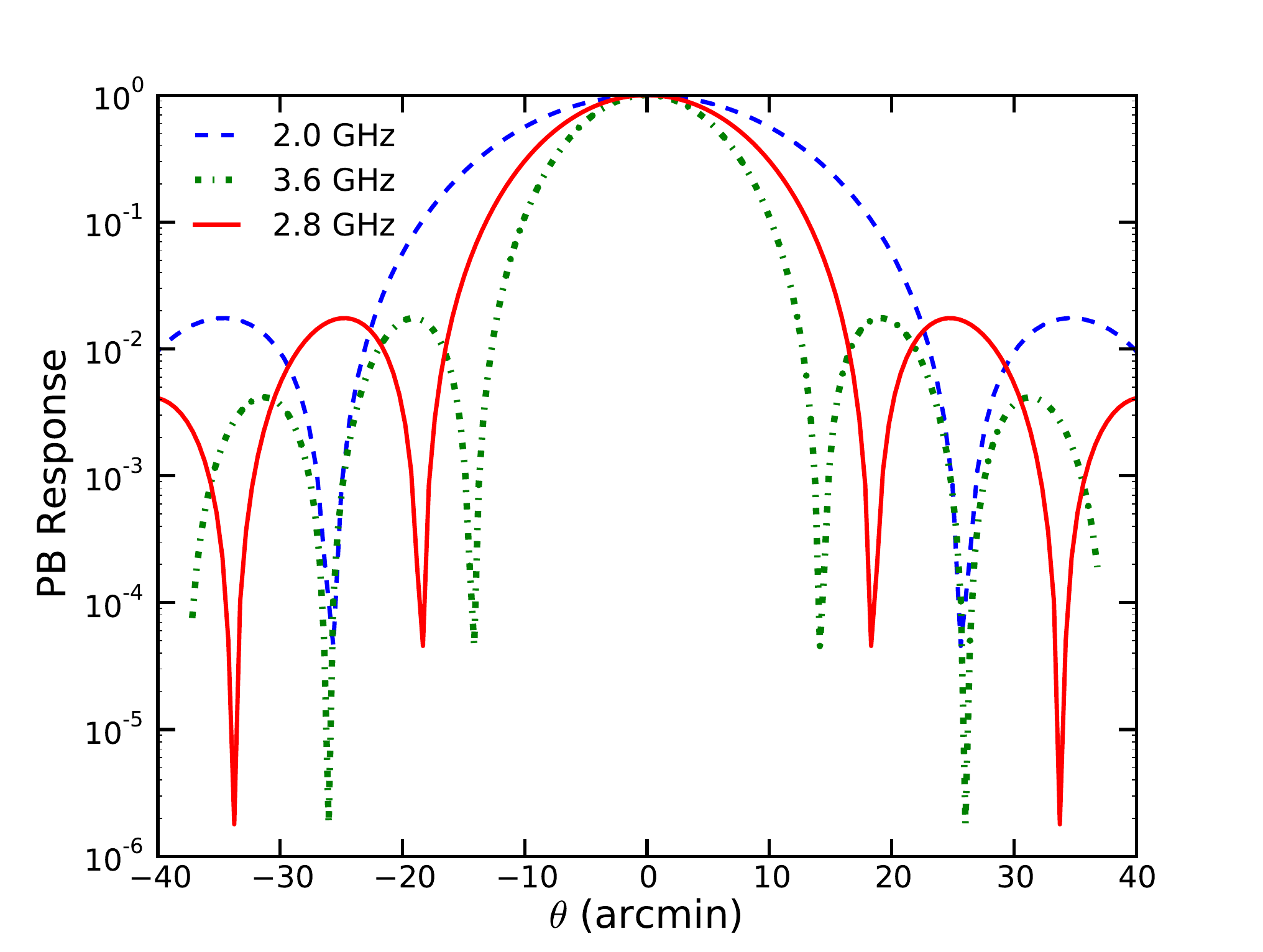}
\caption{The primary beam response at the lowest (blue dashed line), highest (green dot-dashed) and reference (solid red) frequencies.}
\label{fig:PB}
\end{figure}

\begin{figure}[htp]
\centering
\includegraphics[width=3.4in]{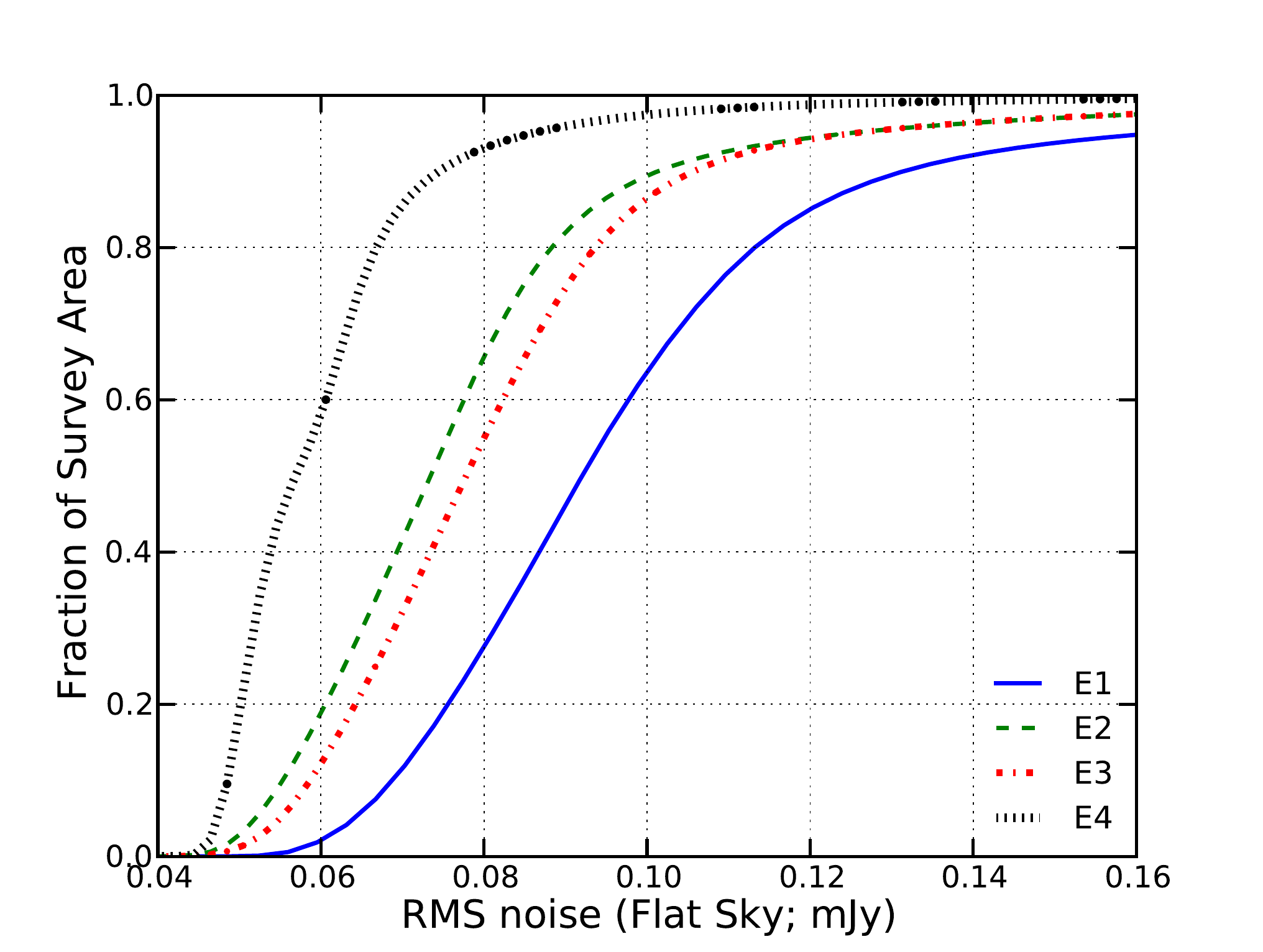}
\caption{The survey area having the rms noise (flat sky) less than or equal to the given sensitivity.}
\label{fig:sensitivity}
\end{figure}

After calibration of the data using the NRAO pipeline, we clipped the visibility amplitudes of all SPWs in the manner described in \S\ref{sec:radio_processing:NRT}.
We then derived and applied a single-round phase-only self calibration solution to every pointing containing a source brighter than 10 mJy in the near-real-time image.
The model for the self calibration was constructed by making a preliminary image with 100 clean iterations using the CASA task {\it clean}.
Self calibration solutions were allowed to break at spectral window and scan boundaries.
Each self calibrated pointing was then imaged by cleaning within the 1\% power point of the primary beam of the lowest frequency, 2.0 GHz, 
using natural weighting in the {\it clean} task in CASA, choosing a 
cellsize of 0.6\arcsec, and 500 clean iterations to minimize clean bias. 
A single Taylor term ($nterms=1$) was used, which was sufficient to achieve the desired dynamic range $\sim10^3$.
We also chose a {\it cyclefactor} of 4.5 to minimize imaging arifacts.
Basic quality checks were made by visually inspecting the images for each pointing.
Using Briggs weighting with a {\it robust} parameter of zero, we re-imaged those pointings in which imaging artifacts persisted.
All the 4800$\times$4800 pix$^2$ cleaned single-pointing images at 2.8 GHz were then sub-imaged to 2400$\times$2400 pix$^2$.
2400 pixels corresponds to 42\%, 15\% and 1\% of the power point of the primary beam at 2.0 GHz, 2.8 GHz and 3.63 GHz respectively (Figure~\ref{fig:PB}).
The median synthesized beam obtained for all three epochs is $3.1\arcsec\times2.0\arcsec$, except for E1R1, where the median synthesized beam size is $3.8\arcsec\times2.1\arcsec$ likely due to increased RFI.
The mean rms noise per epoch is given in Table~\ref{tab:radio_observations}.
The cumulative plot for rms noise (flat sky) as a function of survey area is shown in Figure~\ref{fig:sensitivity}.
For epochs E2 and E3, more than 90\% of the survey area has rms noise lower than 105 $\mu$Jy, while for E4 it is better than 75 $\mu$Jy.
Epoch E1, having rms noise of 130 $\mu$Jy or better over 90\% of the survey area, is severely affected by low-level RFI compared to other epochs.
Note that, for the first three epochs, neighboring pointings are quite far apart and do not contribute significantly to the sensitivity of the overlap region.
Hence, single pointings were imaged separately and treated independently during the source finding step.

\begin{figure}[htp]
\centering
\includegraphics[width=1.68in]{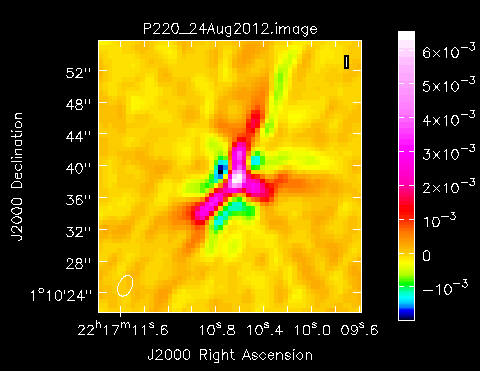}
\includegraphics[width=1.68in]{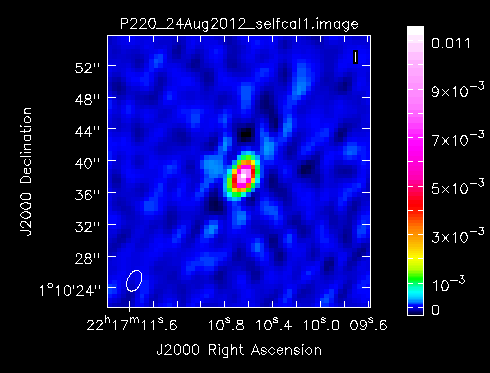}
\caption{The Y-shaped imaging artifact discovered in the first 1.5 hours of observations for the 24Aug2012 (E3R1) epoch.
The image of a bright point source before (left) and after (right) the calibration described in \S\ref{sec:radio_processing:final:imaging}.
The field of view in both panels is identical and equal to 35\arcsec. 
In the left panel, the colorbar runs from $-2.0$ mJy to $+6.5$ mJy, while in the right panel it runs from $-0.5$ mJy to $+12.0$ mJy.}
\label{fig:Y}
\end{figure}

For the 24Aug2012 (epoch E3, region R1) epoch, the first 1.5 hours of observations have been affected by a systematic phase distortion.
About 80 target pointings and six phase calibrator (J2212+0152) scans have been thus affected.
When the calibrated data is imaged, the phase distortion manifests as a three-pronged artifact such that 
point sources appear extended along three axes characteristic of the Y-shape of the VLA configuration (left panel of Figure~\ref{fig:Y}).
We looked at the phase gain solutions per antenna derived from one of the affected phase calibrator scans and plotted it as a function of x- and 
y-positions (east-west and north-south) of the respective antennas in the array.
The center of the array has higher gain phase (degrees) than the arms, and along the arms, it decreases monotonically, in general.
This is equivalent to a bowl-shaped or lenticular phase screen above the array.
The exact cause of the phase distortion is unknown, but it may arise due to RFI, unfavorable weather, the ionosphere, or some internal error in the antenna phase recording.
While the true phases can be restored through self calibration, not all our affected pointings have bright sources to facilitate this. 
Therefore, for the affected pointings, we used the corresponding phases and amplitudes from the 18Jul2012 (epoch E2, region R1) observations as a model for self calibration. 
The image of a bright source after such a calibration is applied, is shown in the right panel of Figure~\ref{fig:Y}.

\subsubsection{Source Catalogs}\label{sec:radio_processing:final:catalogs}
The AIPS task RMSD was used to create an rms image for each individual pointing for epochs E1, E2 and E3, and the resulting background images were supplied as input to SAD.
Source-finding was then carried out with SAD to find sources with SNR$>$5.
For sources present in adjacent pointings, only the catalog entries were retained in which the sources are closest to their respective pointing centers.
Thus, the cataloged sources are within 9\arcmin~(within 40\% of the 2.8 GHz primary beam) from their nearest pointing centers.
Approximately 10$^9$ synthesized beams span the images from our 4-epoch survey, from which we expect 500, 2, and $<$1 noise peaks above 5$\sigma$, 6$\sigma$, and 7$\sigma$ respectively.
While we used 5$\sigma$ source catalogs for the near-real-time transient search, we followed the recommendations of \cite{frail2012}, to filter our new source catalogs and keep only SNR$>$7 sources 
during the final processing.
With the goal of variable and transient search in mind, we further retained only probable point sources\footnote{Sources having the 
ratio of their integrated flux densities to peak flux densities less than 1.5 were selected as probable point sources.
During this step of filtering the source catalogs, we also rejected probable false positives associated with imaging artifacts in the vicinity of bright sources. 
Sources in the catalog that were elongated by a factor of 2.5 larger than the ratio of the major axis to minor axis of the convolved beam were rejected. 
All cataloged objects within 1.6\arcmin~of $>$10 mJy sources were rejected, retaining the $>$10 mJy sources themselves.
In order to filter out false positives with flux densities greater than 10 mJy, all cataloged objects within 1.6\arcmin~of $>$50 mJy sources were rejected, retaining the $>$50 mJy sources themselves.
This reduces our transient search area by less than one square degree, but gets rid of almost most false positives.
Some sidelobes identified as such by eye, were also rejected from the source catalogs.
This false positive rejection step eliminates 438 objects out of 4205 unique objects in our list of probable point sources.
Additional 35 sources were discarded as being resolved after manual inspection of a subset of images from our survey.}.
The resulting catalogs from each epoch were then merged into a single point-source catalog (PSC).

\begin{figure}[htp]
\centering
\includegraphics[width=3.5in]{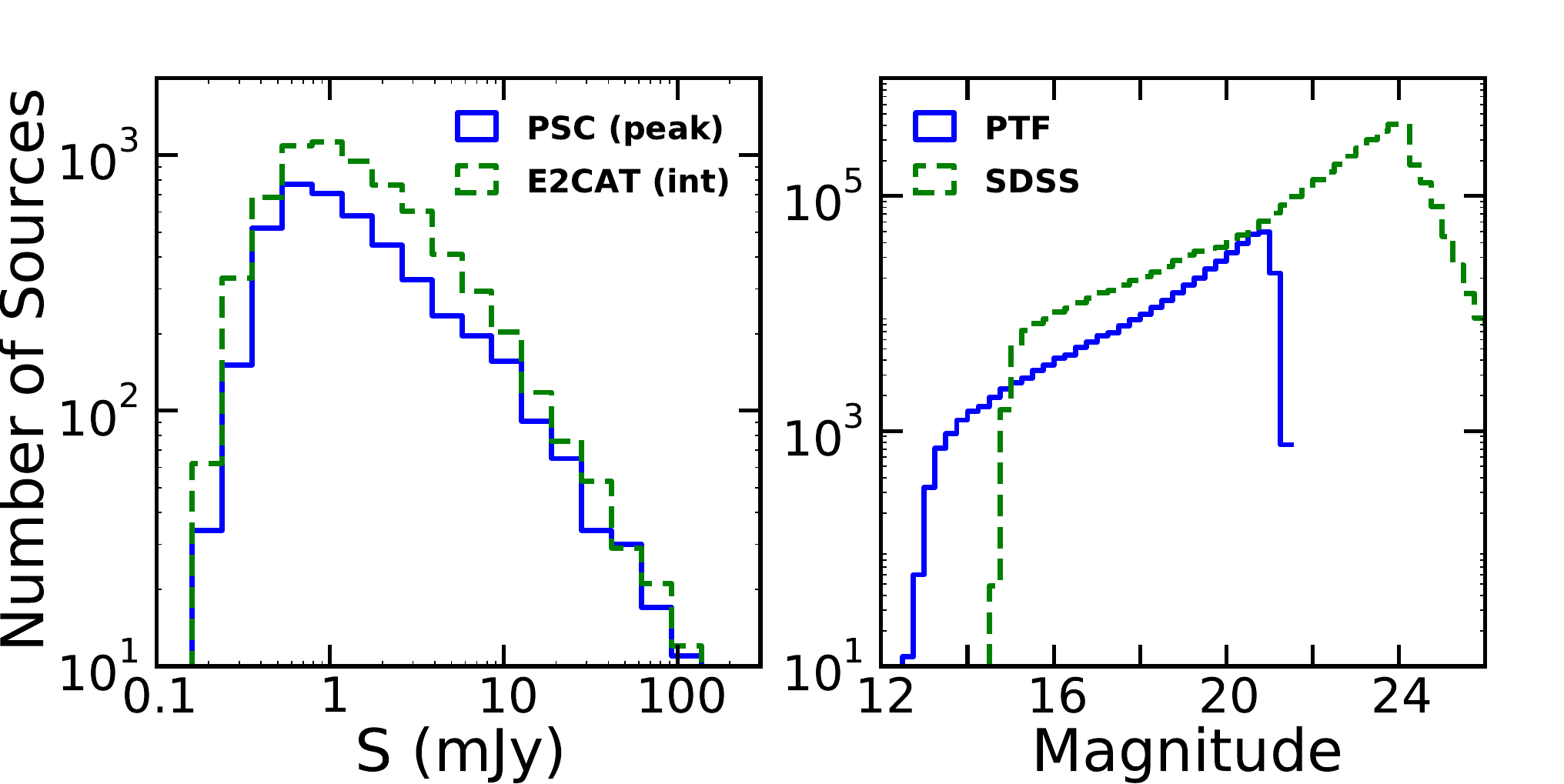}
\caption{{\it Left}: Histograms of the peak flux densities in epoch E2 of sources in the PSC and integrated flux densities for all sources in E2CAT.
{\it Right}: PTF R band and the SDSS (Stripe 82 co-add) r band magnitudes of the optical sources considered in this work. The SDSS histogram between magnitudes 15--21 
is higher than the PTF histogram due to multiple entries of sources in SDSS in this magnitude range.
Below magnitude of 16, the inconsistency is due to the different saturation levels of the SDSS and PTF CCDs.}
\label{fig:S_hist}
\end{figure}

In many cases, it was found that the peak flux densities of sources in the PSC, as reported in the original SAD catalogs, was not in agreement with the peak 
flux densities as seen in the images.
Since accurate peak flux densities are critical for any variability and transient search, we replaced the SAD peak flux density values for all objects in the PSC 
with the corresponding peak pixel values in the single-epoch images.
The uncertainties in the peak flux densities of sources in the PSC were taken directly from their corresponding counterparts in the SAD catalogs or 
by measuring the image rms noise in the vicinity of the sources.
The flux densities of all sources in the PSC and their associated uncertainties were corrected for the primary beam at 2.8 GHz using parameters from the PBCOR task in AIPS.
The PSC thus contains peak flux densities and the uncertainties of 3652 point sources having SNR$>$7 in any one of the four epochs.
The histogram in the left panel of Figure~\ref{fig:S_hist} shows the distribution of flux densities from epoch E2 of sources in the PSC.
We used this PSC for our variability and transient search (\S\ref{sec:radio_transients}).

\subsubsection{Data Release}\label{sec:radio_processing:final:DR}
As part of a preliminary data release\footnote{http://tauceti.caltech.edu/stripe82}, we provide to the astronomical community 
single-epoch images and our PSC for the four epochs of radio observations of the survey region.
Additionally, we release a 5$\sigma$ catalog (E2CAT) of sources from epoch E2.
E2CAT consists of 6846 sources after a preliminary elimination of false positives (probable sidelobes) as done with the PSC, and described in \S\ref{sec:radio_processing:final:catalogs}.
The histogram of the integrated flux densities recorded in E2CAT are shown in the left panel of Figure~\ref{fig:S_hist}.
The final data release, consisting of calibrated UV data, a deep mosaic made from the co-addition of all four epochs, and a deep source catalog, will be done in a later publication (Horesh et al., in prep).

\section{Optical Data Processing}\label{sec:optical_processing}
The PTF near-real-time reduction and transient search was carried out using a pipeline hosted at the Lawrence Livermore Berkeley Laboratory.
Transient search was done by subtracting an older reference (deep co-add) image from the new observations.
For further details on the PTF near-real-time operations, see \S2.1 of \cite{smith2011}.
The Stripe 82 data processed with the IPAC pipeline \citep{laher2014} and photometrically calibrated \citep{ofek2012} was used to make light curves reported in this work.
The processed data and catalogs are available online at the NASA/IPAC Infrared Science Archive\footnote{http://irsa.ipac.caltech.edu/applications/ptf} as part of PTF first data release.
The histogram of the R band magnitudes of PTF sources considered in this work are shown in the right panel of Figure~\ref{fig:S_hist}.
The distribution of r band magnitudes from the SDSS Stripe 82 co-add catalog are also given in Figure~\ref{fig:S_hist}.

\section{Radio Variables and Transients}\label{sec:radio_transients}

In this paper we use the terms ``variables'' and ``transients'' somewhat interchangeably, making note of the fact that the processes listed in 
Table~\ref{tab:transients_summary} are all transient processes.
Here, we use our four survey epochs to carry out two-epoch transient searches on different timescales. 
We are thus probing variability on all timescales that are approximately larger than the duration of each observation (40 seconds per pointing in our case) 
but less than the separation between the two epochs being compared.
Accordingly, in the text below, we refer to a transient search on a ``timescale of X days'' synonymously with ''timescale less than X days``, but note that in 
Table~\ref{tab:transients_summary} we refer to the precise ``timescale of evolution'' since the evolution of the light curves of these transients has been well studied through high-cadence 
targeted observations.

One of the primary goals of our survey was to understand the transient phenomena on timescales of one week, one month, and one year.
For this purpose, we performed a two-epoch comparison of flux densities of sources in our survey using the PSC from \S\ref{sec:radio_processing:final:catalogs}.
Since radio interferometric noise is Gaussian distributed \citep[e.g.][]{condon1998}, we can compare the flux densities of a source between two epochs with the 
statistic $(S_1-S_2)/\sqrt{\sigma_1^2 + \sigma_2^2} = \Delta S/\sigma$ under the null hypothesis that they are drawn form the same distribution 
(and hence the source is non-variable between the two epochs that are being compared).
From statistical theory we know that this quantity is distributed according to the Student-t distribution \citep{student1908,bevington2003}.
We define a source as being a variable if the t-statistic lies beyond the 95\% confidence interval\footnote{For two degrees of freedom this corresponds to a Gaussian 
probability of more than $\pm$4$\sigma$. For the Gaussian distribution, 4$\sigma$ corresponds to a probability of about
1/16,000, while the number of measurements in our variability analysis (few thousand point sources multiplied by four epochs) is about 15,000. 
For our variability analysis we could have used the $\chi^2$ statistic as previous studies have done \citep[e.g.][]{ofek2011,mooley2013}, but 
for cases where the number of degrees of freedom are small, Student's t is preferred.
See also chapter 4 and appendix C of \cite{bevington2003}. }, i.e. if,

\begin{equation}
  V_s = \Big|\frac{\Delta S}{\sigma}\Big| \geqslant 4.3
\label{eqn:var_2epoch}
\end{equation}

The choice of the confidence interval was motivated by the need for high reliability of the variable sources found, while making a possible compromise on the completeness.
This issue of completeness and reliability is taken into account while calculating the rates of variable sources in \S\ref{sec:summary:variability}.
For the two-epoch comparison of variables selected using equation~\ref{eqn:var_2epoch}, we use the modulation index as the measure of variability.

\begin{equation}
  m = \frac{\Delta S}{\overline{S}}
\label{eqn:modidex}
\end{equation}

where $\overline{S}$ is the mean of the flux densities, $S_1\pm\sigma_1$ and $S_2\pm\sigma_2$, in the two epochs being compared.
We note that the definition of modulation index used here is proportional to the one used by \cite{mooley2013}, but larger by a factor of $\sqrt{2}$.
The modulation index is related to the fractional variability, $f_{\rm var}$, by the following equation.

\begin{equation}
  m = 2\times\frac{S_1/S_2-1}{S_1/S_2+1} = 2\times\frac{f_{\rm var}-1}{f_{\rm var}+1}
\label{eqn:fvar}
\end{equation}

\subsection{Near-Real-Time Search}\label{sec:radio_transients:nrt}

Initially, during the near-real-time transient and variability search, we used variability statistic $V_s$ (equation~\ref{eqn:var_2epoch}), and further 
selected sources with fractional variability larger than 50\%.
As soon as a new epoch was observed, it was compared with all previous epochs.
Among $\sim$6000 sources in our near-real-time 5$\sigma$ single-epoch catalogs, we found $\sim$500 to be variables.
We note here, beforehand, that following the final data processing, we found that most of these candidates were falsely identified 
as variable due the RFI-induced amplitude calibration problem identified in \S~\ref{sec:radio_processing:final:rfi}.
 
The shortlist of follow-up candidates was generated from the near-real-time catalogs.
For further filtering of variable candidates, we used multi-wavelength archival data and PTF, and inspected the radio image cutouts by eye.
It turned out that the PTF data were not very useful for this purpose.
As we show in \S\ref{sec:optical_properties}, there is little overlap between variable radio sources and sources that vary in the optical, and almost all these jointly 
variable sources are AGN.
Optical spectra from the SDSS BOSS program \citep{dawson2013} were especially useful in filtering our list of candidates. 
In general, for those radio transients having BOSS spectra, we filtered out sources showing AGN features in their optical spectra. 
WISE colors were also used to eliminate possible AGN.
Preference was given to sources that are undetected in archival radio data.

We also compared our catalogs with those from past surveys at 1.4 GHz, viz. FIRST \citep{white1997} and VLA-Stripe 82 \citep{hodge2011}, and
selected $\sim$10 point sources with implied spectral indices more than 2.5 or less than -2.5 between 3 GHz and 1.4 GHz.
This was motivated by the search for transients such as supernovae and tidal disruption events \citep[Swift J1644+57-like; ][]{zauderer2011,zauderer2013} which evolve on a timescale of years. 
However, most of these candidates turned out to be either GPS sources or flaring AGN (see below).

Our final list of candidates for follow-up consisted of 20 objects that displayed a factor of two or more fractional variability between any two epochs, or were 
new bright sources compared with previous surveys at 1.4 GHz. 

\subsection{Search for Variables and Transients after Final Data Processing}\label{sec:radio_transients:final}


After correcting for the aberrant complex gains during the final data processing (\S\ref{sec:radio_processing:final}), we carried out 
a careful search for variables and transients bearing in mind the possibility of gain compression (\S\ref{sec:radio_processing:final:rfi}) in the target fields.
To maintain consistency during our variability and transient search, we only compared epochs E1, E3, and E4 with epoch E2 in order to probe variability on 
timescales of one month, one week, and 1.5 years respectively.
Note that among the first three epochs, E2 was the least affected by RFI.

For each source in the PSC, we calculated the statistical quantity $V_s$ using equation~\ref{eqn:var_2epoch}.
There appeared to be a small deviation of this statistic from the expected distribution (Student's t distribution).
This may be a result of low-level gain compression in the target fields.
Although the epoch-to-epoch flux densities are not significantly discrepant, we addressed this issue by applying small, relative, corrections to the ratios 
of and the differences in the flux densities of sources between the two epochs being compared.

\begin{figure}[htp]
\centering
\includegraphics[width=3.5in]{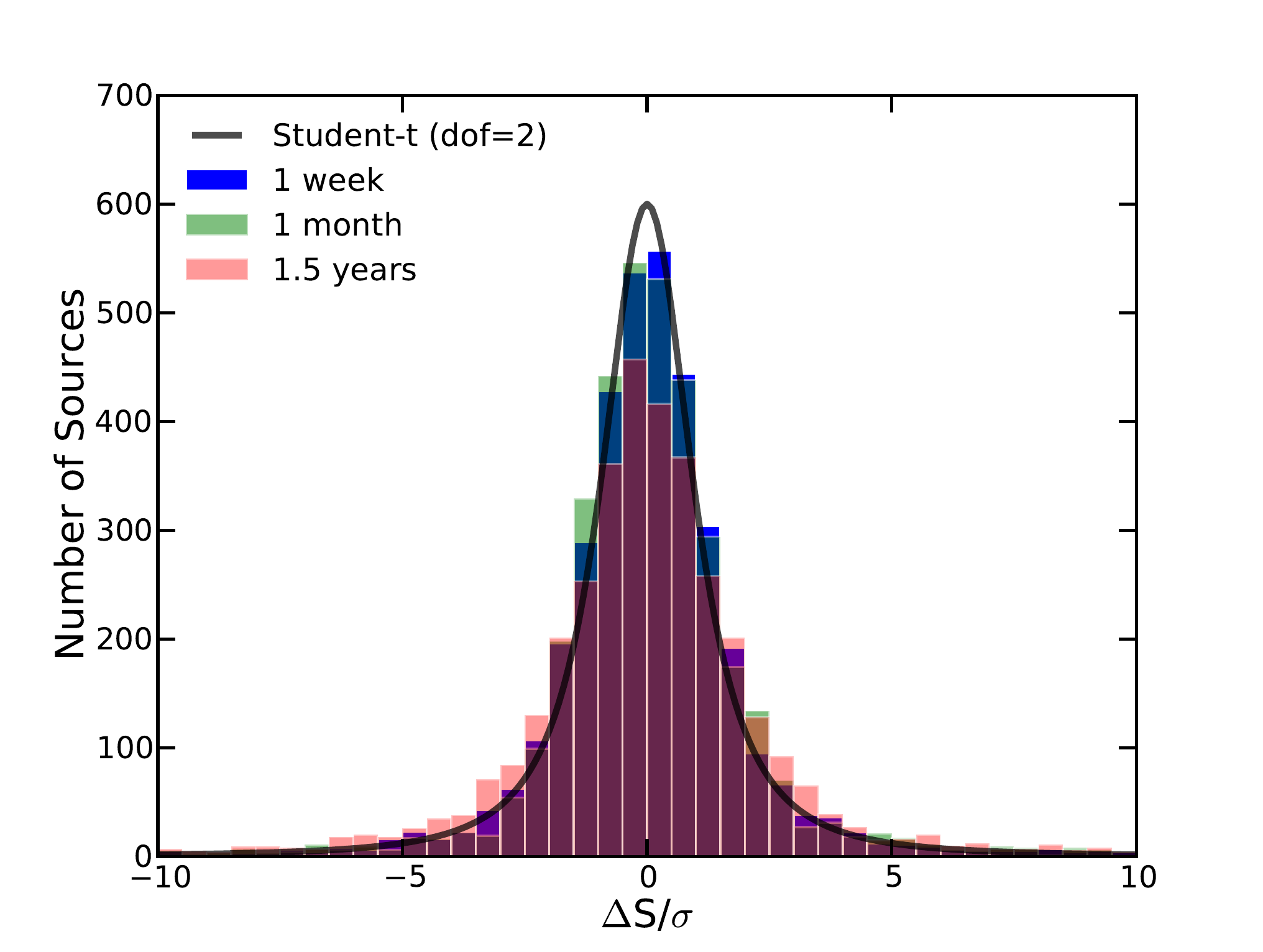}
\caption{A histogram of the variability statistic, $V_s$, for sources in the PSC. 
After the application of the corrections to the flux densities mentioned in \S\ref{sec:radio_transients}, 
$V_s$ agrees with the expected distribution, Student's t.}
\label{fig:varstat}
\end{figure}

For the one-week and one-month comparison, we made multiplicative corrections of 2\%$\pm$3\% (median value with standard deviation) to the ratios of flux densities and 
additive corrections of 20 $\mu$Jy $\pm$ 50 $\mu$Jy to the differences in the flux densities of sources in the PSC.
These corrections were derived for the two-epoch comparisons by plotting the ratios and differences of flux densities of sources as a function of time of observation 
and demanding the ratios and differences to be centered at unity and zero respectively.
During the comparison between epochs E4 and E2, we found it necessary to make two independent sets of corrections to the ratios of flux densities.
In the first set, multiplicative corrections of 3\%$\pm$4\% were seen to be correlated with the declination of the sources.
This discrepancy of flux densities in epoch E4 with respect to epoch E2 is a result of the the OTF survey design and the near-real-time imaging process, and 
will be described in detail in CNSS Paper II.
In short, during the OTF observations carried out for epoch E4, the primary beam moved by 4\arcmin~within every scan; this was not accounted for in the imaging step.
The resulting fractional change in flux density of a source depends on its position in the primary beam.
This effect is manifested as a sinusoidal pattern when the flux density ratios are plotted as a function of source declinations.
The second set of corrections consist of multiplicative corrections of 4\%$\pm$4\% to the ratios of flux densities and may be attributed to gain compression.
Additive corrections of 10 $\mu$Jy $\pm$ 30 $\mu$Jy were applied to the differences in the flux densities between epochs E4 and E2.
With the application of these small, first-order, corrections we ensure to not over-fit the data and remove real variability in flux densities of the sources in the PSC.
The histograms showing the resultant distributions of the variability statistic, $V_s$, is shown in Figure~\ref{fig:varstat}.
It is evident that, after the corrections, $V_s$ agrees well with the Student's t distribution (however, see note on the 1.5 year distribution below).

To determine additional sources of artificial variability, we ran further intensive tests with manual flagging and imaging. 
We found variations of up to a few percent in the flux densities of sources just by choosing different values for the input parameters to the CASA {\it clean} task.
Specifically, these parameters are: cleaning iterations ({\it niter}) and Taylor terms (nterms).
We also noticed variations by changing the values provided to the {\it clipminmax} parameter in the {\it flagdata} task.
Additionally, we know that the flux density absolute calibration (we used 3C48 as the flux density standard calibrator in our survey) is usually accurate 
to only 3--5\% \citep[e.g.][]{ofek2011,thyagarajan2011,mooley2013}.

\begin{figure*}[htp]
\centering
\includegraphics[width=7in]{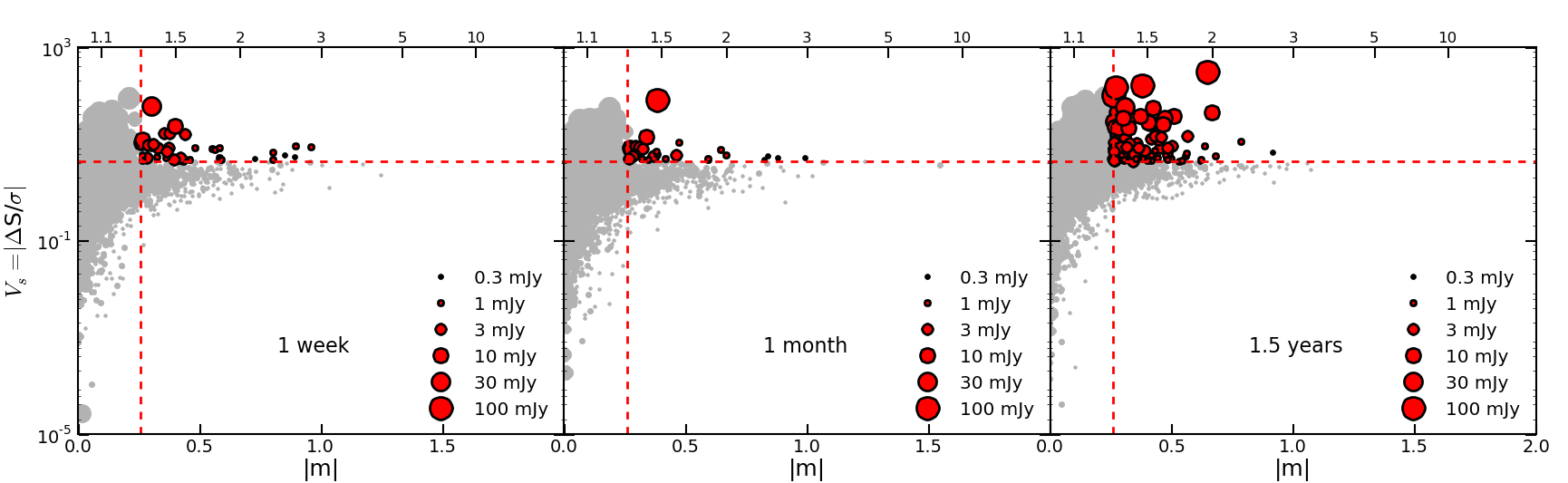}
\caption{The variability statistic, $V_s$, as a function of the modulation index, $m$, for sources in the PSC for the various timescales considered in this work.
The dashed red lines indicate our selection criteria for variables. 
Filled gray circles denote sources that are not variable while red circles have been selected as variables. 
The sizes of the circles indicate the mean flux densities of the sources in the two epochs being compared.
We find 38, 31, and 96 variable sources on timescales of one week, one month, and 1.5 years respectively, amounting to a total of 142 variable sources.
The upper x-axis in each of the three panels represents the fractional variability as given by equation~\ref{eqn:fvar}.
See \S\ref{sec:radio_transients} for details.}
\label{fig:varstat_m}
\end{figure*}

All in all, due to gain compression and other possible causes of artificial variability mentioned above, we adopt a conservative criterion for selection of 
our variable sources: fractional variability greater than 0.3 ($|m|>0.26$).
Our final variability selection criteria are $V_s\geqslant4.3$ and $|m|>0.26$.
The plots of the $V_s$ versus $m$ for sources in the PSC after applying the corrections mentioned earlier in this subsection are shown in Figure~\ref{fig:varstat_m}.

For the sources in the PSC satisfying the variability selection criteria, we manually inspected the image cutouts and their fitted Gaussian 
parameters, viz. major axes ($b$), minor axes ($a$), the peak flux densities ($S_{\rm peak}$) and the integrated flux densities ($S_{\rm int}$), in order to ensure that they are not resolved. 
For sources detected at a high significance level (SNR$\gtrsim$15), we used the criteria: $S_{\rm int}/S_{\rm peak}\lesssim1.1$, $b\lesssim1.1\times${\tt BMAJ}, and 
$a\lesssim1.1\times${\tt BMIN}, where the variables in typewriter text denote the major and minor axes of the synthesized beam. 
For the sources detected at low significance, we relaxed the integrated-to-peak flux density ratio to 1.3--1.5.
This procedure for selecting unresolved is motivated by the standard techniques used for deep radio surveys \citep[e.g. ][]{huynh2005,franzen2015}.
While this approach \citep[c.f.][]{hodge2013} errs on the conservative side, we prefer to identify real variables rather than false positives caused by resolution effects.

We thus find 142 variables among 3652 sources in our PSC, or, in other words, about 3.8\% of the sample has fractional variability more than 30\% on timescales $<$1.5 years.
The number of point sources varying on week, month, and 1.5 year timescales are 38, 31, and 96 respectively, or 1.0\%, 0.8\% and 2.6\% respectively.
These sources are listed in Table~\ref{tab:transients}.
A glance at Figure~\ref{fig:varstat} suggests that the variability statistic, $V_s$, comparing epochs E2 and E4 (probing variability on a timescale of 1.5 years) deviates from the 
Student's t distribution in the tail.
This may be due to some calibration errors from the near-real-time processing that we have not tried correcting for, and therefore the fraction of variables on a 1.5-year timescale, 2.6\%, 
is likely an upper limit.
Note that the variable sources are distributed uniformly on the sky with no pattern indicative of particular region(s) of the sky being affected by calibration errors.

\begin{figure}[htp]
\centering
\includegraphics[width=3.5in]{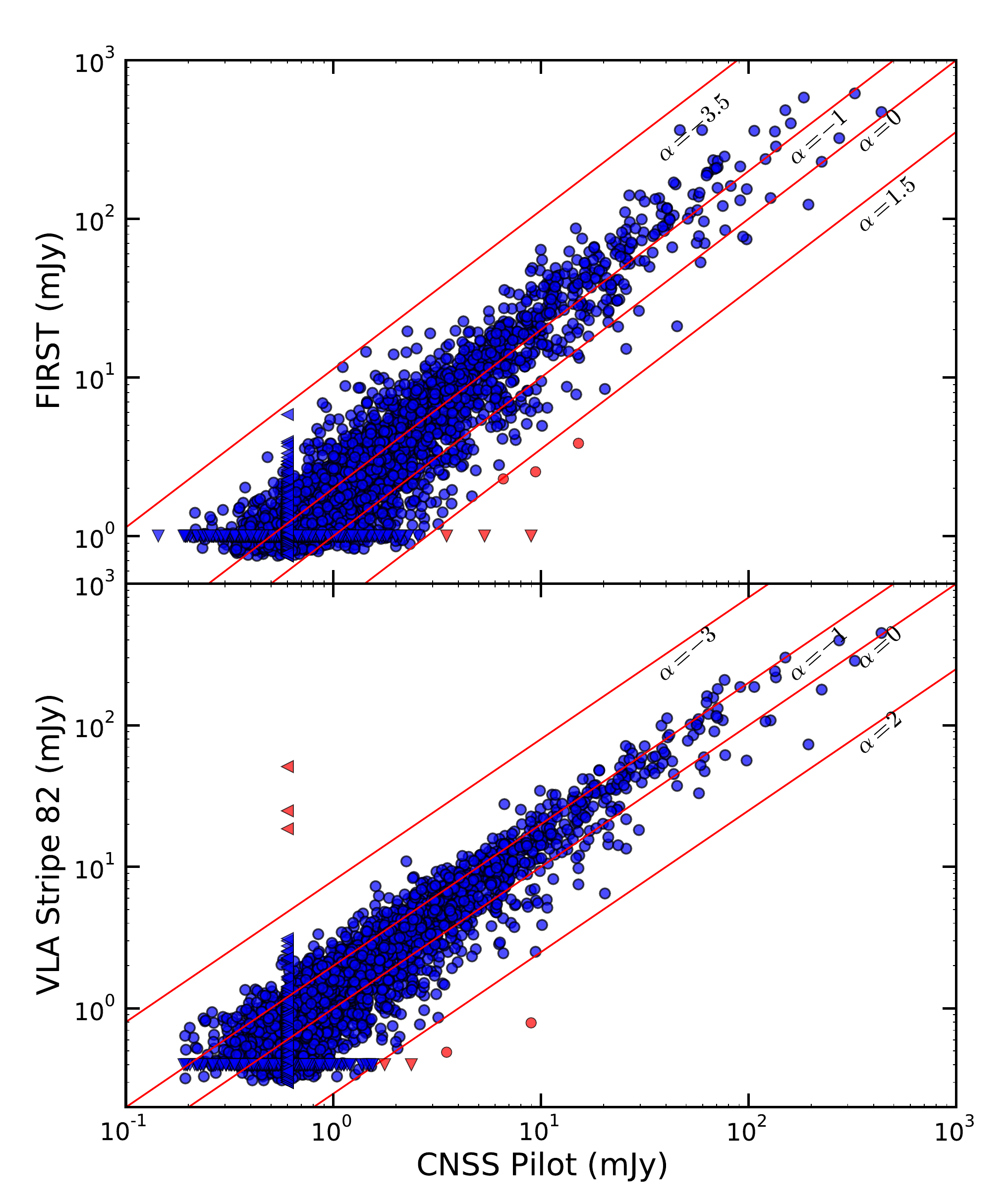}
\caption{A comparison of the 3 GHz peak flux densities of sources in the PSC with the 1.4 GHz peak flux densities from the FIRST (top) and VLA-Stripe 82 (bottom) surveys. 
these two surveys (Figure~\ref{fig:var_CNSS_FIRST_Hodge}).
Triangles denote upper limits to the peak flux density.
Red color denotes sources selected as variable candidates by our search criteria.
See \S\ref{sec:radio_transients:final} for details.}
\label{fig:var_CNSS_FIRST_Hodge}
\end{figure}

Since our 50 deg$^2$ survey region contains rich archival data from the FIRST and VLA-Stripe 82 surveys, we extended our transient search 
to timescales of up to $\sim$20 years.
We compared the 3 GHz peak flux densities of sources in the PSC with the 1.4 GHz peak flux densities from these two surveys (Figure~\ref{fig:var_CNSS_FIRST_Hodge}).
The mean spectral index\footnote{We define the spectral index, $\alpha$, as $S\propto\nu^\alpha$.}, $\alpha_{\rm mean}$, appears to be between $-0.5$ and $-1$, as expected \citep[e.g.][]{randall2012}.
With the view of finding only the extreme variable and transient candidates, we searched beyond implied spectral indices of $-$1$\pm$2.5 (i.e. less than $-3.5$ or greater than 1.5) 
with respect to the FIRST survey, and beyond $-$0.5$\pm$2.5 with respect to the VLA-Stripe 82 survey.
The threshold of $\pm$2.5 in spectral index is somewhat arbitrary; it is motivated by our need for filtering out most of the sources and selecting only a few, extreme, objects.
Our search resulted in 11 unique objects.
Two candidates are resolved out (closely-separated radio lobes) in our data; one candidate is identified as such due to its erroneously absent in VLA-Stripe 82 catalog (but present in the image); 
and one candidate, VTC220456-000147, is identified as a variable also on a timescale of 1 month.
Accordingly, out of the eight genuine variables\footnote{Note that any extragalactic source variable on a timescale even as large as a decade will be a point source 
for an angular resolution of 1\arcsec, given the light travel distance. Here, we have presented only isolated point sources in our final list of variables.}, 
seven have been listed in Table~\ref{tab:transients} in the timescale $<$20 years section.

Below, we discuss five candidates\footnote{We adopt the following naming convention for our variable and transient sources: VTC$hhmmss+ddmmss$, 
where ``VTC'' stands for ``VLA transient candidate'' and the right ascension and declination are both rounded at the least significant digit.} 
in detail, that are meant to be broadly representative of our complete sample of 142 variables and transients.
Additionally, we discuss about two transients found serendipitously.
A summary of all the variable and transient sources found in this work is given in Table~\ref{tab:transients}, and a summary of radio follow-up observations is given in Table~\ref{tab:radio_followup}.

\subsection{Transients On Timescales $<$1 Week}\label{sec:radio_transients:one_week}

\subsubsection{VTC225411-010651: AGN flare}\label{sec:radio_transients:one_week:VTC225411-010651}

\begin{figure}[htp]
\centering
\includegraphics[width=3.4in]{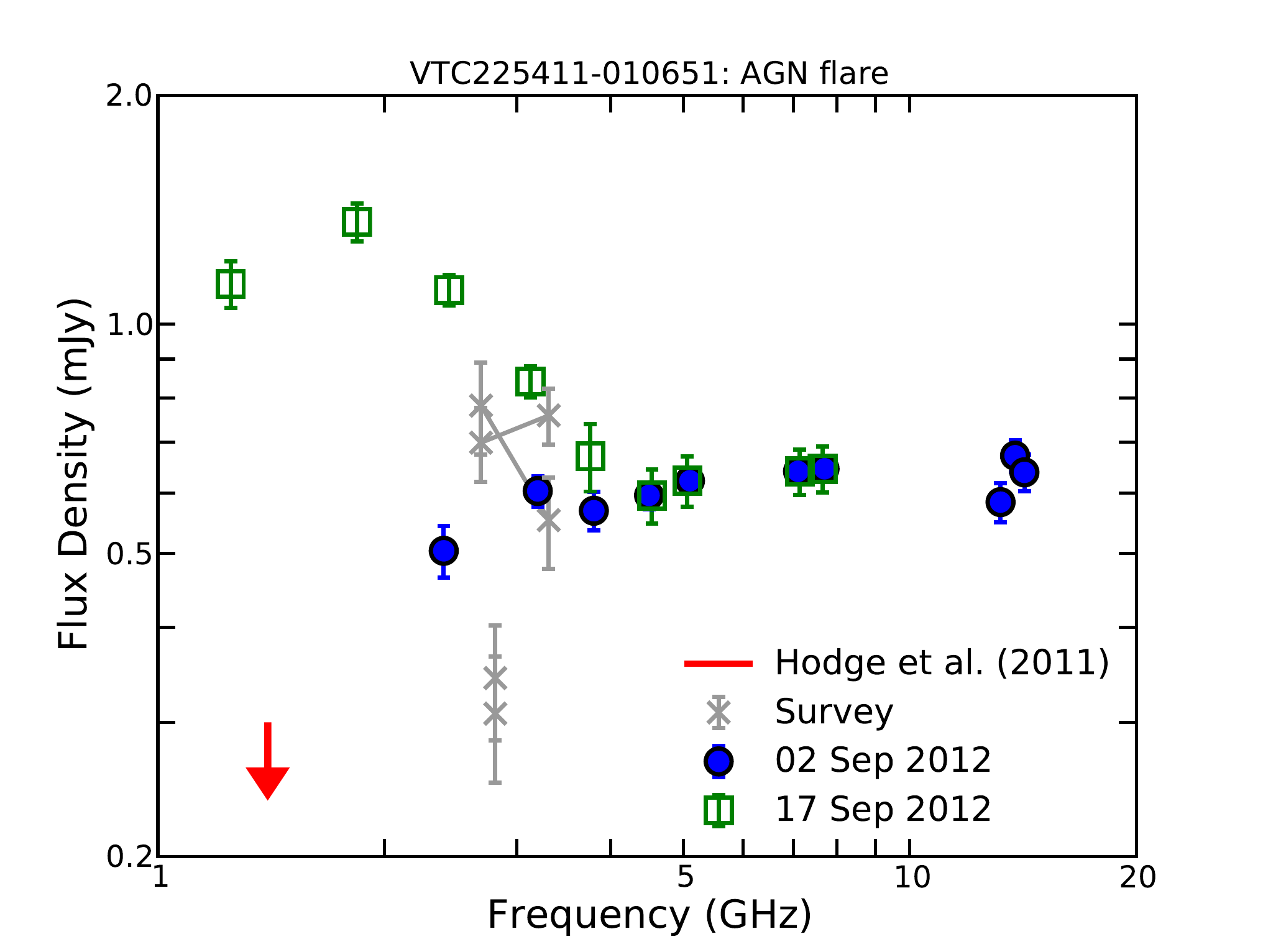}
\caption{Radio continuum spectrum of VTC225411-010651 at two follow-up epochs (blue circles and green squares). 
The 3$\sigma$ upper limit at 1.4 GHz from \cite{hodge2011} is shown in red. 
The measurements from the four survey epochs are shown in grey. 
The first two epochs have spectral indices of $-1.68\pm0.95$ and $0.39\pm0.68$ within the S band respectively and are indicated by grey crosses joined by 
straight lines between the two frequencies used for spectral index measurement. 
See \S\ref{sec:radio_transients:one_week:VTC225411-010651} for details.}
\label{fig:VTC225411-010651_RadioSpectrum}
\end{figure}

This variable source has flux densities 
0.644$\pm$0.063, 0.782$\pm$0.050, 0.300$\pm$0.056, and 0.300$\pm$0.052 mJy in epochs E1 to E4, and in the first two epochs it 
has spectral indices $-1.68\pm0.95$ and $0.39\pm0.68$ respectively at 3 GHz.
At the location of VTC225411-010651, the 1.4 GHz peak pixel values from the FIRST and \cite{hodge2011} surveys are $0.30\pm0.17$ and $0.12\pm0.10$ mJy, respectively. 
To investigate further, we obtained follow-up observations of this object between 1--15 GHz in two epochs (on 02 Sep and 17 Sep 2012; 10 days and 25 days respectively after the epoch E3R2) with the VLA.
The continuum radio spectra are shown in Figure~\ref{fig:VTC225411-010651_RadioSpectrum}.
The first follow-up observation reveals a flat spectrum source in the 2--15 GHz frequency range.
The second observation indicates a spectrum peaked at about 1.5 GHz with the optically thin part having a spectral index of $-0.99\pm0.16$. 
The spectrum is flat beyond 4 GHz.
Sub-mm (100 GHz) follow-up observations on 09 Sep 2012 with CARMA\footnote{All CARMA data were reduced with MIRIAD and VLA follow-up data with the NRAO CASA pipeline 1.2.0} 
gave a non-detection with a 3$\sigma$ upper limit of 2.1 mJy.

There is no optical counterpart in PTF, but a faint SDSS DR7 \citep{adelman2009} source having $r\simeq23.3$ mag. lies 1.8\arcsec ~away from the radio source position of VTC225411-010651.
The photometric redshift from SDSS is $0.64\pm0.14$.
Assuming a 1.4 GHz quiescent flux density of 0.2 mJy, we can calculate the radio-to-optical flux density ratio for the host galaxy to be $log(S_{\rm 1.4 GHz}/S_g)=2.1$, which 
is typical of radio-loud AGN \citep[e.g.][]{padovani2011}.
WISE colors ($W1-W2=0.37\pm0.29$ mag, $W2-W3=3.90\pm0.51$ mag) of the host galaxy are consistent with a LINER / (U)LIRG / spiral galaxy \citep{cutri2012,wright2010}.

The flat spectrum of this source in the first follow-up epoch indicates a jet with unresolved knots.
Assuming a 3 GHz quiescent flux density of 0.3 mJy, we obtain of radio luminosity of $5.3\pm2.5\times10^{23}$ erg s$^{-1}$ Hz$^{-1}$ at the SDSS photometric redshift.
It is thus likely that this source is a low-luminosity radio-loud AGN, in which case, the peaked spectrum in the second follow-up 
epoch can be naturally explained as an intrinsic AGN flaring phenomenon.
The flaring spectrum in Figure~\ref{fig:VTC225411-010651_RadioSpectrum} is reminiscent of the shock-in-jet model that has been extensively used 
to explain flaring in bright quasars \citep{marscher1985,turler2000,fromm2011}.
The strong variability of VTC225411-010651 in the S band between epochs E2 and E3 of our survey indicates that the flares in this AGN evolve on a timescale of less than one week.
We note that current radio variability surveys are primarily focused towards the monitoring of blazars and bright quasars, where the flares at GHz frequencies evolve 
typically on timescales of a few months \citep[e.g.][]{hovatta2008,richards2011}, and that VTC225411-010651 presents the first ever direct evidence of shocks 
propagating down the jet in sub-mJy AGN on a relatively short timescale.

\subsection{Transients On Timescales $<$1 Month}\label{sec:radio_transients:one_month}

\subsubsection{VTC223612+001006: RS CVn flare}\label{sec:radio_transients:one_month:VTC223612+001006}
\begin{figure}[htp]
\centering
\includegraphics[width=3.7in]{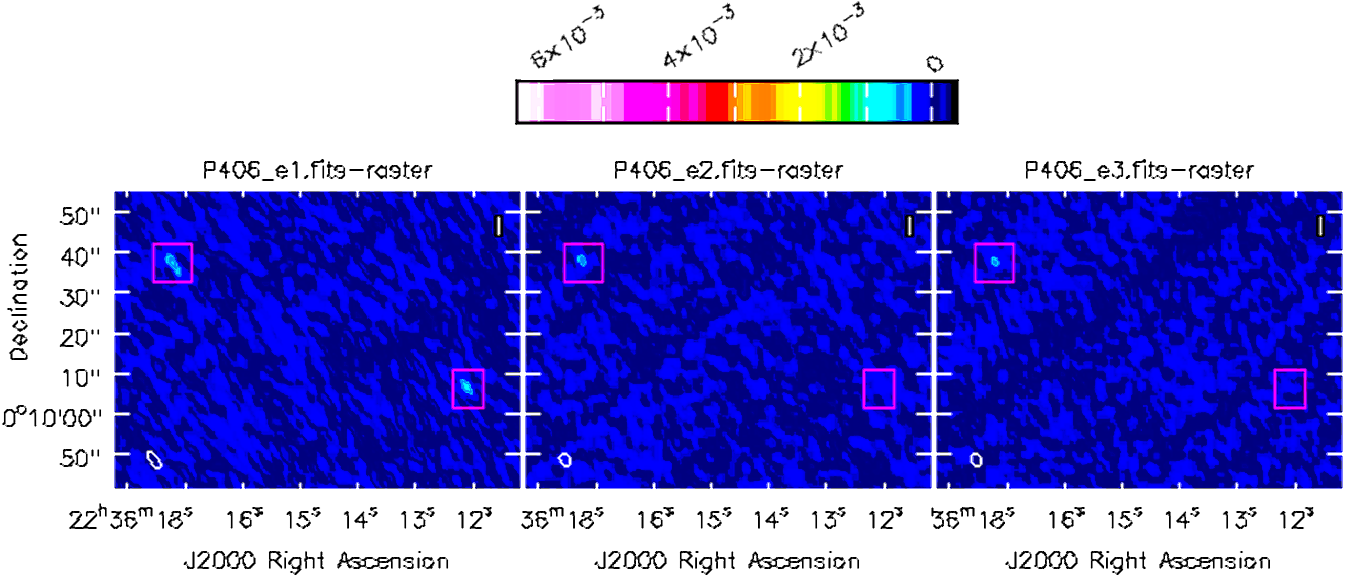}
\includegraphics[width=3.5in]{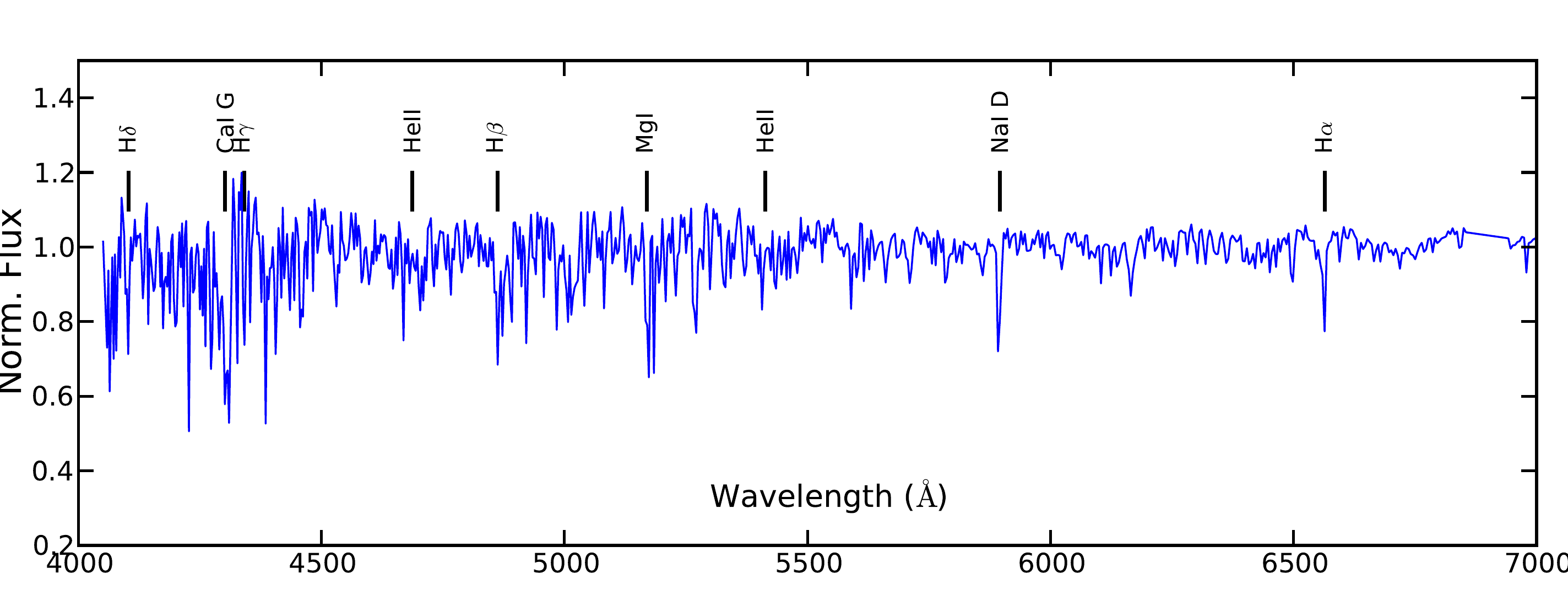}
\caption{{\it Top}: The radio image cutouts from three epochs, E1, E2 and E3, showing VTC223612+001006 and a comparison 0.5 mJy persistent source. 
The transient is detected with an SNR of 9 in the first epoch, and is below SNR of 4 in the other two epochs. 
The colorbar has units of Jy.
{\it Bottom}: The continuum-normalized optical spectrum of VTC223612+001006 observed with ESI at KeckII on 13 Sep 2012. The spectrum has been binned with 20 data points in each bin. We classify this star as a late G to early K subgiant or giant. See \S\ref{sec:radio_transients:one_month:VTC223612+001006} for details. }
\label{fig:1222af}
\end{figure}

VTC223612+001006 is detected as a transient on 13 Jul 2012 (epoch E1, regions R1).
The peak pixel values at the location of the transient in epochs E1--E4 are 0.804$\pm$0.086, 0.271$\pm$0.071, 0.179$\pm$0.073, 0.098$\pm$0.057 mJy respectively.
The image cutouts near the source location for the three epochs are shown in Figure~\ref{fig:1222af}.
A comparison source about 1.5\arcmin ~away from this transient candidate has relatively stable peak flux densities, 0.604, 0.584, 0.551, and 0.488 mJy in epochs E1--E4 
respectively, indicating that the variability of the candidate is secure.
In the FIRST and \cite{hodge2011} surveys, the peak pixel values at the location of the transient candidate are $0.13\pm0.11$ and $0.16\pm0.06$ mJy respectively.

The nearest optical counterpart of VTC223612 is the 10th V-band magnitude star HD 214129 ($\sim$1\arcsec ~away), documented as a K0IV star by \cite{torres2006}, and as a G5/6III star by \cite{kharchenko2009}.
HD 214129 is a known visual binary and possibly a triple stellar system \citep[e.g.][]{mason2001}, and has been detected in the X-rays by ROSAT \citep[1RXS J223612.5+001008; ][]{voges1999}.
HD 214129 is saturated in PTF and SDSS, precluding the study of optical variability of this Galactic radio transient.
We obtained an optical spectrum for this star using the Echelle spectrograph (ESI) at KeckII on 13 Sep 2012, which suggests a spectral type similar to K0IV.
The spectrum does not show any strong emission lines within our spectral coverage between 4000--10000 \AA. 
Figure~\ref{fig:1222af} shows the binned optical spectrum between 4000--7000 \AA.
The narrow absorption line profiles and absence of Lorentzian wings indicate that the star is a subgiant or a giant.
Fitting a blackbody to the photometric data from the SDSS, NOMAD and WISE catalogs gives an effective temperature of $5000\pm200$ K.

If this were a main sequence star, it would have a photometric parallax distance of $d\sim65$ pc, while if it is a giant, $d\sim700$ pc.
A sub-giant can have any value between these extremes.
An independent constraint on the distance comes from the space velocity.
Using the proper motion and radial velocity, $\mu_\alpha, \mu_\delta=127.3, -39.4~\mbox{mas yr$^{-1}$}$ and $v_r=-3.3~\mbox{km s$^{-1}$}$ \citep{hog2000,torres2006}, we can 
calculate the space velocity (heliocentric) for $d\sim65$ pc and $d\sim700$ pc as $v\sim40$ and $v\sim440~\mbox{km s$^{-1}$}$ respectively \citep{johnson1987}.
Typically for stars we expect heliocentric space velocities between 0--100 \mbox{km s$^{-1}$} \citep[e.g.][]{dehnen1998}.
Therefore, we expect $40<d\lesssim160$.
Here, we adopt a distance of 100 pc.

We used a 1 keV collisionally-excited plasma (APEC) model and an absorbing Galactic hydrogen column of $N_H=5.05\times10^{20}~\mbox{cm$^{-2}$}$ (derived from the {\it nh} task in {\it HEASoft}) 
to convert the ROSAT/PSPC countrate from 1RXS (0.23 counts s$^{-1}$) in the 0.1--2.4 keV energy band to an unabsorbed flux of $3.2\times10^{-12}~\mbox{erg cm$^{-2}$ s$^{-1}$}$ using 
WebPIMMS\footnote{https://heasarc.gsfc.nasa.gov/cgi-bin/Tools/w3pimms/w3pimms.pl}.
This corresponds to an X-ray luminosity of $3.8\times10^{30}~\mbox{erg s$^{-1}$}$ at 100 pc.
The X-ray luminosity, the X-ray-to-optical flux ratio of 1.4$\times$10$^{-3}$, and the spectral type all suggest that HD 214129 is an RS CVn-type binary.
Accordingly, we expect the quiescent 3 GHz radio luminosity of the transient to be $L_R \lesssim 10^{15 \pm 1}$ erg s$^{-1}$ Hz$^{-1}$ \citep[][; however, strictly speaking, 
their L$_X$-L$_R$ relationship is valid for 5 GHz flux density]{benz1994,gudel2002}.
From epoch E4 we get the 3$\sigma$ upper limit on the quiescent 3 GHz flux density of this transient as $2.4\times10^{15}$ erg s$^{-1}$ Hz$^{-1}$.
Using a distance of 100 pc, we can estimate the flaring radio luminosity in epoch E1 as $1.1\times10^{16}$ erg s$^{-1}$ Hz$^{-1}$.
From the binarity, the optical spectrum, radio and X-ray luminosities, X-ray hardness ratios from ROSAT, and the X-ray-to-optical flux ratio, we conclude that this radio transient is 
a flare from an active binary system.


\subsubsection{VTC225707-010238: Flaring Type-I QSO}\label{sec:radio_transients:one_month:VTC225707-010238}
\begin{figure}[htp]
\centering
\includegraphics[width=3.5in]{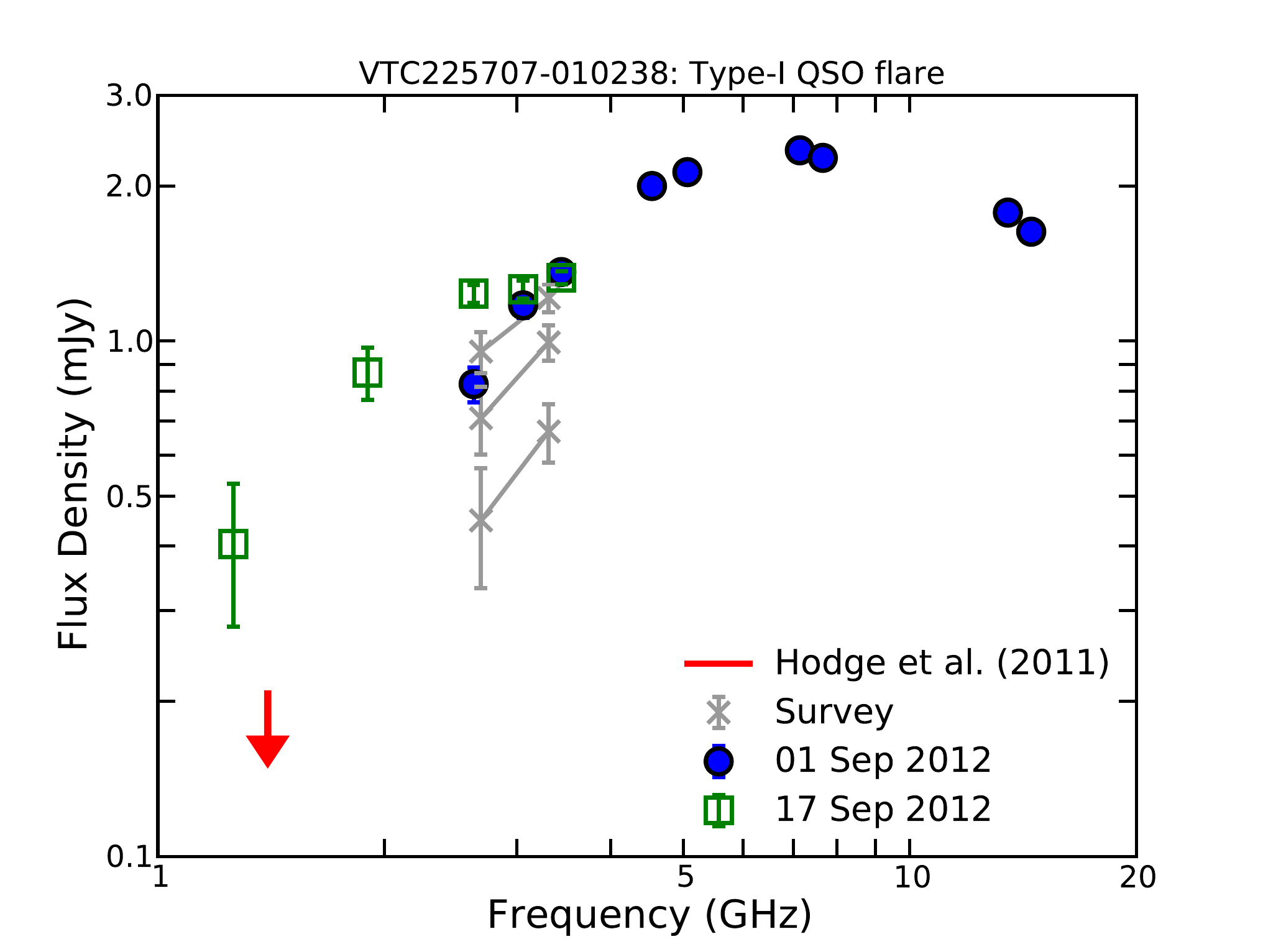}
\includegraphics[width=3.5in]{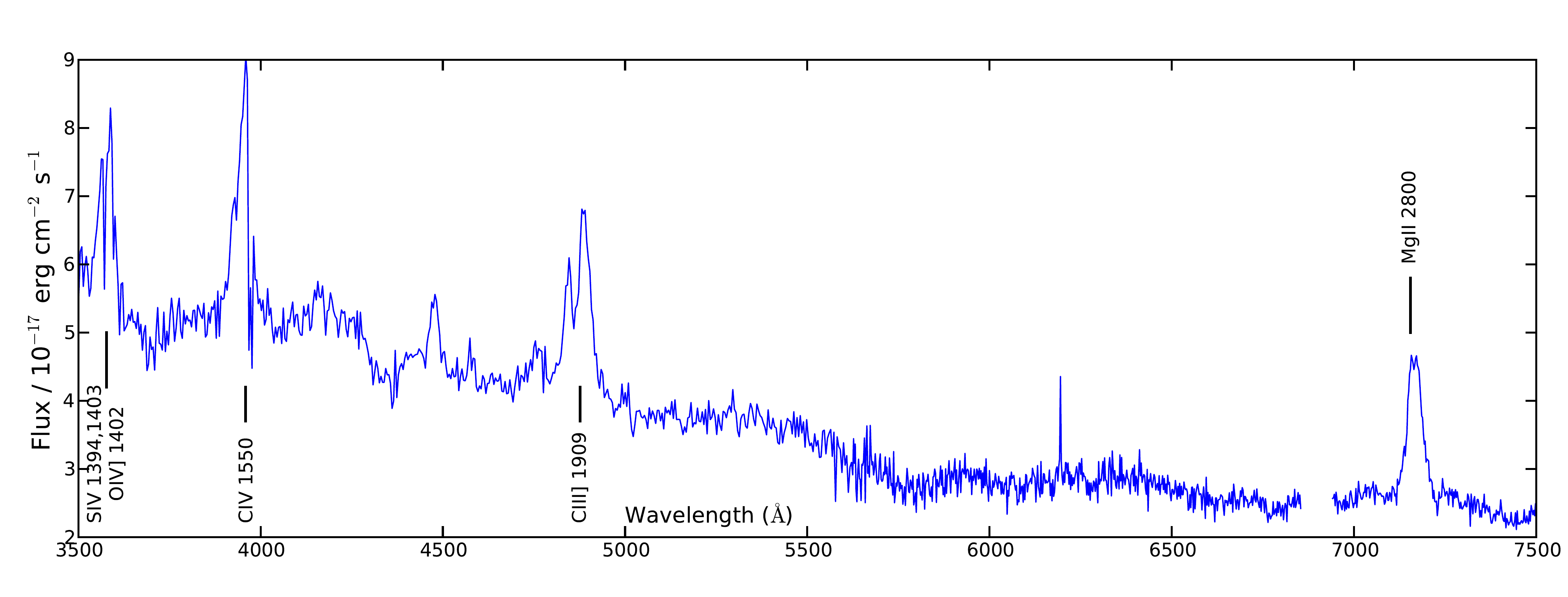}
\caption{{\it Top}: The radio continuum spectrum of VTC225707-010238 at two follow-up epochs (blue circles and green squares). 
The 3$\sigma$ upper limit at 1.4 GHz from \cite{hodge2011} is shown in red. 
The measurements from the first three survey epochs are shown in grey. 
See \S\ref{sec:radio_transients:one_month:VTC225707-010238} for details.
{\it Bottom}: The optical spectrum of VTC225707-010238 observed with LRIS at KeckI on 18 Aug 2012. 
The spectrum is typical of a blue quasar. 
We derive a redshift of 1.56.}
\label{fig:1222y}
\end{figure}

VTC225707-010238 is a persistent but variable source with flux densities 0.602$\pm$0.073, 1.218$\pm$0.069, 0.914$\pm$0.065, and 1.192$\pm$0.056 mJy in the four epochs of our survey.
The spectral indices within the S band for the first three epochs are $1.93\pm1.42$, $1.16\pm0.53$ and $1.64\pm0.82$.
The 1.4 GHz peak flux densities from the FIRST and \cite{hodge2011} surveys are $0.17\pm0.14$ and $0.17\pm0.07$ mJy respectively.
We followed up this transient with the VLA on 01 Sep 2012 (2--15 GHz) and 17 Sep 2012 (1--4 GHz).
The continuum radio spectra are shown in the top panel of Figure~\ref{fig:1222y}.
Both follow-up spectra are peaked at a few GHz.
On 01 Sep 2012, the peak is 2.5 mJy at 7 GHz, while the 17 Sep 2012 spectrum is likely peaked at 3 GHz with a flux density of 1.5 mJy.
This might suggest that this object is a flaring Gigahertz-peaked spectrum (GPS; young AGN) source.
Sub-mm (100 GHz) follow-up observations on 09 Sep 2012 with CARMA gave a non-detection with 3$\sigma$ supper limit of 2.1 mJy.

The optical counterpart of VTC225707-010238 is a blue point source in SDSS having $r=20.1$ mag.
The optical light curve from PTF shows intra-day small amplitude (sub-magnitude) variability.
Follow-up spectroscopy with KeckI LRIS on 18 Aug 2012 (bottom panel of Figure~\ref{fig:1222y}) shows that this is a quasar at a redshift of 1.56.
We therefore conclude that this transient is a flaring type-I quasar.

\subsection{Transients On Timescales $<$1.5 Year}\label{sec:radio_transients:one_year}

\subsubsection{VTC224730+000008: Flaring AGN}\label{sec:radio_transients:one_year:VTC224730+000008}

VTC224730+000008 is a variable source with flux densities 284.2$\pm$1.2, 272.9$\pm$0.8, 260.3$\pm$1.0, and 535.2$\pm$0.4 mJy in the four epochs of our survey.
Its 1.4 GHz peak flux densities from the FIRST and VLA-Stripe 82 surveys are $317.1\pm0.1$ mJy and $388.7\pm0.1$ mJy respectively.
The optical counterpart of VTC224730+000008 is a blue quasar, classified as a BL Lac-type object.
Blazar-like optical variability is indeed evident in the PTF and SDSS photometric data.
Optical spectra from the SDSS BOSS program indicate that the redshift is z$\sim$1.
Given this information, we conclude that VTC224730+000008 is an intermediate-redshift blazar.

\subsection{Transients On Timescales $\lesssim$20 Years}\label{sec:radio_transients:gt_one_year}

\subsubsection{VTC233002-002736: Renewed activity of Type-II QSO}\label{sec:radio_transients:gt_one_year:VTC233002-002736}
\begin{figure}[htp]
\centering
\includegraphics[width=3.5in]{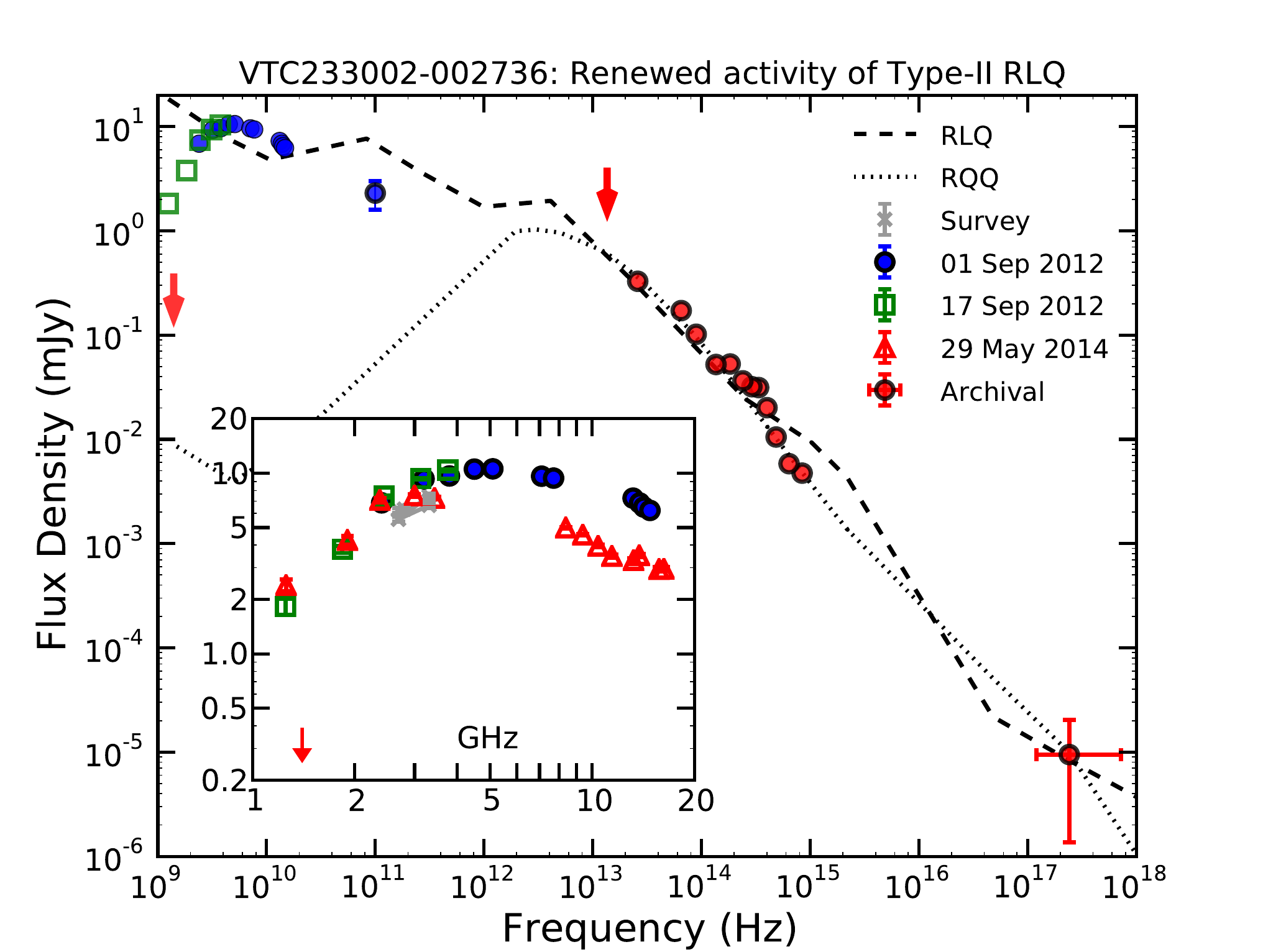}
\includegraphics[width=3.2in]{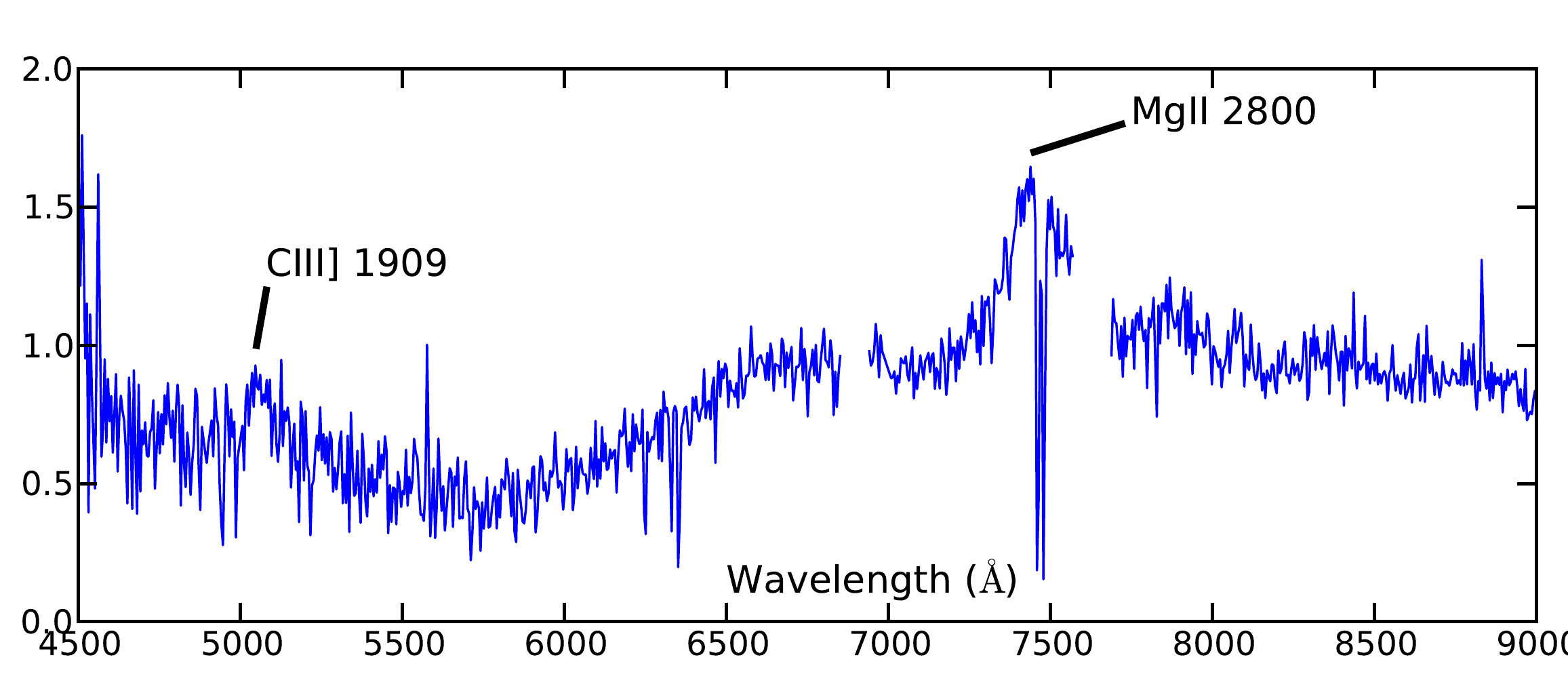}
\caption{{\it Top}: The multi-frequency continuum spectrum of VTC233002-002736. 
Archival photometry is shown as filled red circles.
The 3$\sigma$ upper limit at 1.4 GHz from the FIRST survey and the upper limit from WISE W4 filter are shown as red arrows.
Blue circles and green squares indicate the first and second follow-up observations.
Black dashed and dotted lines are average spectra for radio-loud and radio-quiet quasars from \cite{elvis1994}.
The {\it inset} shows the zoom-in of the follow-up radio continuum spectra from the three epochs.
Red triangles show data from the third follow-up epoch, and the measurements from the four survey epochs appear in grey.
{\it Bottom}: The optical spectrum of VTC233002-002736 observed with DIEMOS at KeckII on 14 Sep 2012. 
The y-axis is flux in arbitrary units.
Broad emission lines indicate a quasar at a redshift of 1.65.
See \S\ref{sec:radio_transients:gt_one_year:VTC233002-002736} for details.}
\label{fig:VTC233002-002736}
\end{figure}

This is a transient with respect to the FIRST survey.
Its flux density is 5.492$\pm$0.157, 5.342$\pm$0.143, 5.742$\pm$0.147, 5.510$\pm$0.073 mJy in our four survey epochs, E1--E4, with spectral indices $1.42\pm0.23$, $0.86\pm0.22$, and $0.63\pm0.21$ in 
epochs E1--E3, within the S band.
The flux density in the FIRST survey at the location of VTC233002-002736 is $0.34\pm0.13$ mJy (mean epoch 1999.2).
This implies an order of magnitude or more increase in flux density at 1.4 GHz over the past decade, and relatively stable flux density at 3 GHz over the past 2 years.
We obtained follow-up observations with the VLA (1--15 GHz) on 01 Sep 2012, 17 Sep 2012, and 29 May 2014.
These observations reveal a Gigahertz-peaked spectrum source with an optically thick spectral index of 2.1 between 1--3 GHz.
On 01 Sep 2012, the spectrum peaks at 5 GHz with a flux density of about 10 mJy, and the optically thin spectral index is $-0.6$ between 7--15 GHz.
In the 29 May 2014 observations, spectral flattening is observed between 2--15 GHz with respect to 01 Sep 2012 and the spectral peak appears to be at 3 GHz, suggesting a significant evolution 
in the spectrum beyond 3 GHz.
The radio continuum spectra from the survey and follow-up observations are shown in the inset of the top panel of Figure~\ref{fig:VTC233002-002736}.
Follow-up observations from CARMA at 100 GHz on 09 Sep 2012 give a detection at $2.3\pm0.7$ mJy.

The SDSS counterpart of VTC233002-002736 is a reddish point source with $r=21.3$ mag.
The detection in the PTF is only marginal, and no meaningful optical variability information can be obtained.
The SDSS light curve between 1998--2008 reveals sub-magnitude variability, typical of AGN, on shortest timescales of a few weeks.
VTC233002-002736 also has an X-ray counterpart in XMM-Newton \citep{lamassa2013}.
The multi-frequency continuum spectrum of VTC233002-002736 is shown in the top panel of Figure~\ref{fig:VTC233002-002736}.
The optical to mid-infrared part of the spectrum agrees well with a type-II quasar template from \cite{polletta2007}.
A comparison between the radio and optical flux densities ($R\simeq2.4$) indicates that this is a radio loud quasar.
We obtained a follow-up optical spectrum on 14 Sep 2012 with DIEMOS at Keck II (Figure~\ref{fig:VTC233002-002736}, bottom panel).
The spectrum shows broad CIII] and MgII spectral lines allowing us to calculate the redshift, $z=1.65$.

From the nature of the radio spectrum, the timescale of its evolution, and the nature of the host galaxy, we conclude that VTC233002-002736 is due to renewed jet-activity from a type-II radio loud 
quasar, where the GPS spectrum is indicative of a young jet.
The order-of-magnitude increase in flux density with respect to the FIRST survey could be indicative of an enhanced accretion phenomenon leading to an intensified jet.
The flattening of spectral index at GHz frequencies between 2012 and 2014 may imply cessation in the increased accretion episode and/or the interaction of the jet with the circum-nuclear material in the host galaxy.

\subsection{Transients Found Serendipitously}\label{sec:radio_transients:serendipitous}
This subsection describes two transients that were found while manually inspecting archival radio images and analyzing our 5$\sigma$ source catalogs. 
Being below the cutoff values that we have used, our variability criteria are not able to recover them.
These two exemplars highlight the possibility that there are many more transients in our pilot survey data than have been reported in this work.

\subsubsection{VTC221515-005028: Extreme variability of flat-spectrum AGN}\label{sec:radio_transients:serendipitous:VTC221515-005028}
\begin{figure}[htp]
\centering
\includegraphics[width=3.4in]{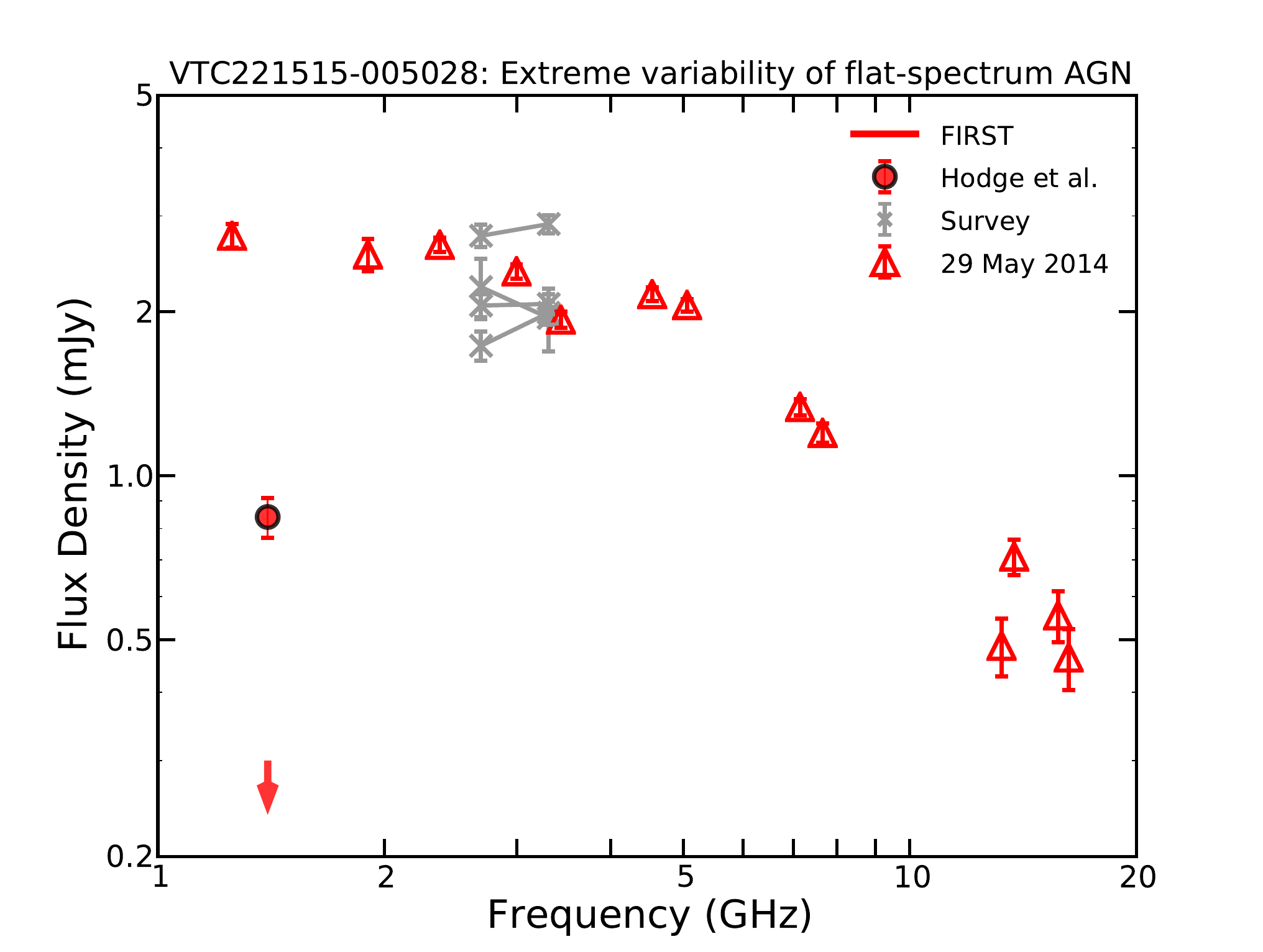}
\caption{The radio continuum spectrum of VTC221515-005028 at the follow-up epoch 29May2014 (red triangles). 
The 3$\sigma$ upper limit from the FIRST survey is shown with the red arrow and the detection from the \cite{hodge2011} survey is marked by a red circle.
he measurements from the four survey epochs are shown in grey. 
See \S\ref{sec:radio_transients:serendipitous:VTC221515-005028} for details.}
\label{fig:VTC221515-005028}
\end{figure}

VTC221515-005028 has flux densities 2.569$\pm$0.086, 1.989$\pm$0.062, 1.778$\pm$0.071, 1.787$\pm$0.058 mJy and spectral indices $0.24\pm0.30$, $0.03\pm0.31$, $0.68\pm0.38$, and $-0.45\pm0.72$ 
in epochs E1--E4 of our survey.
It is absent in the FIRST survey (mean epoch 1996.1), where the flux density at that location is $0.25\pm0.10$ mJy, but present in the 1.4 GHz Stripe 82 survey catalog \citep{hodge2011} with 
a peak flux density of 0.79 mJy.
For the transient search on timescales $\lesssim$20 years (\S\ref{sec:radio_transients:gt_one_year}) we assumed the upper limit to the flux density in FIRST as 1 mJy, 
and hence did not recovered this source as a transient using our spectral index criterion.
We carried out radio follow-up observations with the VLA (1--15 GHz) on 29 May 2014.
The radio continuum spectrum are shown in Figure~\ref{fig:VTC221515-005028}.
The source appears to have a flat spectrum between 1--5 GHz and a spectral index of $-1.1$ between 5--15 GHz.

The optical counterpart of VTC221515-005028 is a faint red galaxy having $r\simeq20.8$ mag.
From the PTF light curve it is seen to exhibit stochastic variability with a maximum amplitude of one magnitude in the R band, and having a shortest timescale of about three days.
The SDSS light curve between 1999--2008 reveals sub-magnitude variability and intermittent flares with amplitudes upto $r=2$ mag on $\sim$1 day timescale.
The photometric redshift from SDSS is 0.37.
The radio-to-optical flux density ratio for the host galaxy, $R=log(S_{\rm 1.4 GHz}/S_g)\simeq2.8$, and the radio luminosity at 1.4 GHz, $9.0\times10^{29} ~\mbox{erg s$^{-1}$ Hz$^{-1}$}$, 
suggest a low-luminosity radio-loud AGN.
We thus conclude that VTC221515-005028 is an extremely variable flat-spectrum AGN.

\subsubsection{VTC230241+003450: Flare from dKe Star}\label{sec:radio_transients:serendipitous:VTC230241+003450}

VTC230241+003450 is detected as a 6$\sigma$ source on 15 Aug 2012 (epoch E2, regions R2) and is not detected in the other epochs.
The peak pixel values at the location of this transient in epochs E1--E4 are 0.090$\pm$0.078, 0.422$\pm$0.069, 0.046$\pm$0.076, and 0.127$\pm$0.065 mJy respectively.
In the FIRST and \cite{hodge2011} surveys, the peak pixel values at the location of the transient candidate are $0.13\pm0.11$ and $0.16\pm0.06$ mJy respectively.
The optical counterpart of this transient is a 11th magnitude star, SDSS J230241.41+003450.2, classified as a K4Ve star by \cite{torres2006}.
There is no evidence for binarity.
Fitting a blackbody to the photometric data from SDSS, the NOMAD catalog, and WISE gives an estimate of the effective temperature, 3800$\pm$500 K, and the 
distance, 70$\pm$20 pc (assuming the radius of a main sequence star).
SDSS J2302+00 derive a $v$~sin($i$) of 85 km s$^{-1}$.
This star has a ROSAT counterpart, 1RXS J230240.3+003453, 17\arcsec$\pm$12\arcsec~away, whose hardness ratios are consistent with those of a coronal emitter.
We used a 1 keV APEC model in WebPIMMS to convert the ROSAT/PSPC countrate to a flux of $1.1\times10^{-12}~\mbox{erg cm$^{-2}$ s$^{-1}$}$ in the 0.1--2.4 keV energy band.
The X-ray-to-optical flux ratio is 0.002, and the X-ray luminosity at a distance of 70 pc is $6.4\times10^{29}~\mbox{erg s$^{-1}$}$, both values being consistent with a dKe star.
The quiescent 3 GHz radio luminosity of VTC230241+003450 can therefore be estimated as $6.4\times10^{14\pm1}$ erg s$^{-1}$ Hz$^{-1}$ \citep{benz1994,gudel2002}.
This can be compared with the flaring luminosity, $5.9\times10^{15}$ erg s$^{-1}$ Hz$^{-1}$.
We conclude that this transient is a flare from a dKe star.

\section{Optical Properties of Radio Sources}\label{sec:optical_properties}

\begin{figure}[htp]
\centering
\includegraphics[width=3.6in]{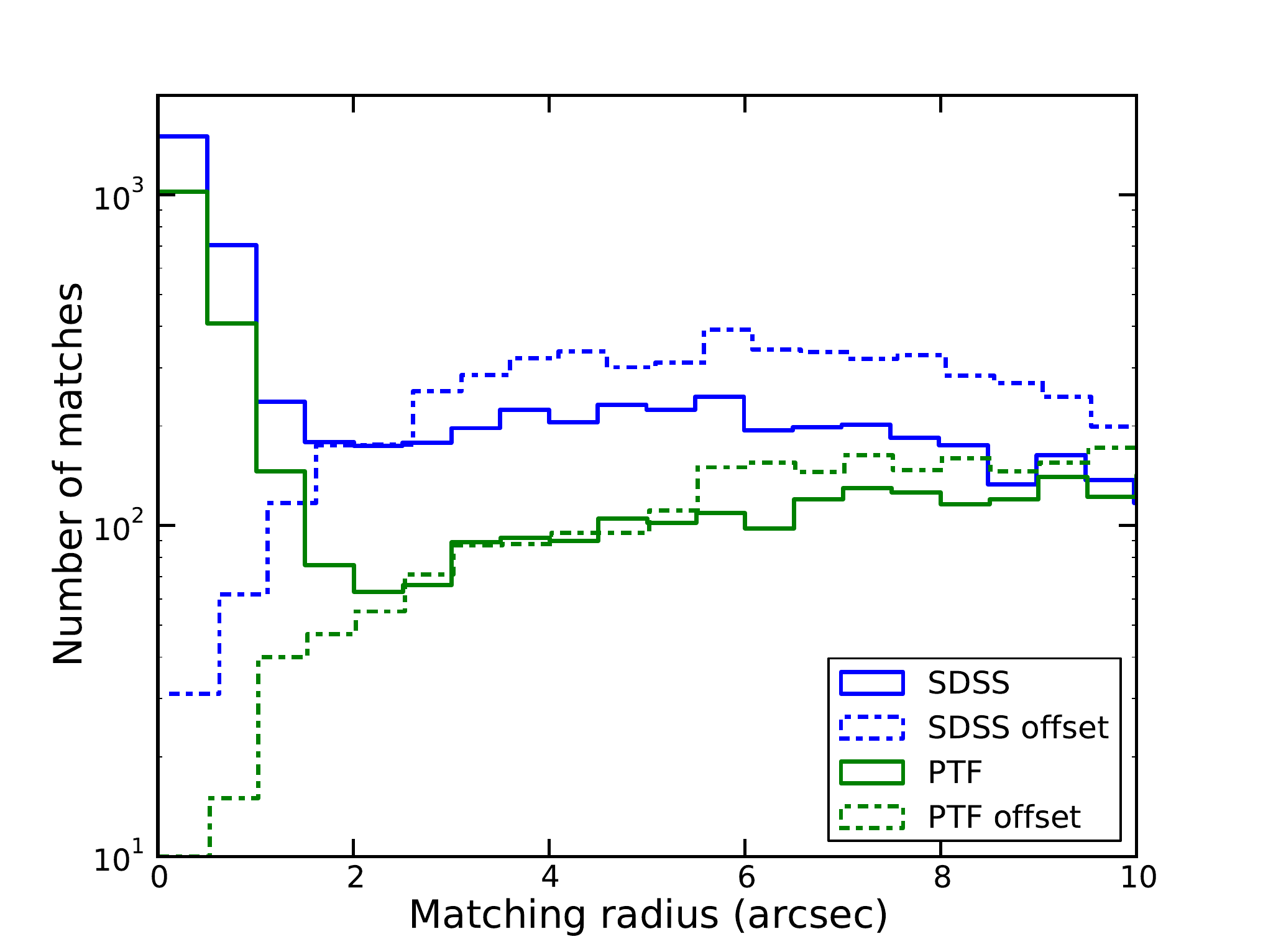}
\caption{The number of optical matches for the radio sources in E2CAT for different matching radii. 
Solid lines indicate the matching process carried out using the actual radio source positions and the 
dot-dashed lines indicate background matches found by offsetting the radio source positions by 1\arcmin ~in 
an arbitrary direction. Blue and green lines show radio versus SDSS and PTF matching respectively.}
\label{fig:matching}
\end{figure}
\begin{figure}[htp]
\centering
\includegraphics[width=3.6in]{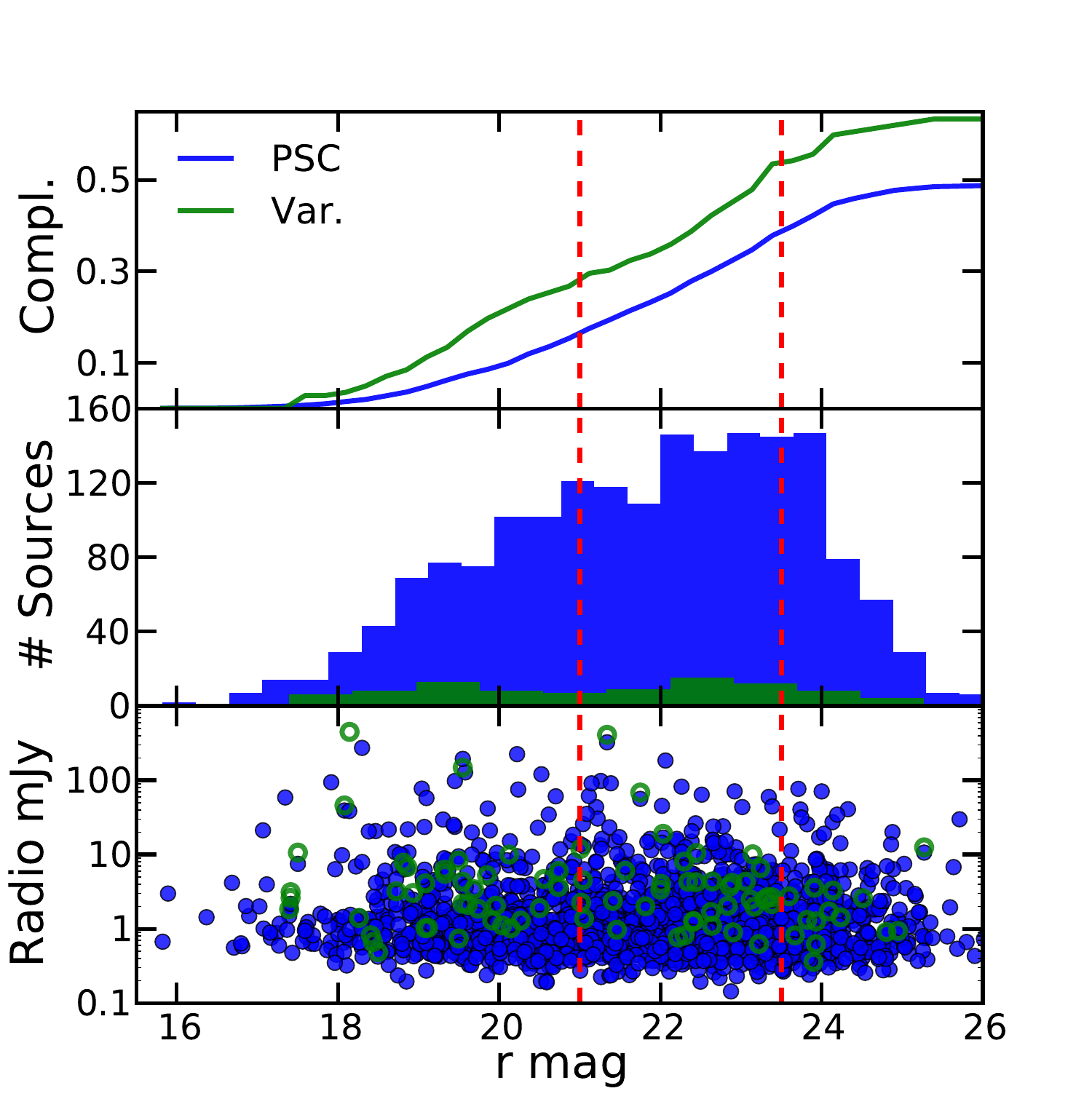}
\caption{Optical counterparts of persistent (blue) and variable (green) radio sources in the PSC. 
The x-axis is r-band magnitude from SDSS. 
The top panel gives the completeness of the radio-optical cross-matching, the middle panel 
shows histograms of persistent and variable radio sources with optical counterparts, and the bottom panel plots the 
radio flux densities versus optical magnitudes. The approximate limiting magnitudes for PTF (per-epoch; $R=21$ mag.) and 
SDSS (Stripe 82 co-add; $r=23.5$ mag.) are shown as red dashed lines.}
\label{fig:radio_optical_quiescent}
\end{figure}

PTF carried out a concurrent optical survey which resulted in the identification of a few hundred thousand sources per epoch down to a limiting magnitude of R$\simeq$21 mag\footnote{Note that, 
for the 60-second snapshots taken by PTF, the CCDs are saturated at $\sim$10--11 mag. This represents the lower limiting magnitude of the optical study carried out here.}.
Our radio survey further benefits from the presence of SDSS deep co-add images containing more than a million objects over the 50 deg$^2$ to r$\simeq$23.5 mag \citep{annis2011}.
To find optical counterparts of radio sources we followed the procedure from \citep{hodge2011}.
We matched each source in the E2CAT with the SDSS and PTF catalogs using a 15\arcsec ~matching radius, and selected only the nearest match.
To understand the false matching rate, we repeated the search by offsetting the radio source positions by 1\arcmin ~in an arbitrary direction.
The resulting number of matches as a function of the matching radius is plotted as a histogram in Figure~\ref{fig:matching}.
Based on these results we choose a matching radius of 1\arcsec ~for SDSS and 1.5\arcsec ~for PTF to achieve a false matching rate of less than 3\% and completeness better than 85\%.
The PTF source positions for the faintest objects are known to have a larger scatter than theoretically expected, and hence the larger matching radius of 1.5\arcsec ~is reasonable.
Using these matching radii, we found the optical counterparts of sources in the PSC.
The corresponding matching fraction in SDSS as a function of the r-band magnitude is shown graphically in Figure~\ref{fig:radio_optical_quiescent}.
49\% of the radio sources in the PSC have an optical counterpart down to the SDSS r-band 
limit\footnote{If all the radio sources (components) from the E2CAT are matched with SDSS sources, then the completeness is much lower, $\sim$35\%.}.
\cite{hodge2011}, using the deep co-added SDSS images (limit i$\sim$23.5 mag), found a matching ratio (within a 1\arcsec~radius) of 44.4\%. 
For radio variable sources, we found a somewhat higher matching fraction of 63\% (Figure~\ref{fig:radio_optical_quiescent}), which is also in agreement with \citeauthor{hodge2011}.

The modest optical matching rate in our medium-wide, medium-deep survey
was not unexpected. \cite{ivezic2002} compared the SDSS 
(limit r$\sim$22 mag) and the FIRST (5$\sigma\sim$1 mJy) surveys over in 1230 
deg$^2$ of the sky and matched about 30\% of the 108,000 FIRST radio
sources.  \cite{mcmahon2002} looked at the 
382,892 FIRST sources in the north Galactic cap and looked for optical
matches using the APM scans of POSS-I plates (R$\sim$20 mag, B$\sim$21.5 mag) for a 18\% 
identification rate (70,000 sources). The optical match rate improves
substantially in deep, narrow surveys. For example, \cite{huynh2008} use Australia
Telescope Compact Array (ATCA) data taken toward the Hubble Deep Field
(HDF) at 20 cm, 11 cm, 6 cm and 3.6 cm, establishing a 66\% matching 
of optical counterparts to I$=$23.5 mag. There is a strong color 
dependence, with the matching rate increasing from the blue to the near 
infrared \citep{bouchefry2007,smolcic2008}. Match rates approach 100\% with the 
use of deep infrared data \citep{bonzini2012}.

\begin{figure}[htp]
\centering
\includegraphics[width=3.5in]{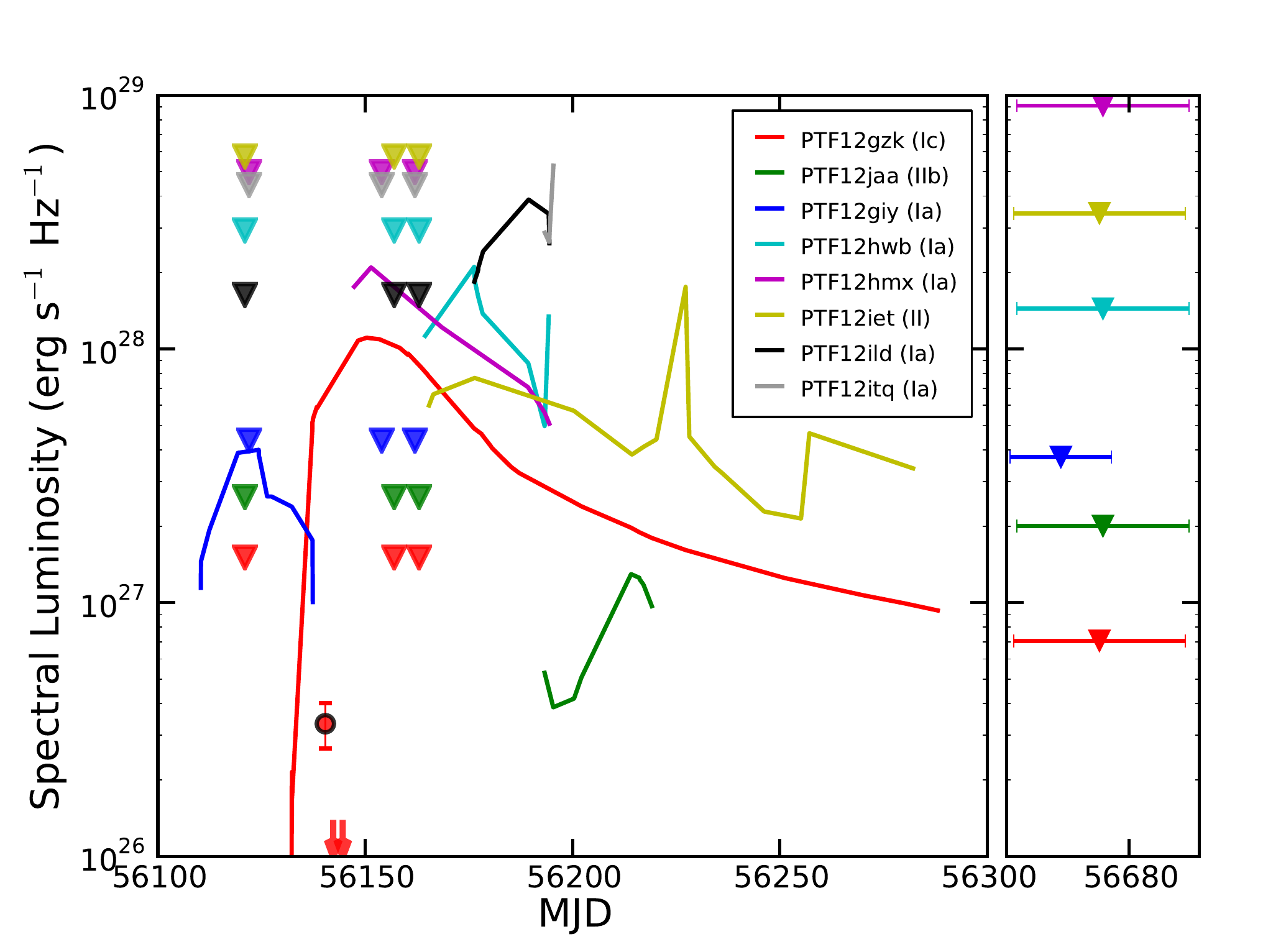}
\caption{Light curves of the 8 spectroscopically-confirmed supernovae found in the PTF survey.
Each supernova is represented by a unique color.
Optical photometric data from PTF and follow-up observations are shown as solid lines.
The 3$\sigma$ radio upper limits from the Jansky VLA survey (four epochs) are shown as downward-pointing triangles.
For PTF12gzk, the radio detection (red circle and errorbar) and 3$\sigma$ upper limits (downward-pointing red arrows) at 5 GHz from \citep{horesh2013} are also marked.
See \S\ref{sec:optical_properties} for details.}
\label{fig:SNe}
\end{figure}
\begin{figure}[htp]
\centering
\includegraphics[width=3in]{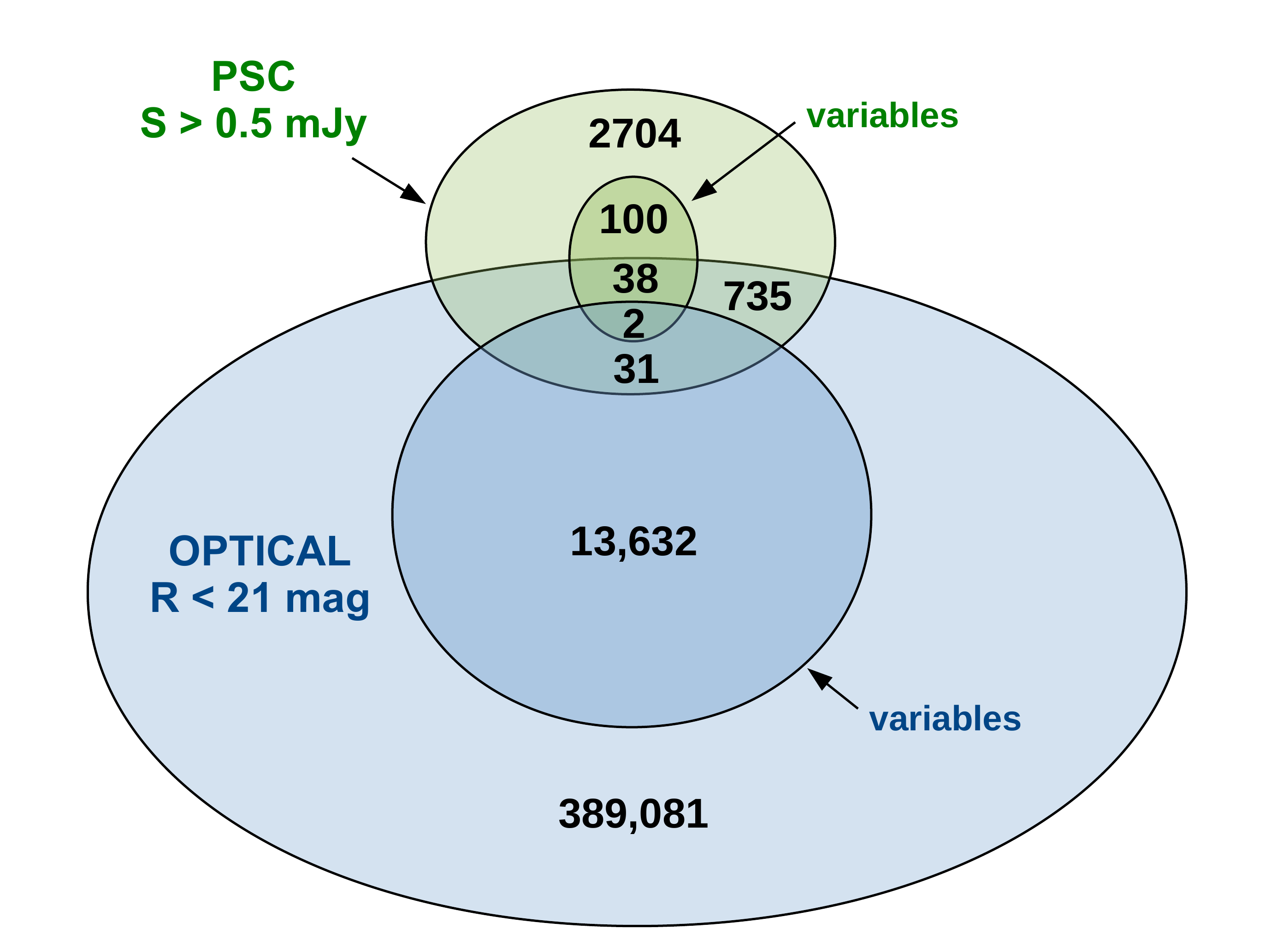}
\caption{Venn diagram showing the number of persistent as well as variable optical sources from PTF and radio sources from the PSC. 
The sets plotted here are not to scale.}
\label{fig:venn}
\end{figure}

The near-real-time optical transient search carried out via image subtraction during the three-month high-cadence PTF observations resulted in approximately 
8 million detections.
Following standard practice, about 0.9 million of these (corresponding to only $\sim$50,000 unique optical sources) were identified in machine learning software as unlikely 
to be image subtraction artifacts, 
and further filtering was carried out using several stringent selection criteria (such as: at least two detections within one hour, no coincidence with stellar counterparts or AGN, etc.) 
and automated classifiers \citep{bloom2012}.
This was followed by human inspection of the subtracted images, light curves, and automated classifications, and the list was narrowed down to 193 candidates for further follow-up.
Out of the 193 candidates, only 10 sources were ultimately followed up spectroscopically, among which are 8 confirmed 
supernovae, PTF12gzk (SN Ic, z$=$0.014), 12jaa (IIb, 0.024), 12giy (Ia, 0.029), 12hwb (Ia, 0.056), 12hmx (Ia, 0.085), 12iet (II, 0.095), 12ild (Ia, 0.17), and 12itq (Ia, 0.22).
The multi-wavelength observations of PTF12gzk have been discussed at length by \cite{ben-ami2012,horesh2013}.
In Figure~\ref{fig:SNe} we plot the optical light curves of these eight supernovae as well as the upper limits to their spectral luminosity in the radio.
The radio non-detections are in accordance with the expected flux densities of Type II, Ic and Ia supernovae (Table~\ref{tab:transients_summary}).
The radio detection and subsequent non-detections of PTF12gzk from \citeauthor{horesh2013} are also shown for reference, and these data emphasize the need for 
deep radio observations for optically-bright supernovae such as the ones generally found by PTF.
After all the survey observations were complete, we compiled optical light curves of sources using the catalogs hosted at IPAC (\S\ref{sec:optical_processing}).
In order to enable robust variability search, we selected only those optical sources having more than 16 reliable observations (unflagged in SExtractor) and R band magnitudes 
between 10 and 23.
There are 402,747 such sources in the PTF database with the 90\% completeness of this sample corresponding to approximately an R band limiting magnitude of 21.
802 (0.2\%) of these sources have counterparts in the PSC (i.e. 22\% of the PSC sources have optical counterparts).
13,667 (3.4\%) are optical variables\footnote{Given the optical light curves, sources having a $\chi^2$ probability less than 1 in 500,000 and variability $>$30\%, i.e. 
a standard deviation of $>$0.28 mag. were selected as optical variables. These criteria were designed to be similar to the radio variability criteria.}.
Only 42 radio sources have optical variable counterparts among the sample of 3652, while only two of these are variable also in the radio.
Thus, given the limiting magnitude of R$=$21 mag. in optical and our source detection threshold of $\sim$0.5 mJy in the radio, we find that 
the overlap between optical variables and radio variable sources is extremely small. 
These demographics are succinctly presented in a Venn diagram in Figure~\ref{fig:venn}.

\begin{figure*}[htp]
\centering
\includegraphics[width=1.3in]{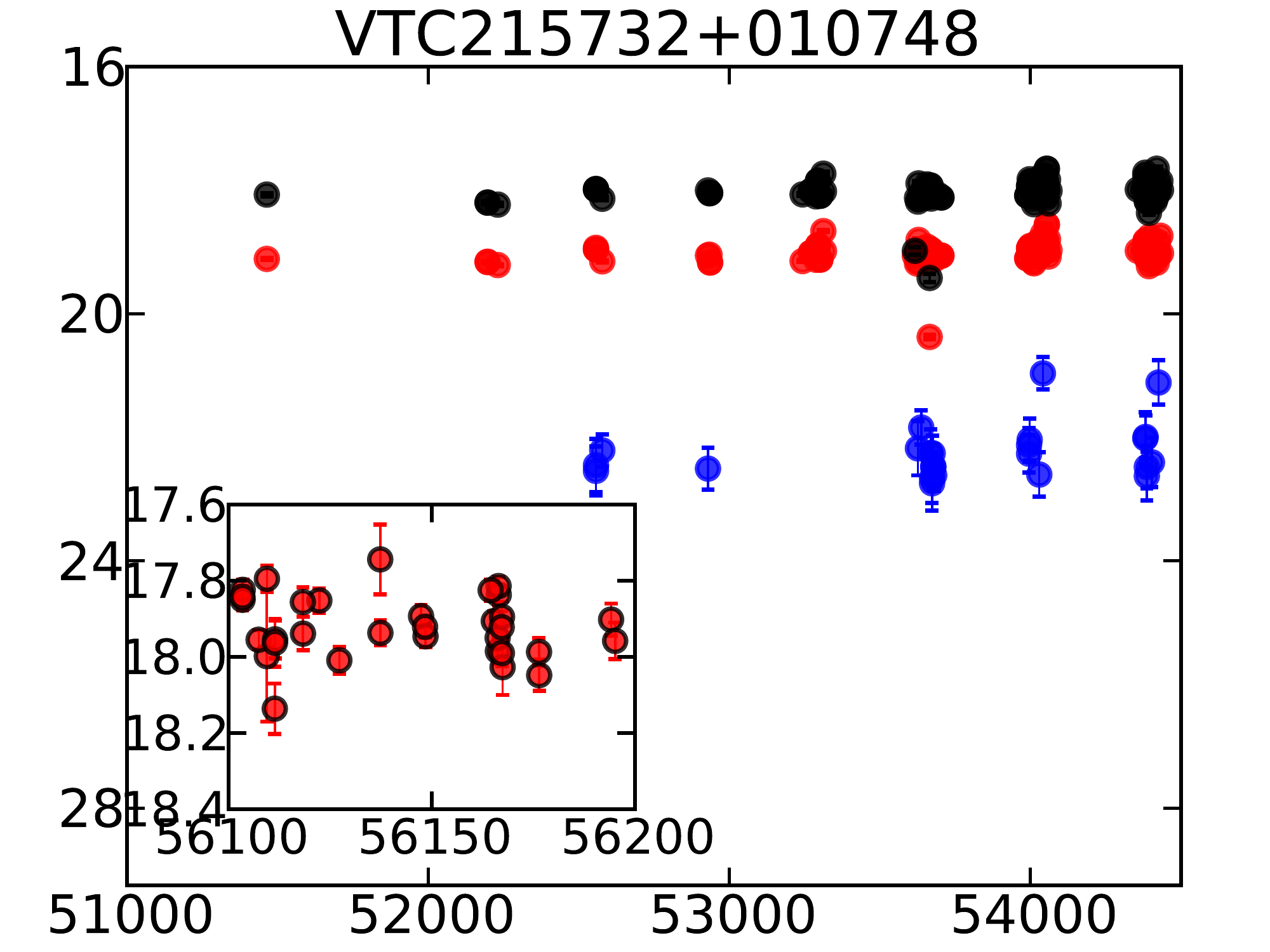}
\includegraphics[width=1.3in]{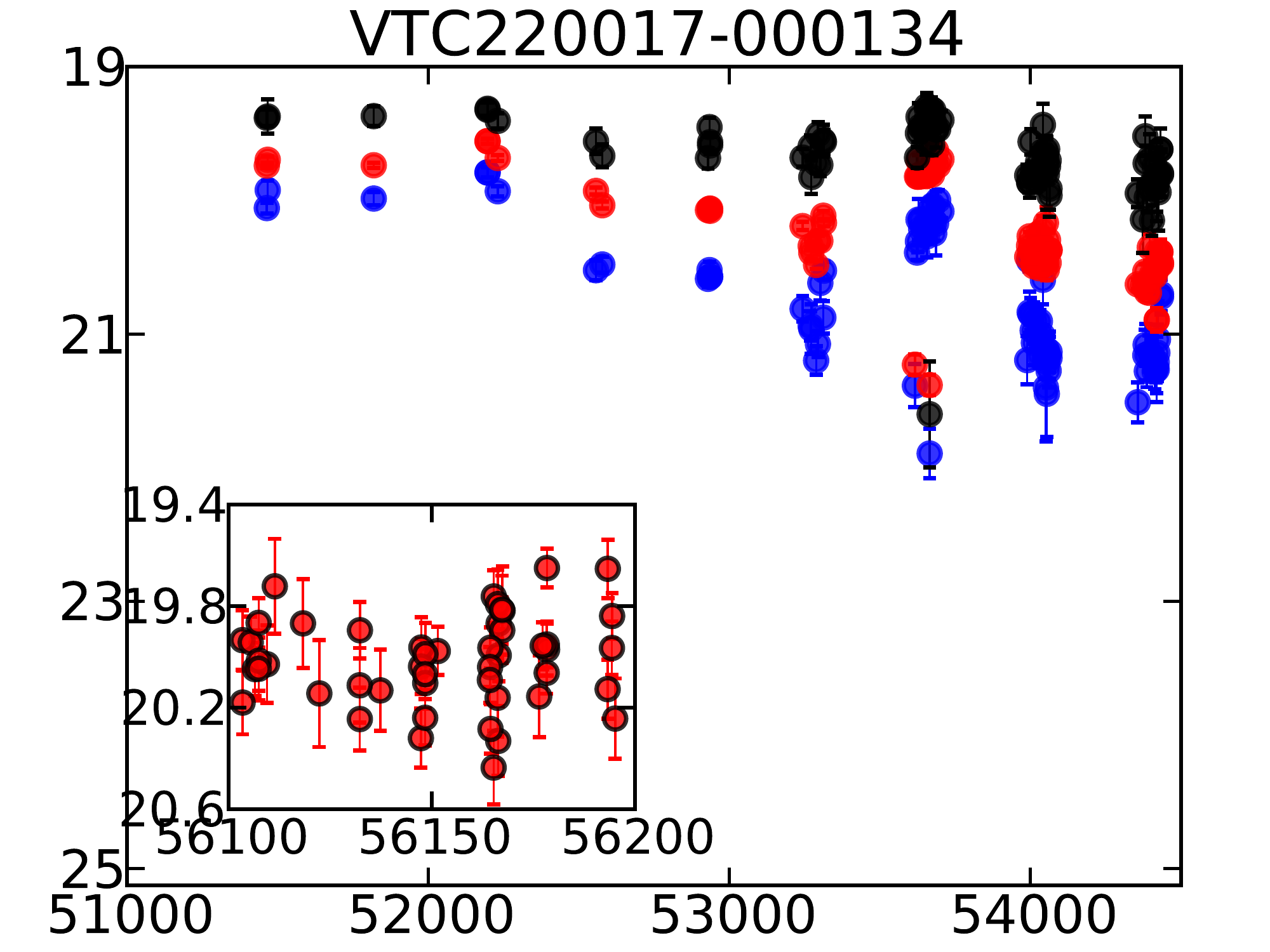}
\includegraphics[width=1.3in]{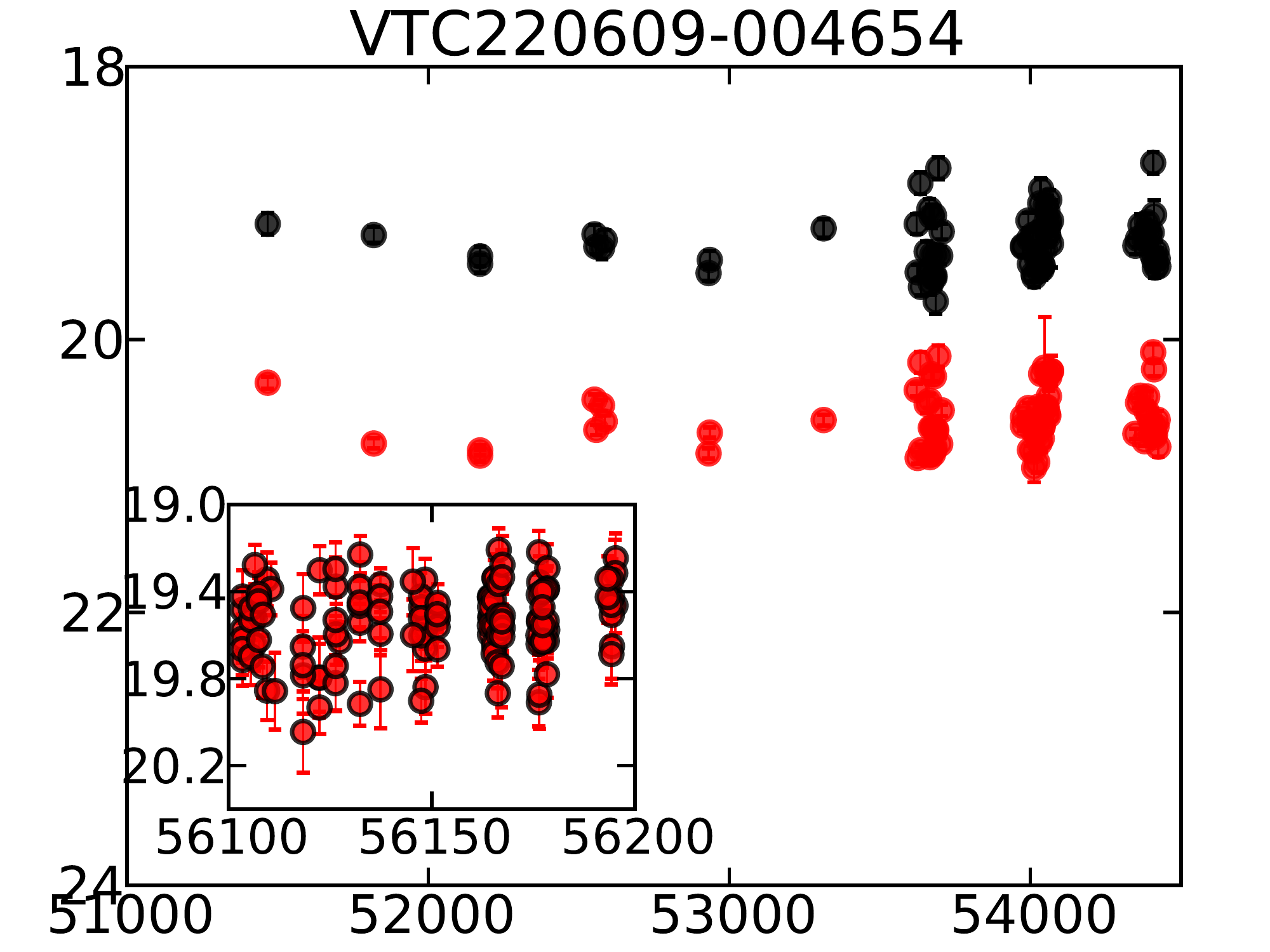}
\includegraphics[width=1.3in]{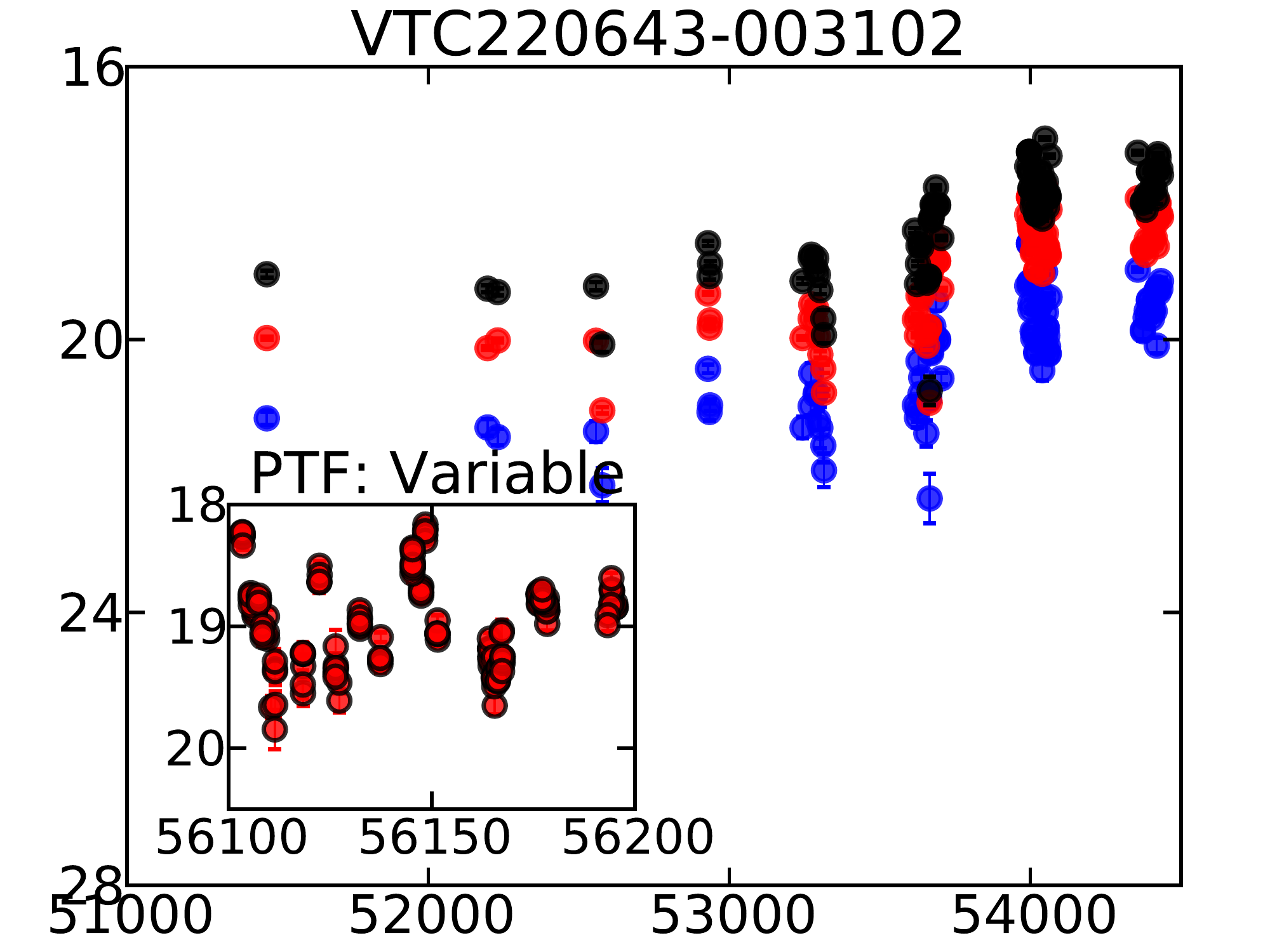}
\includegraphics[width=1.3in]{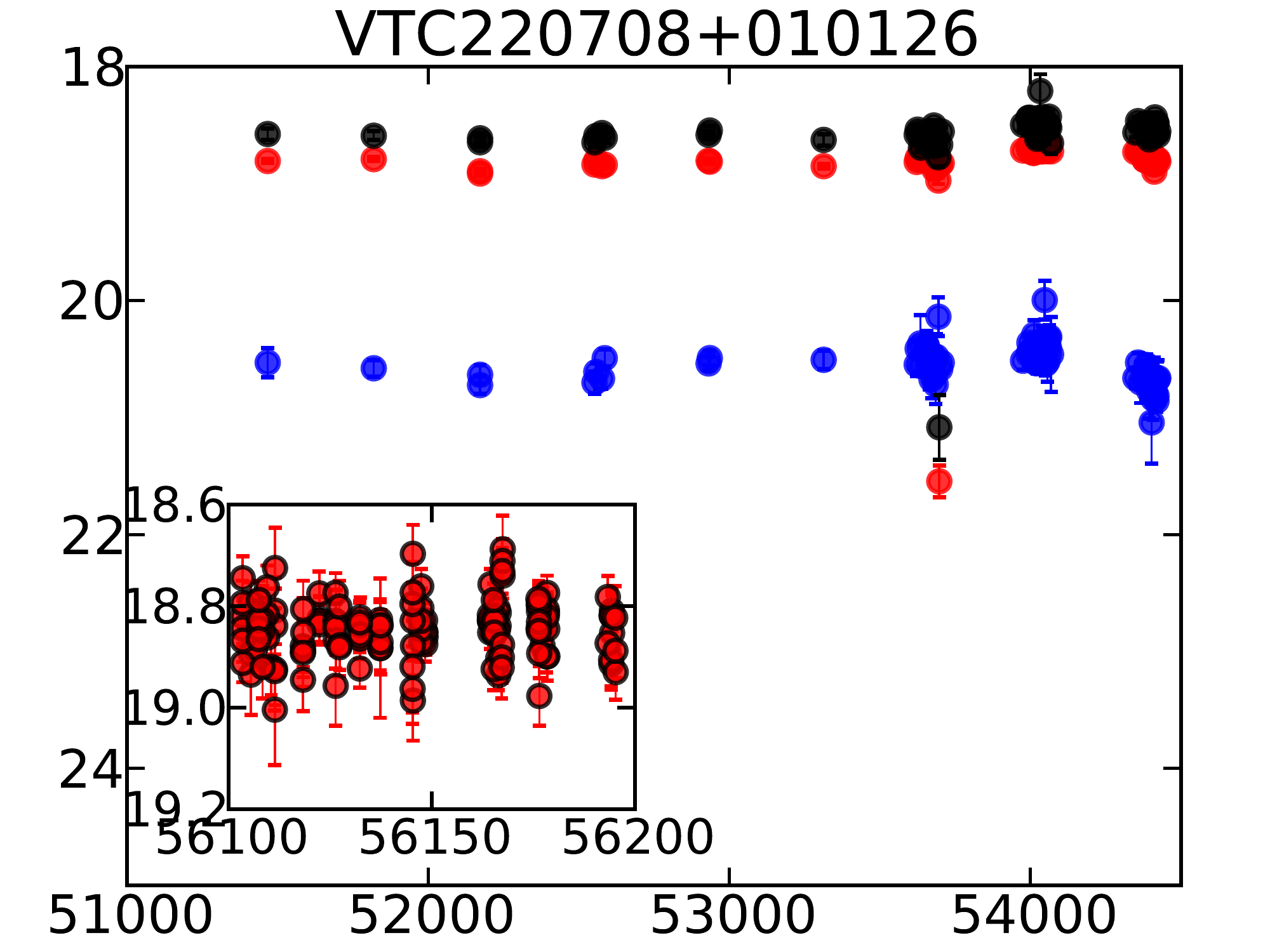}
\includegraphics[width=1.3in]{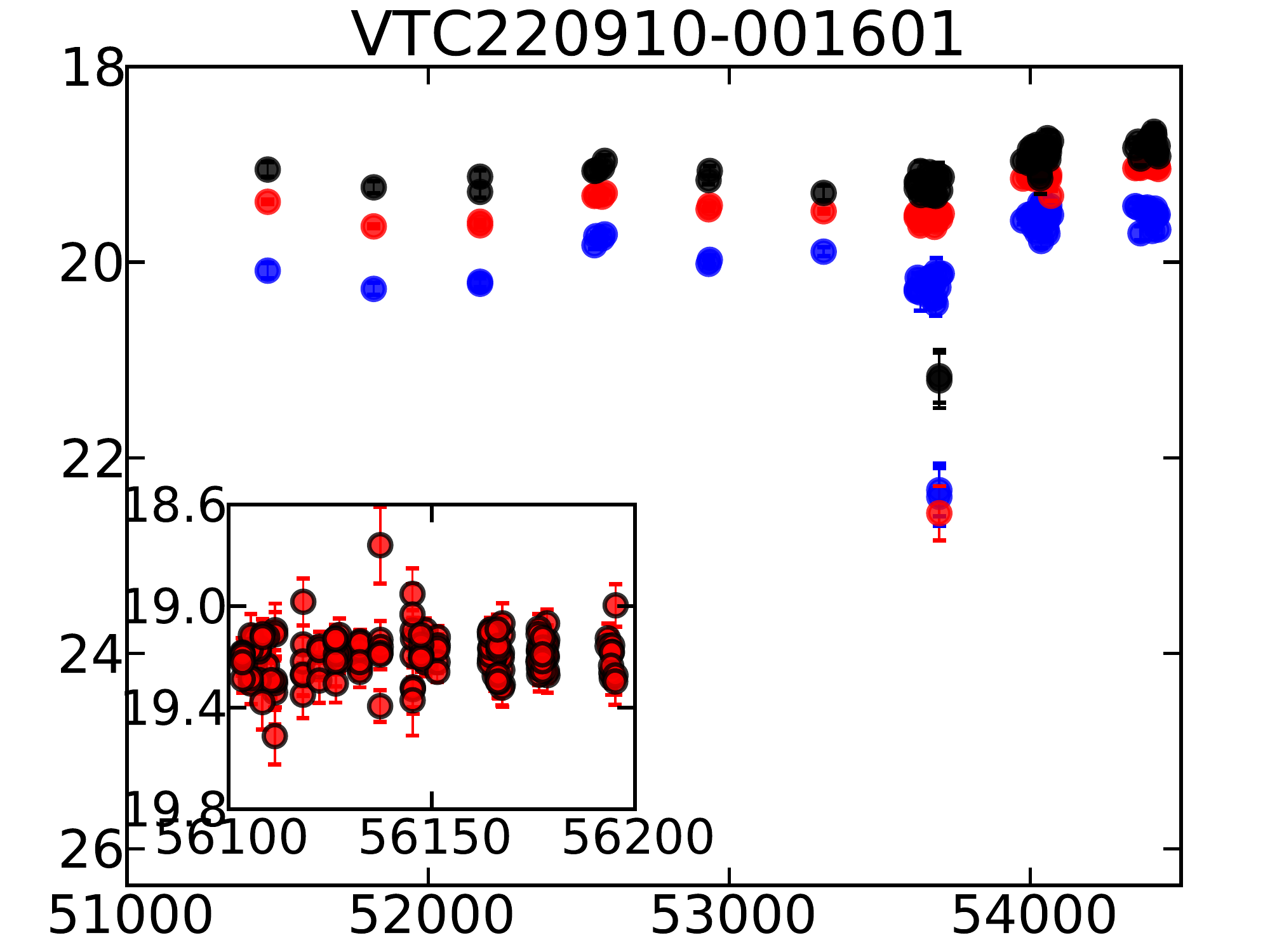}
\includegraphics[width=1.3in]{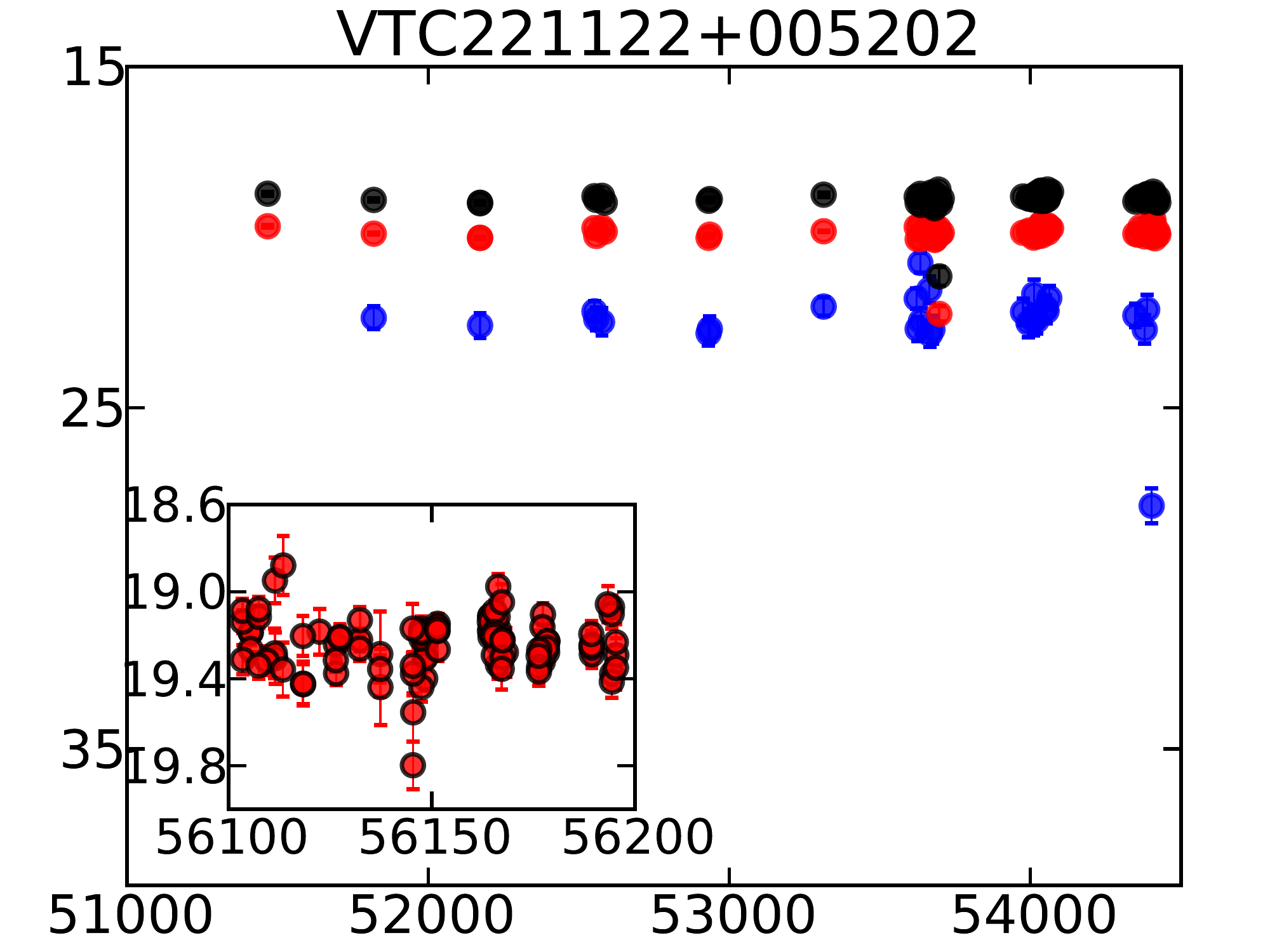}
\includegraphics[width=1.3in]{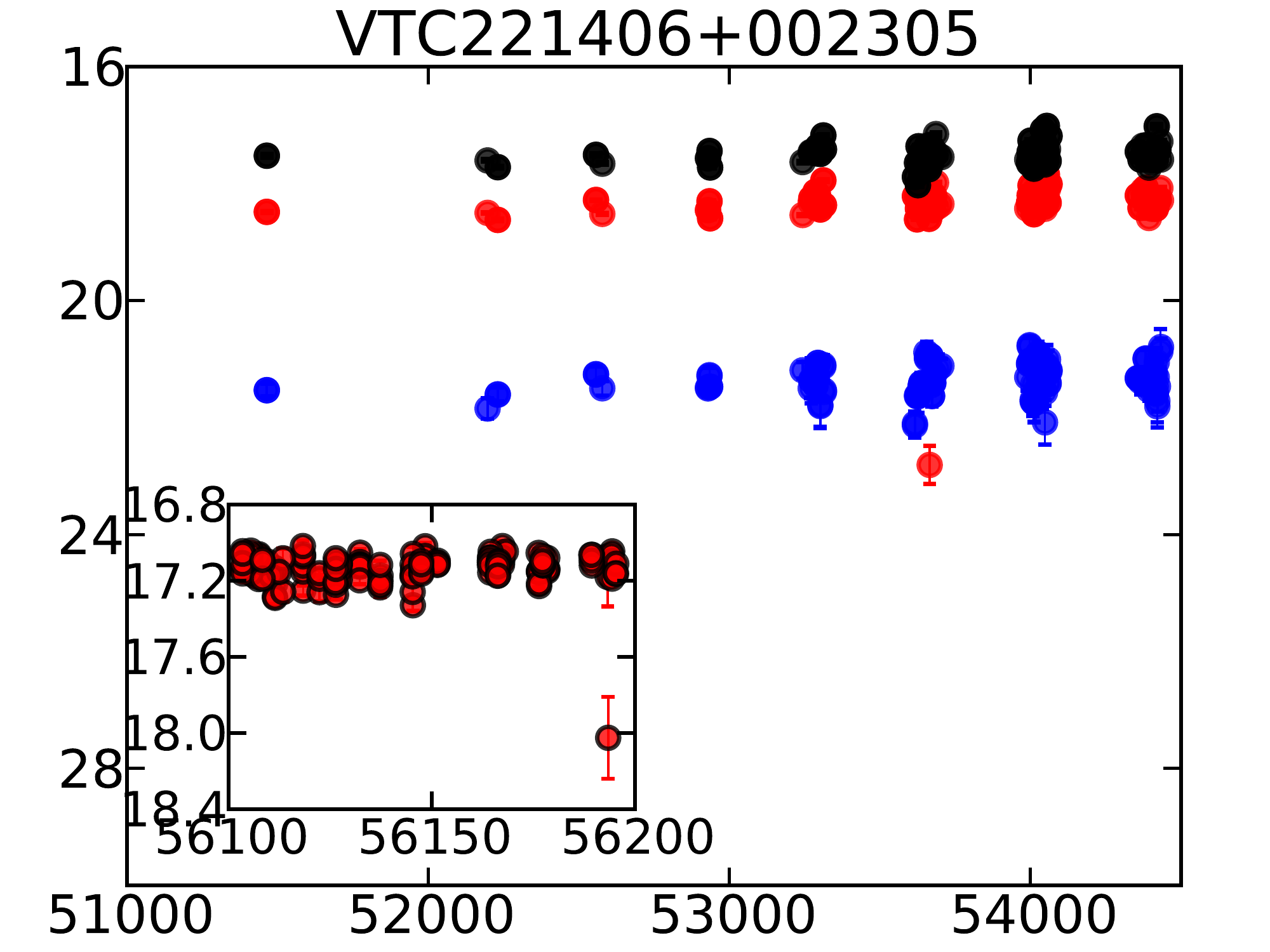}
\includegraphics[width=1.3in]{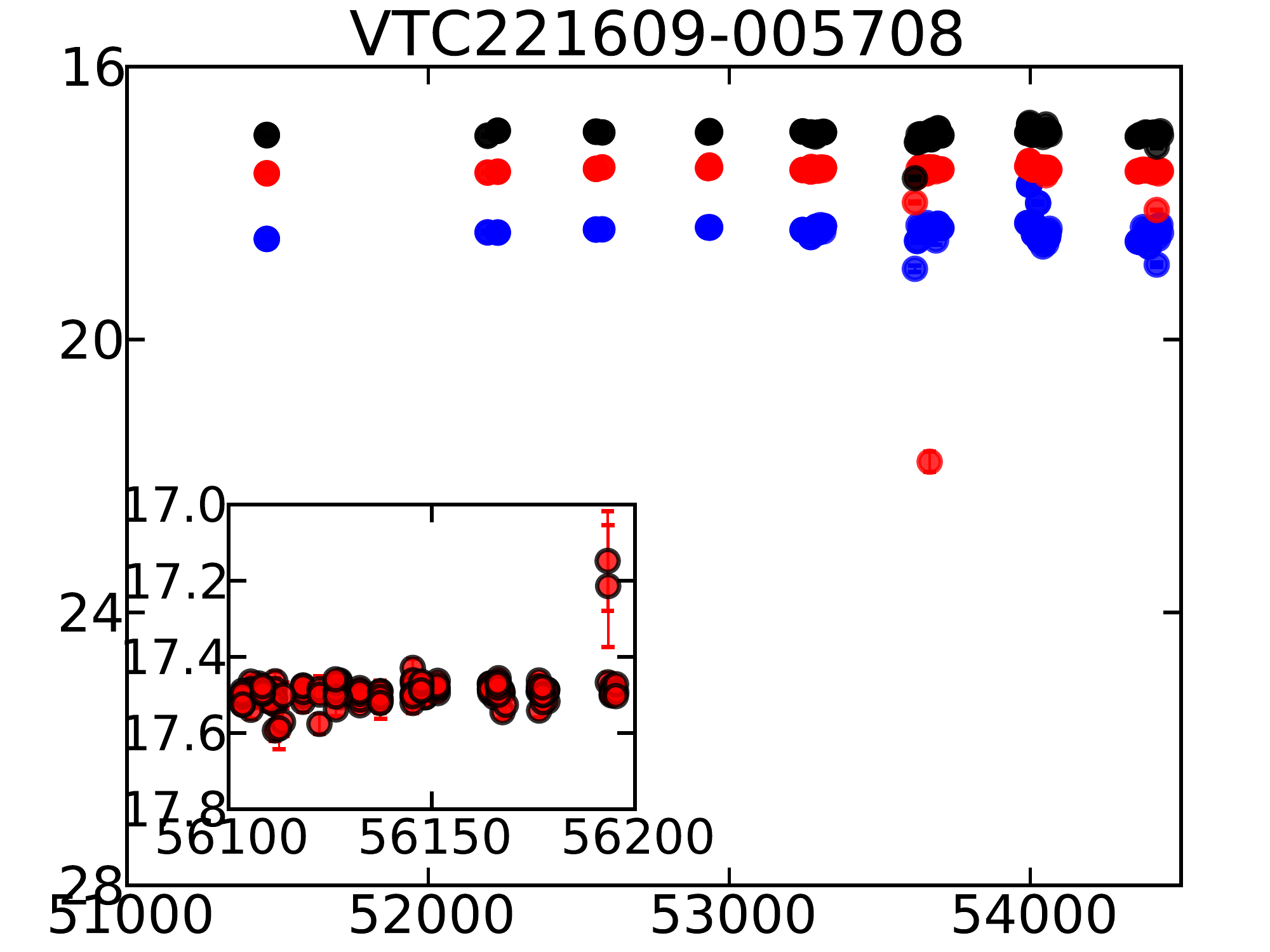}
\includegraphics[width=1.3in]{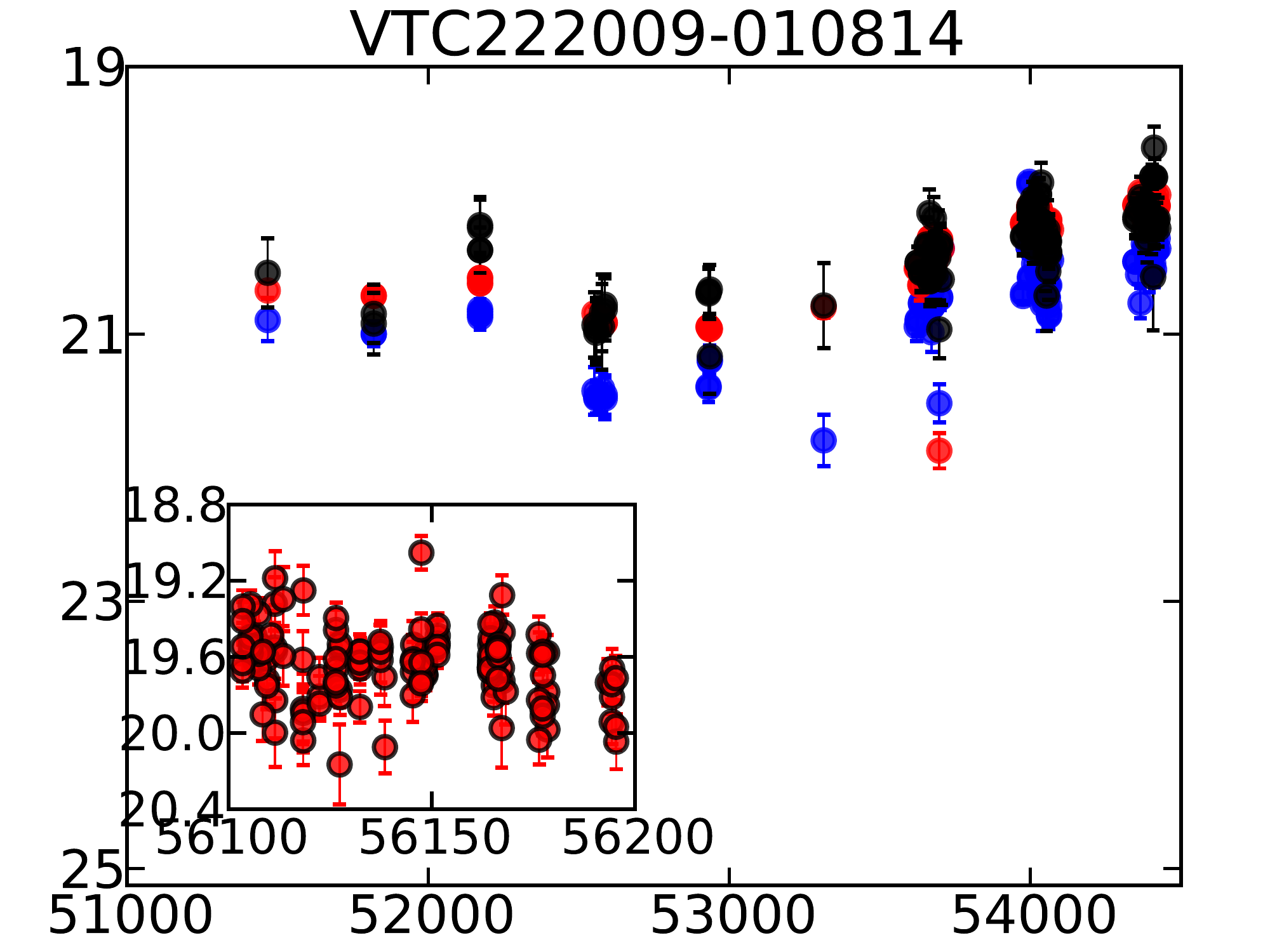}
\includegraphics[width=1.3in]{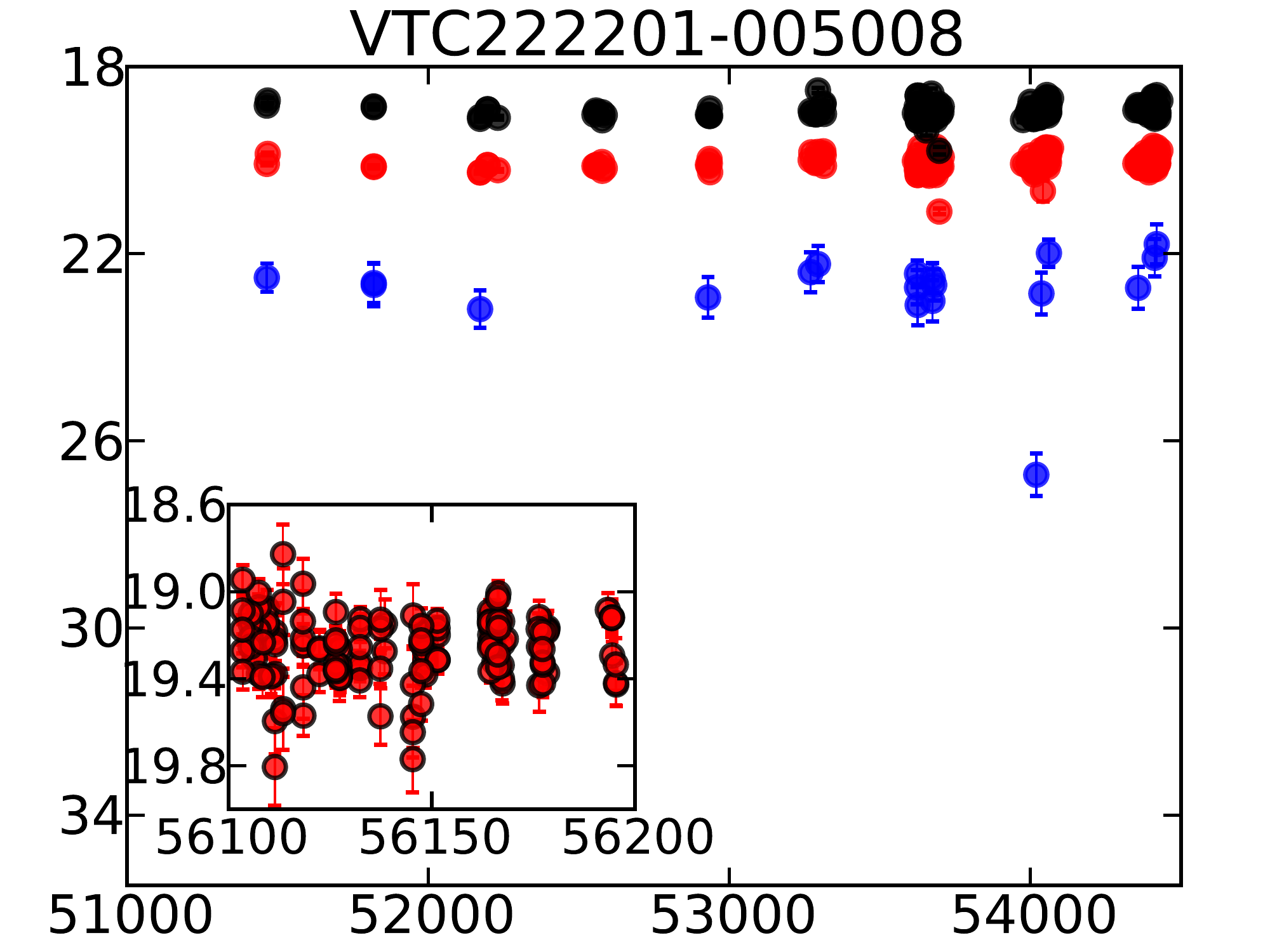}
\includegraphics[width=1.3in]{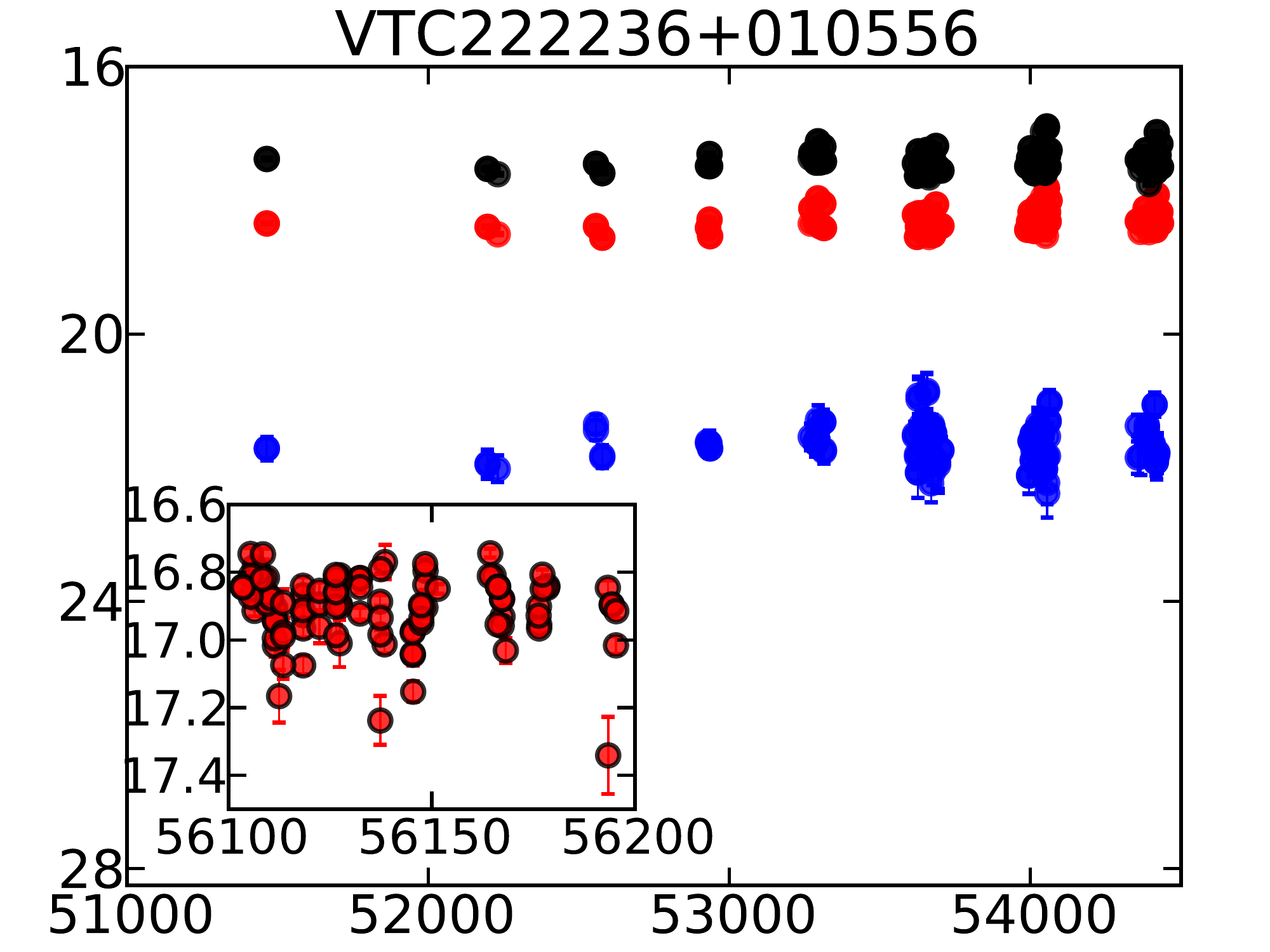}
\includegraphics[width=1.3in]{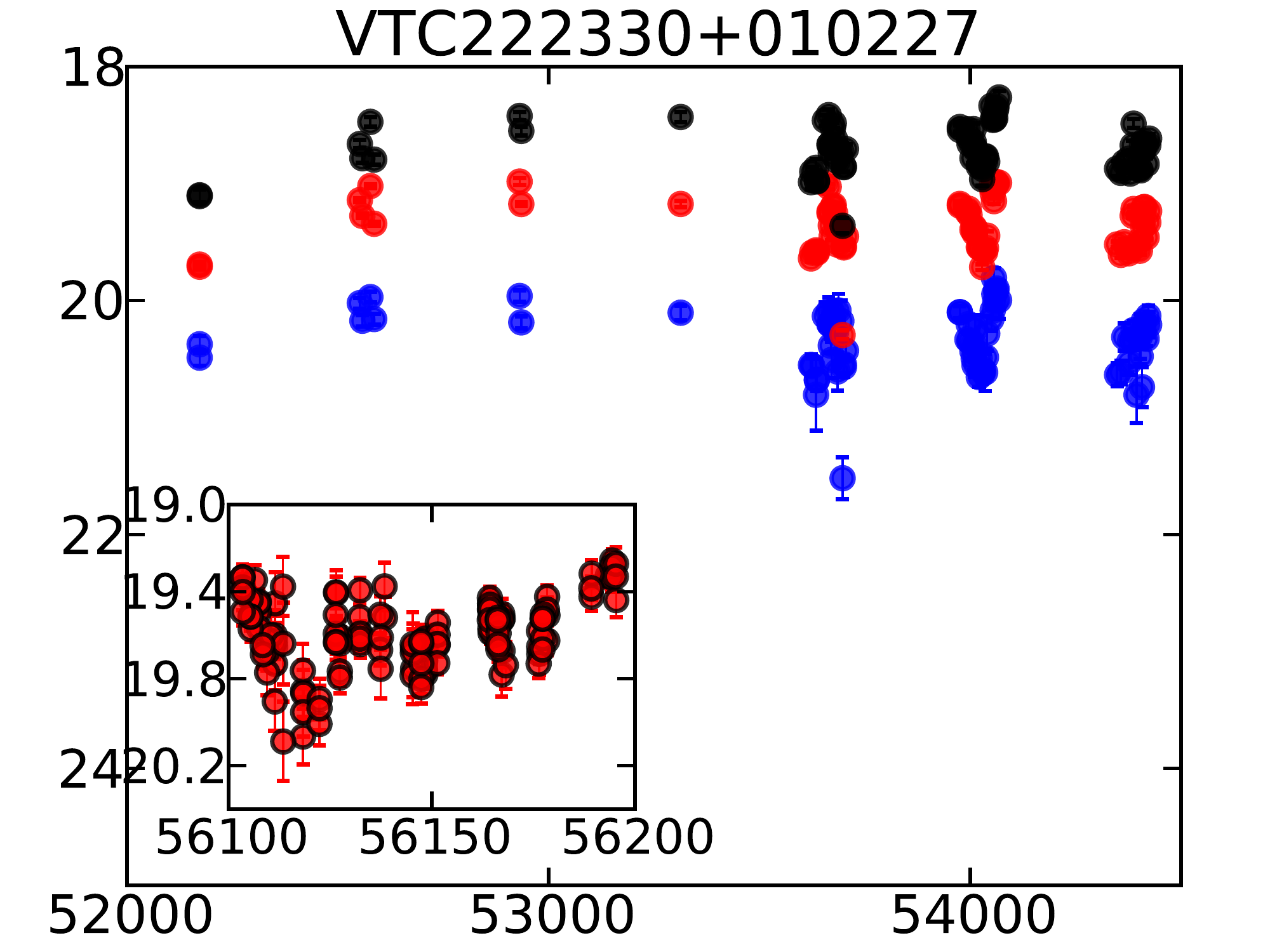}
\includegraphics[width=1.3in]{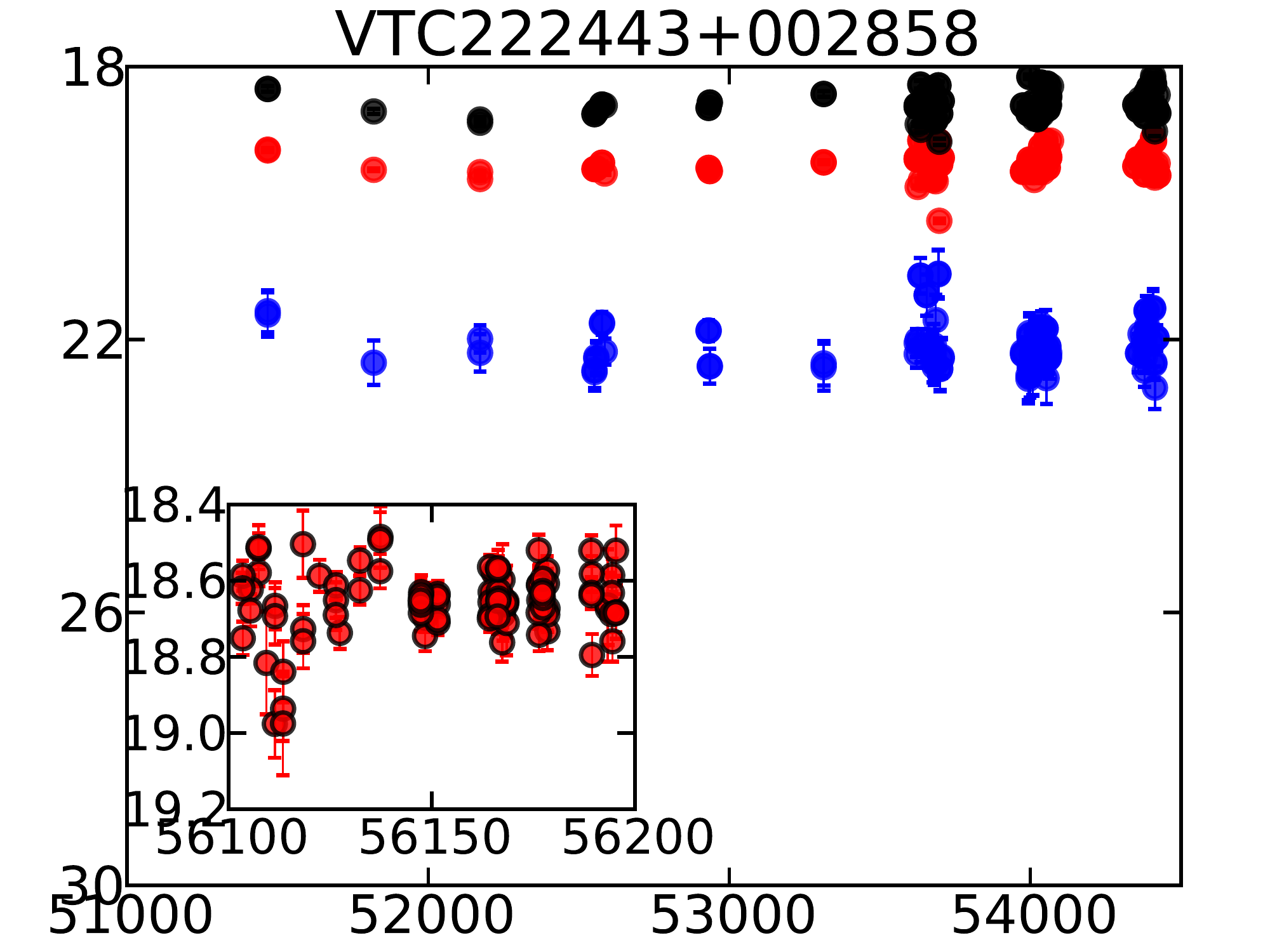}
\includegraphics[width=1.3in]{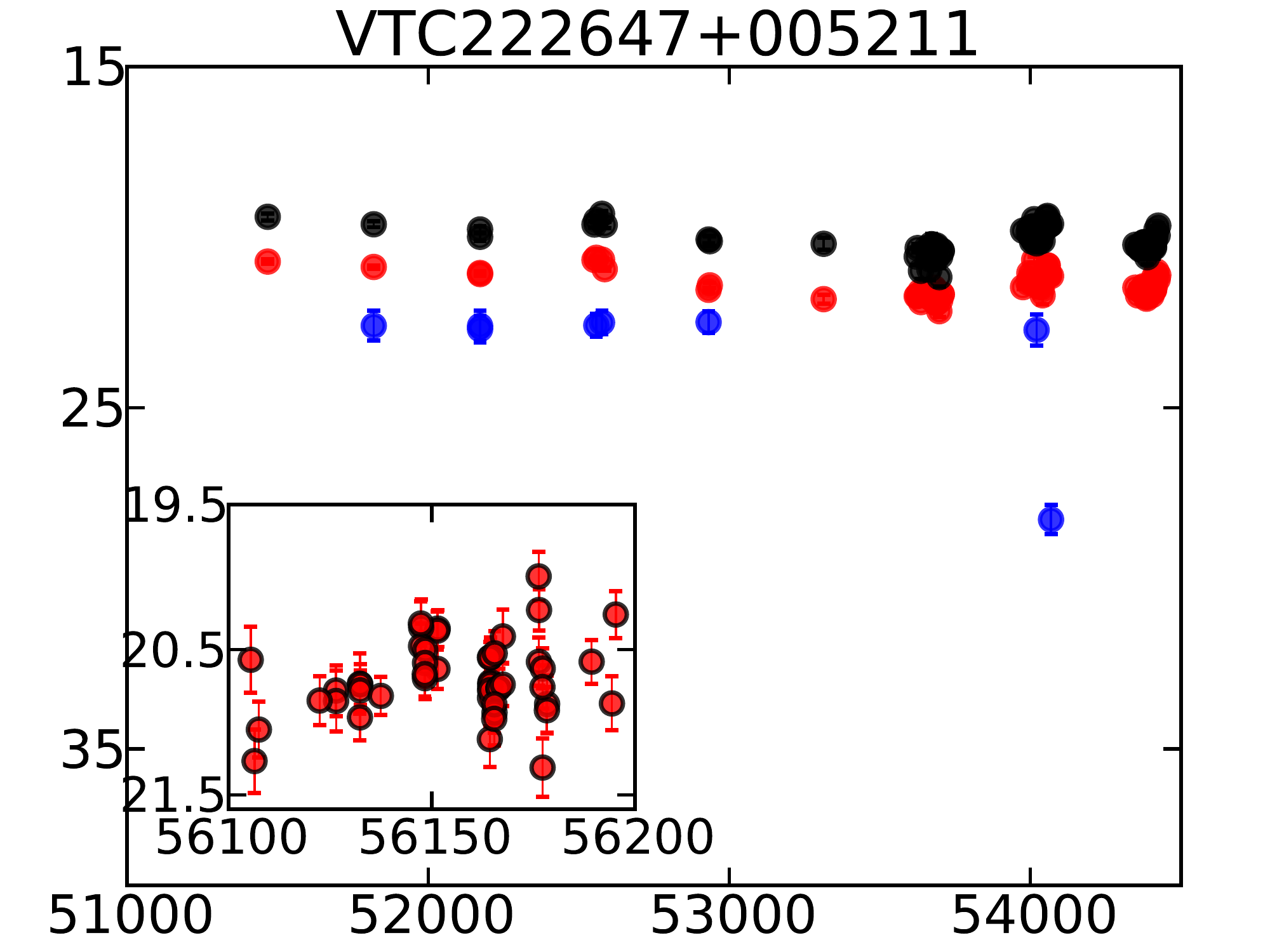}
\includegraphics[width=1.3in]{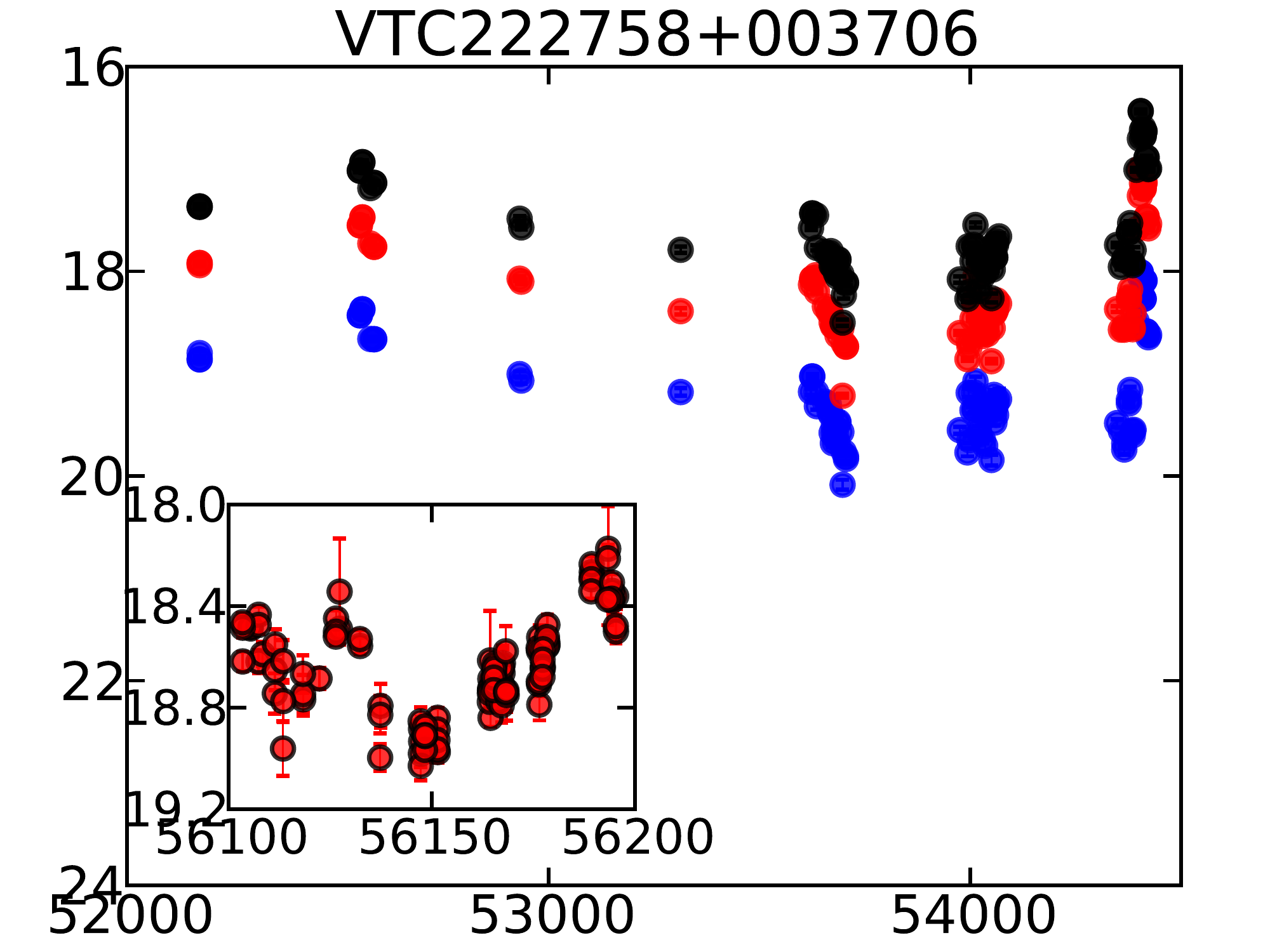}
\includegraphics[width=1.3in]{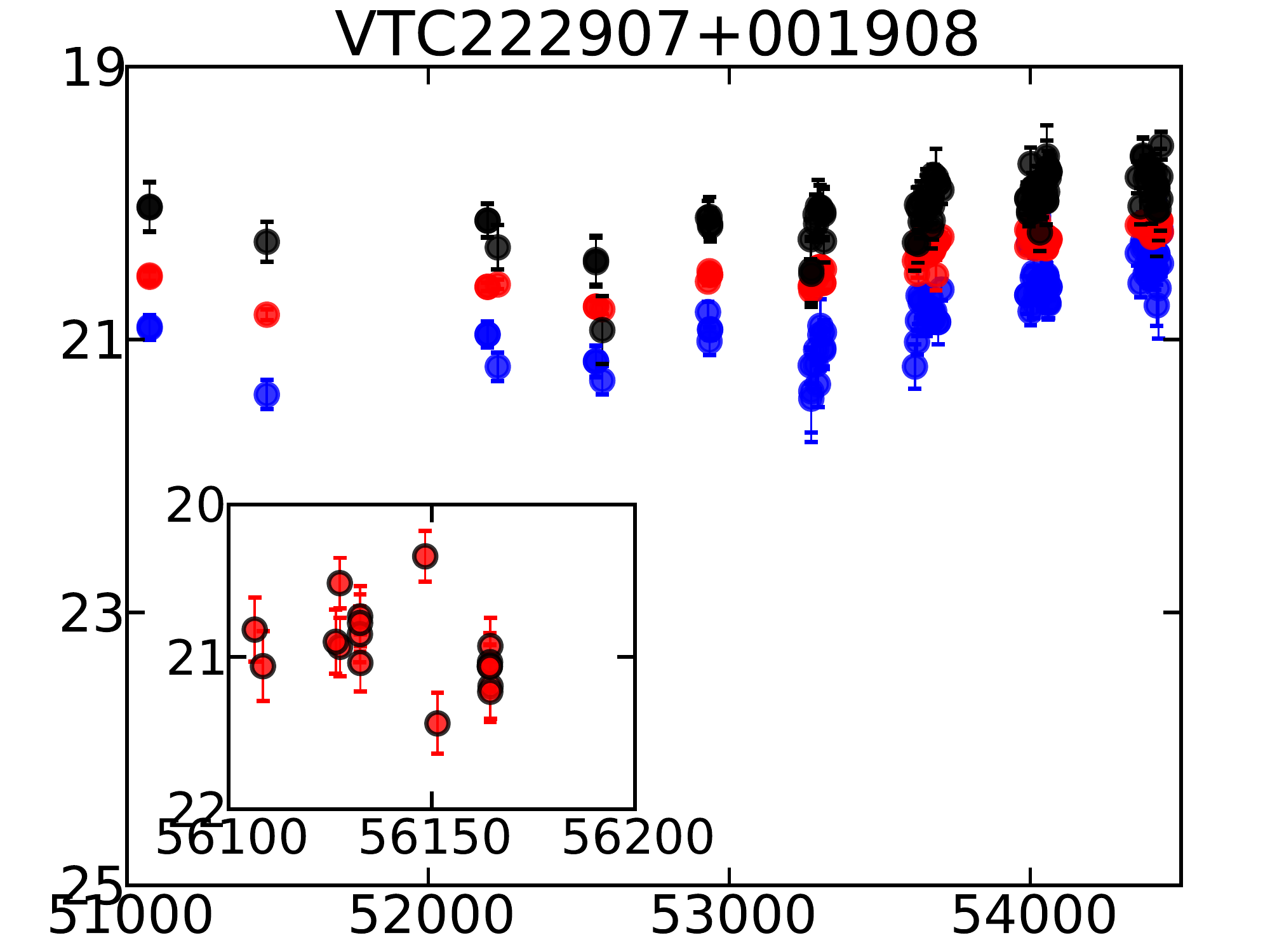}
\includegraphics[width=1.3in]{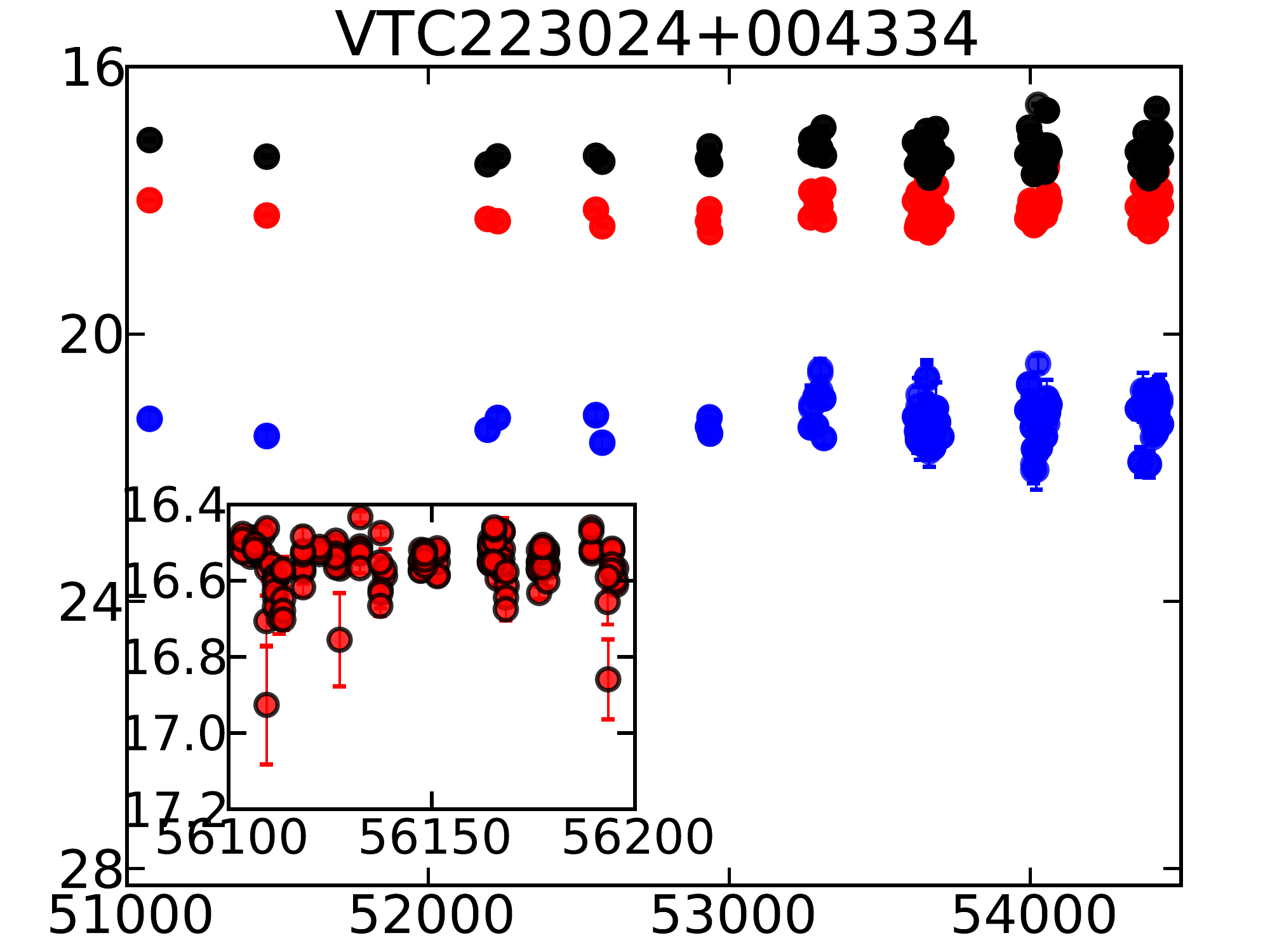}
\includegraphics[width=1.3in]{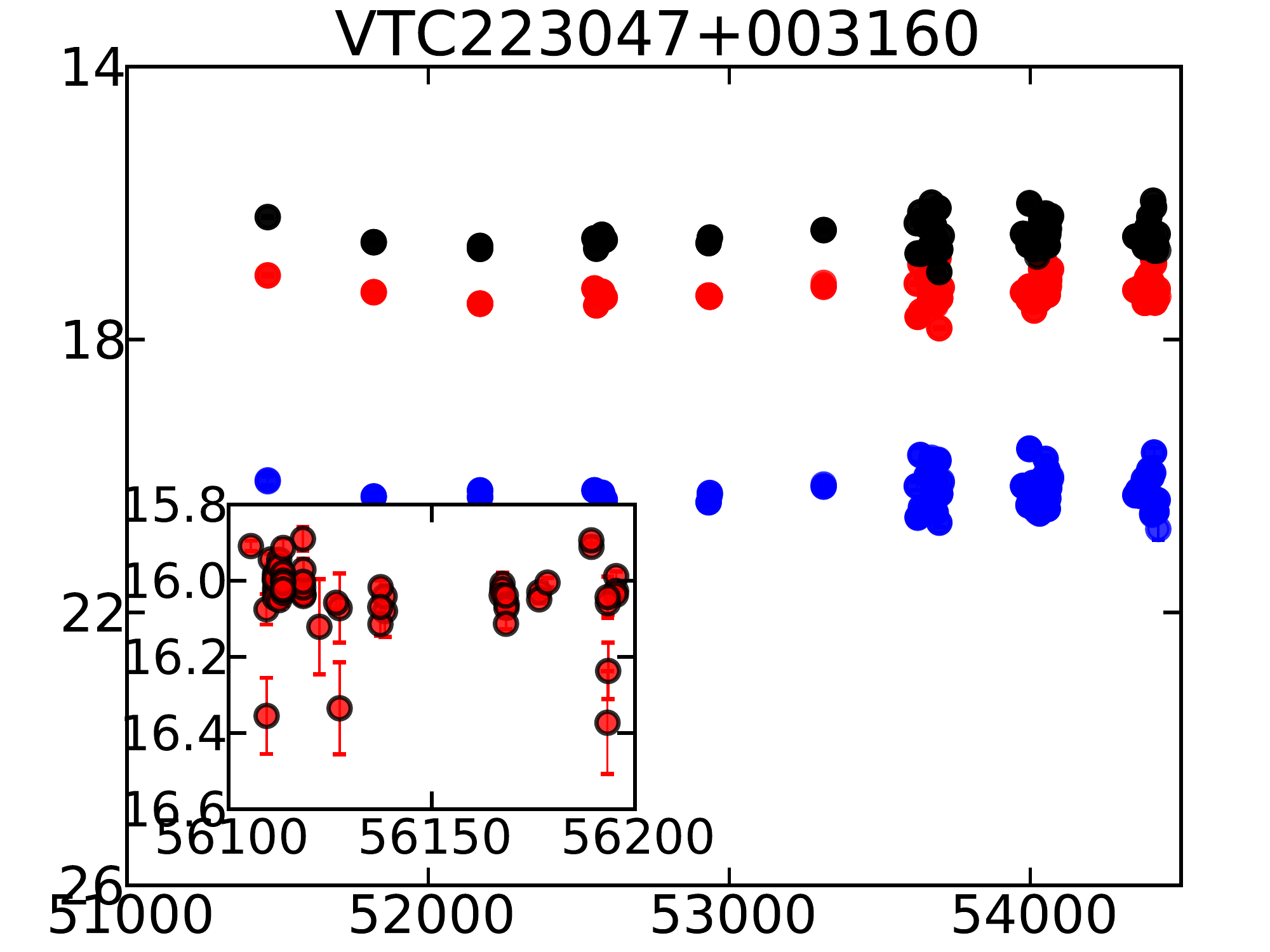}
\includegraphics[width=1.3in]{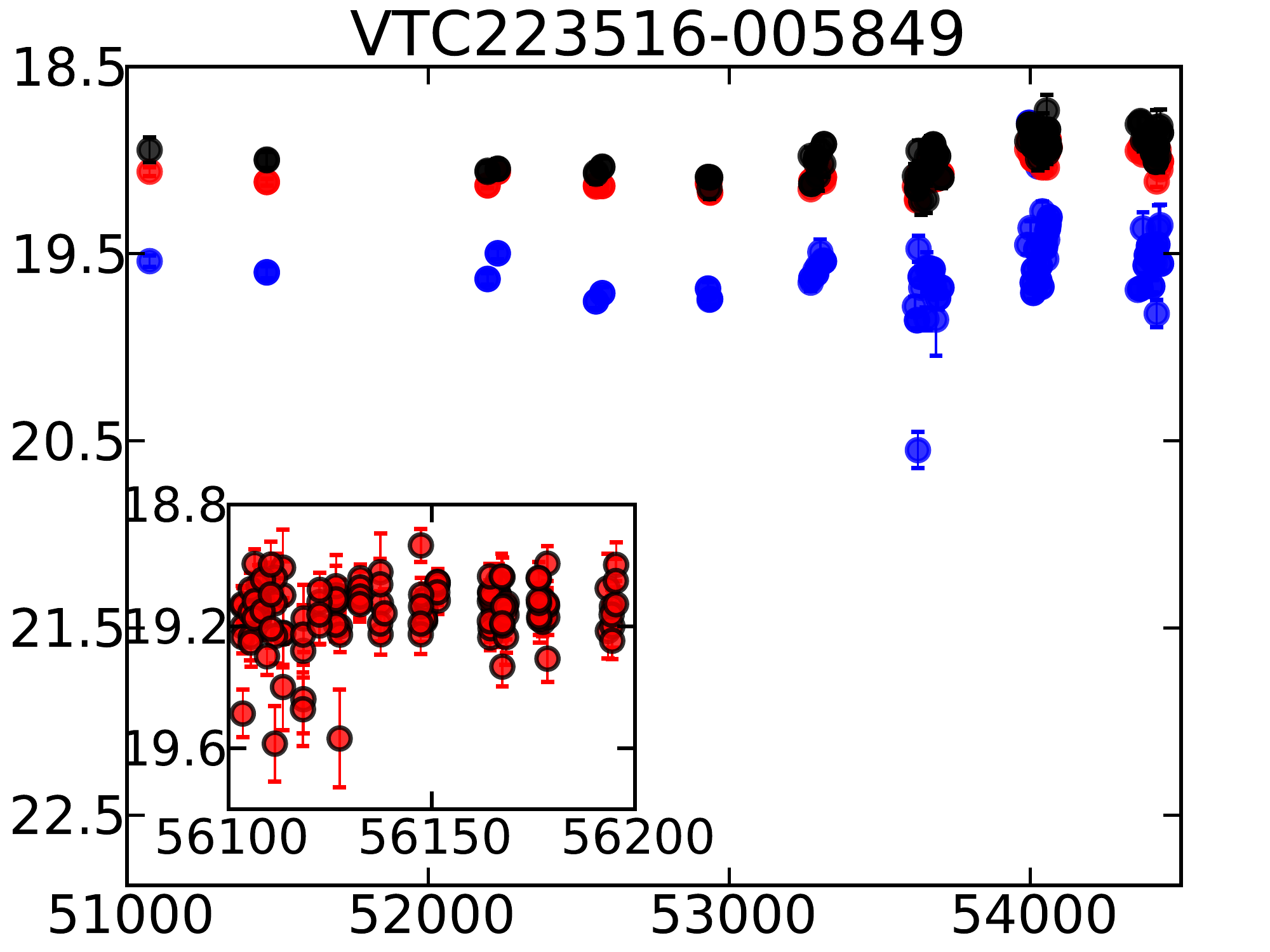}
\includegraphics[width=1.3in]{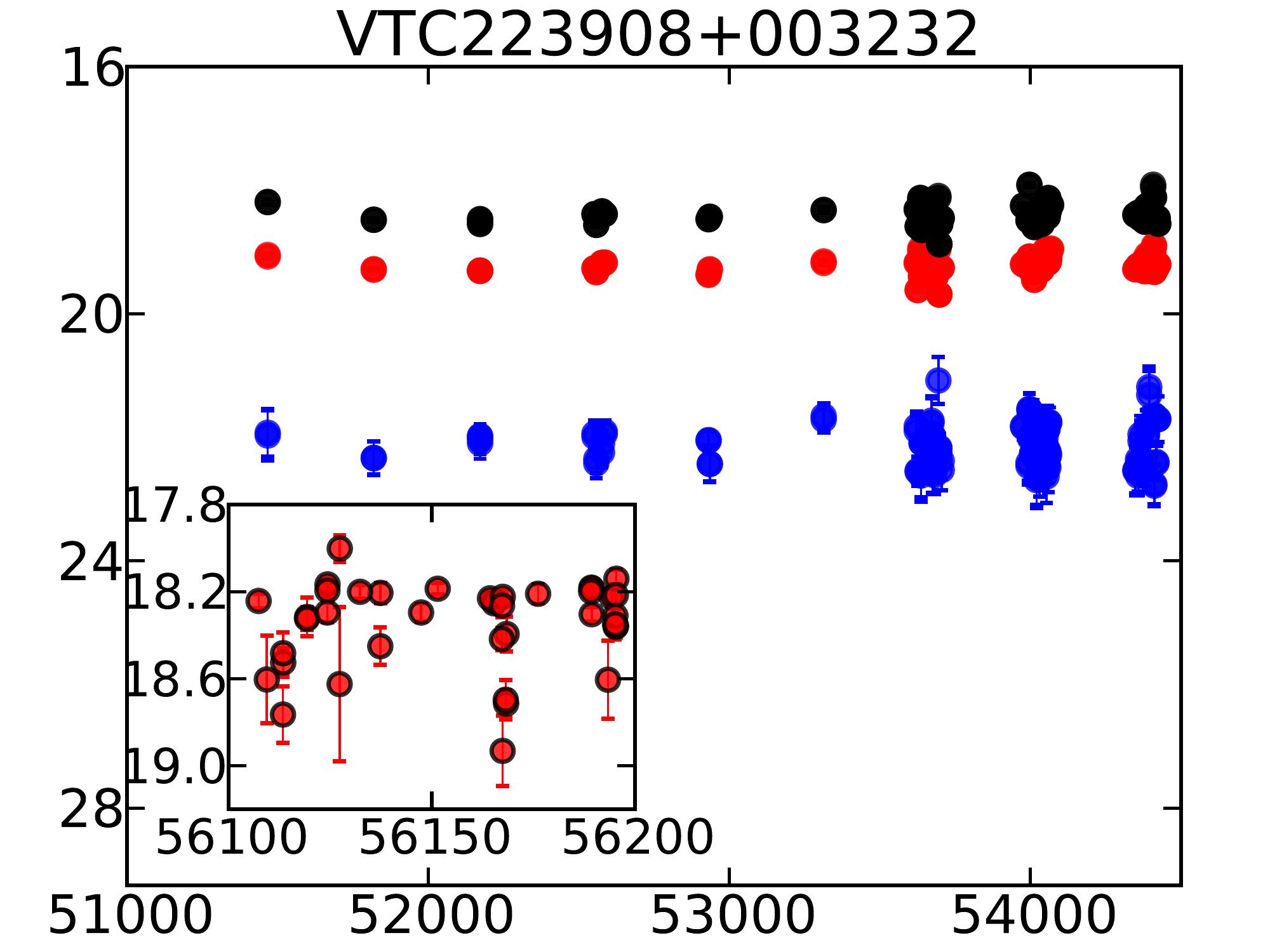}
\includegraphics[width=1.3in]{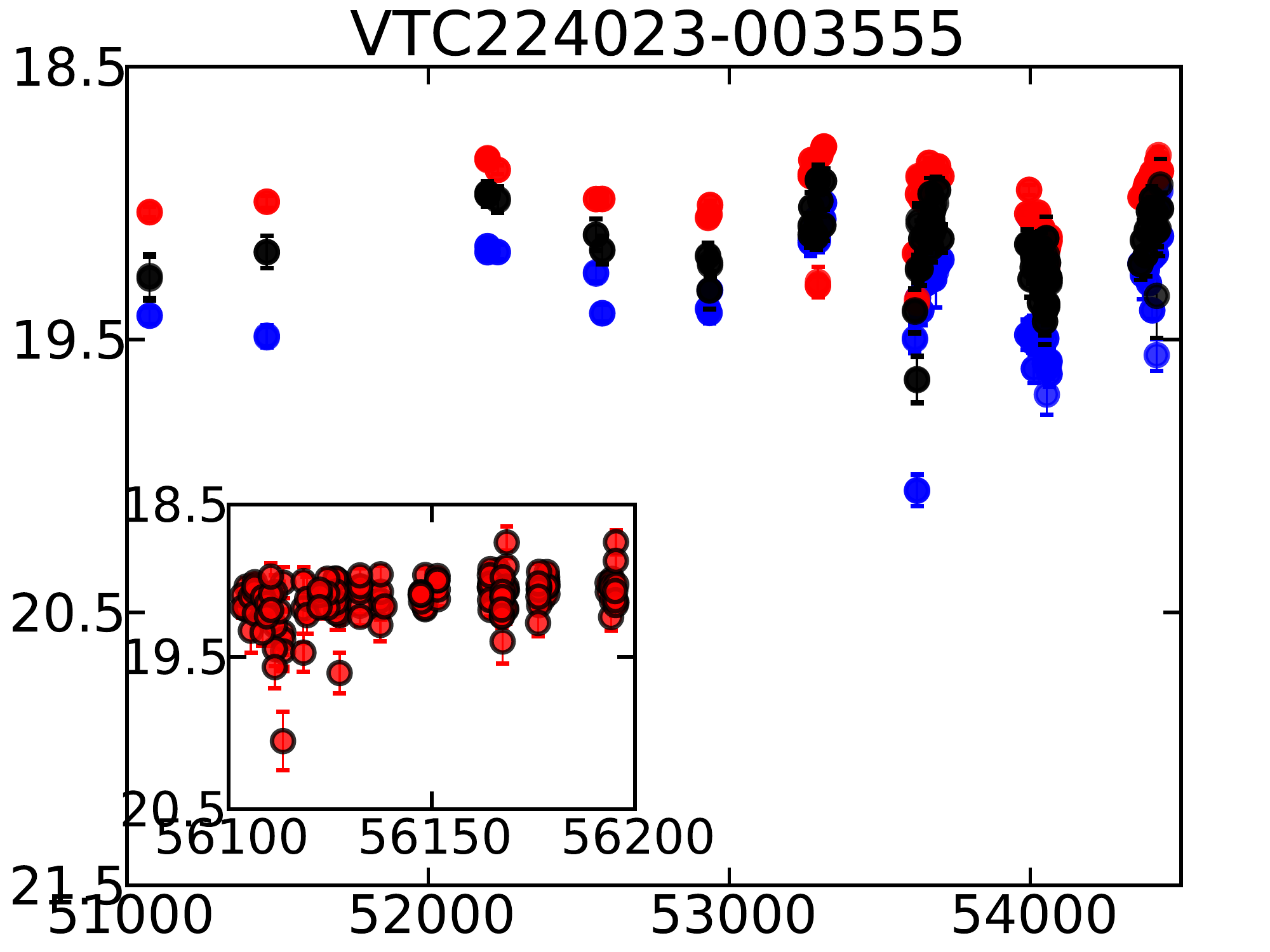}
\includegraphics[width=1.3in]{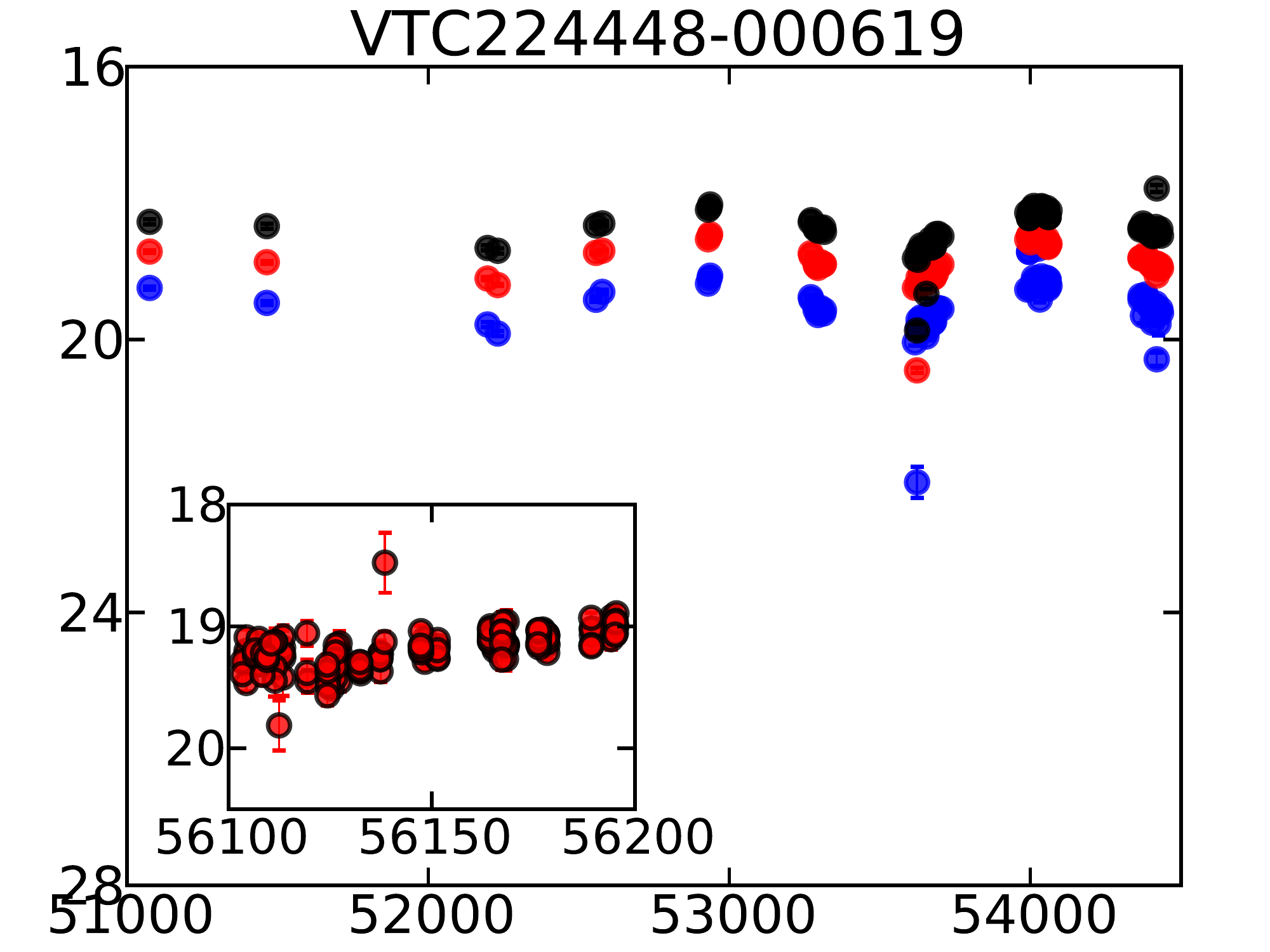}
\includegraphics[width=1.3in]{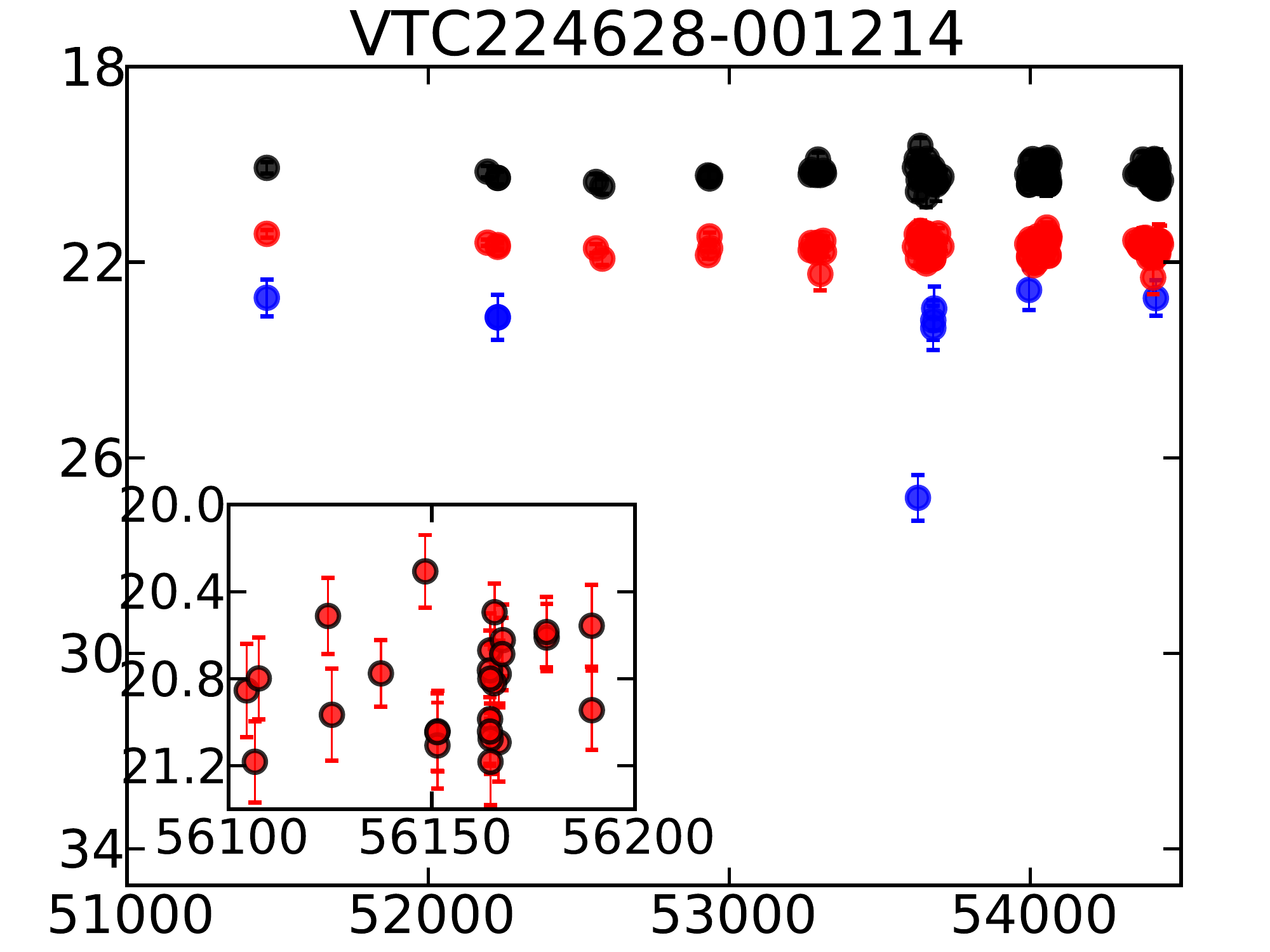}
\includegraphics[width=1.3in]{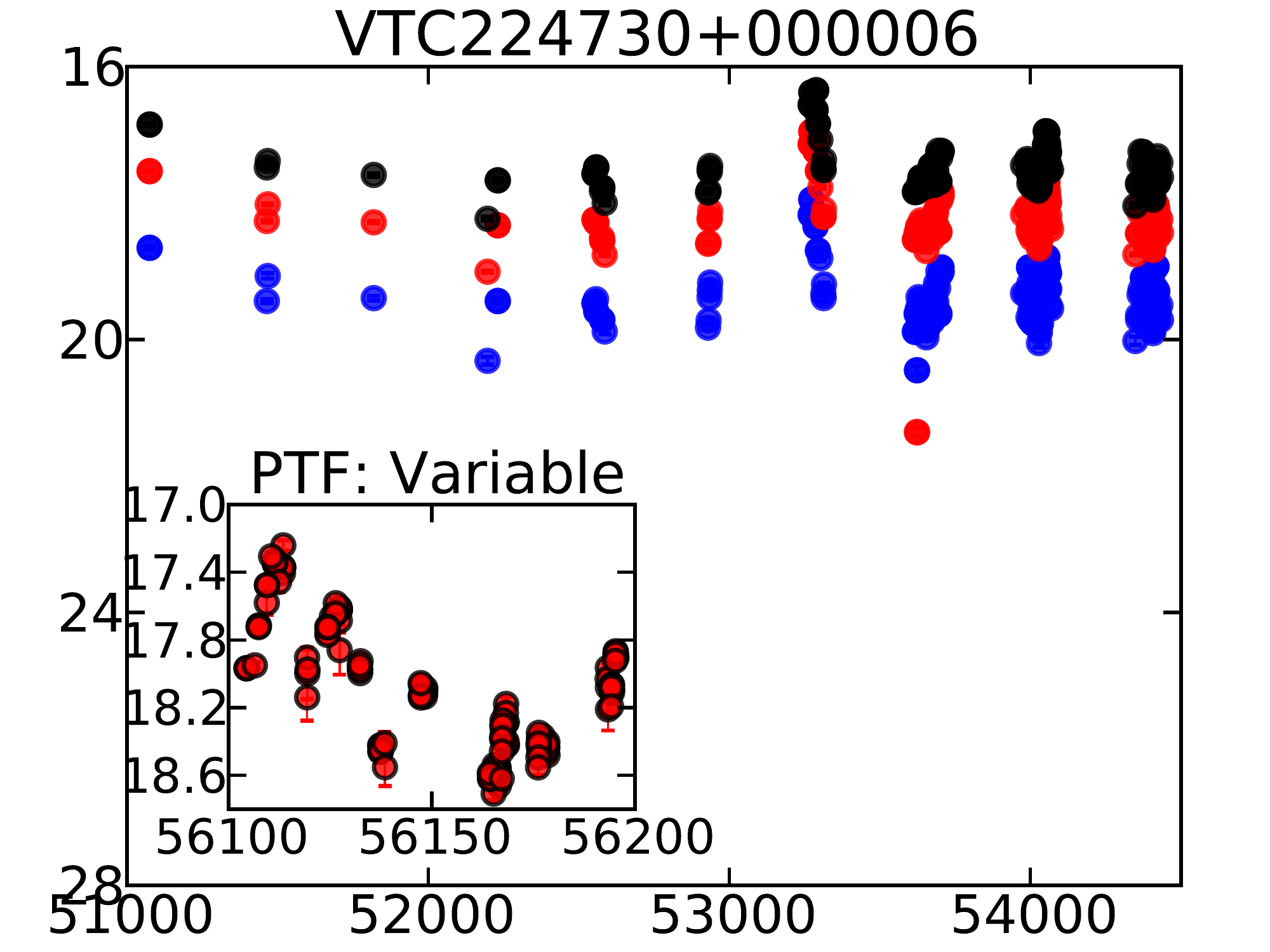}
\includegraphics[width=1.3in]{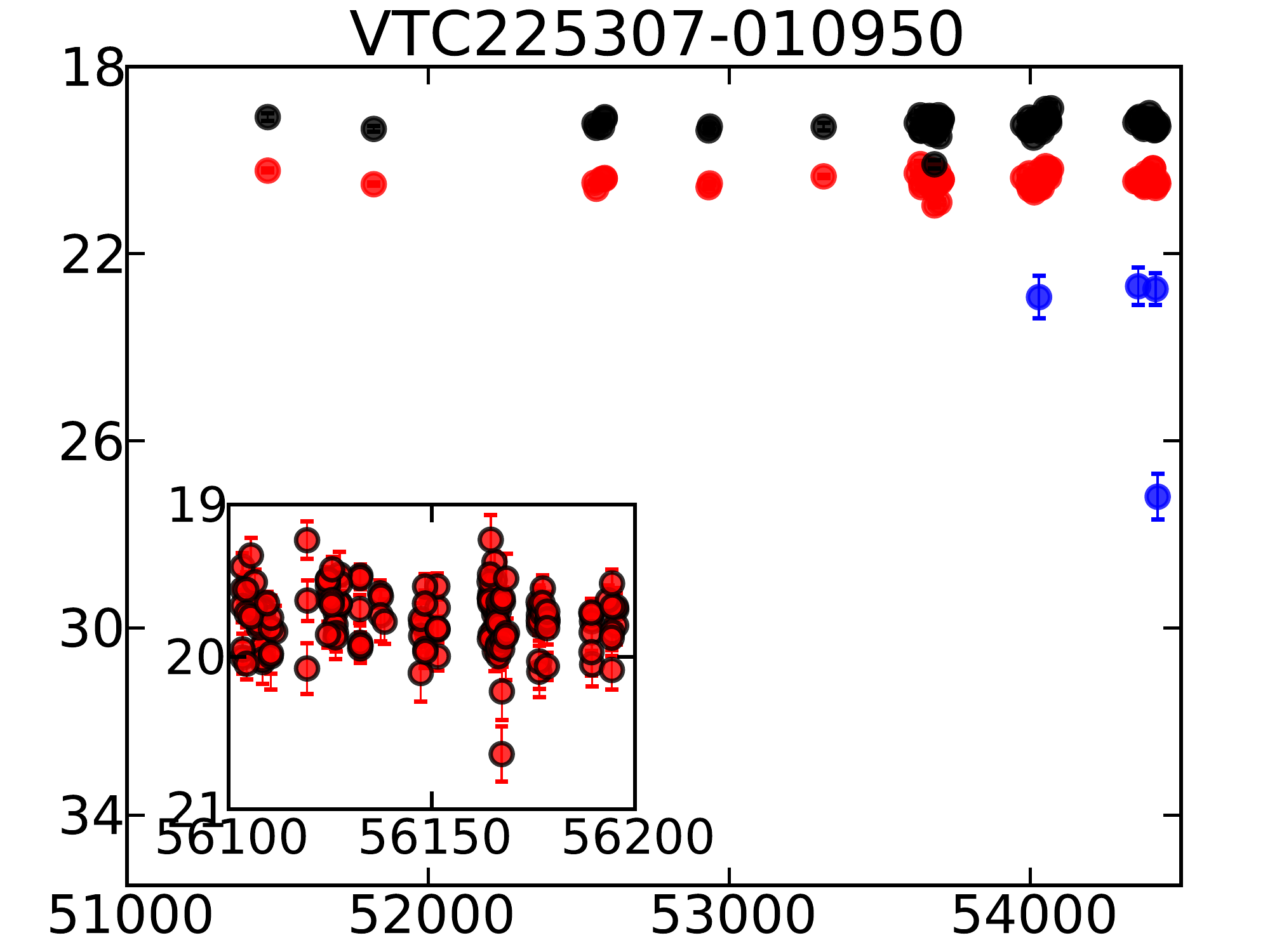}
\includegraphics[width=1.3in]{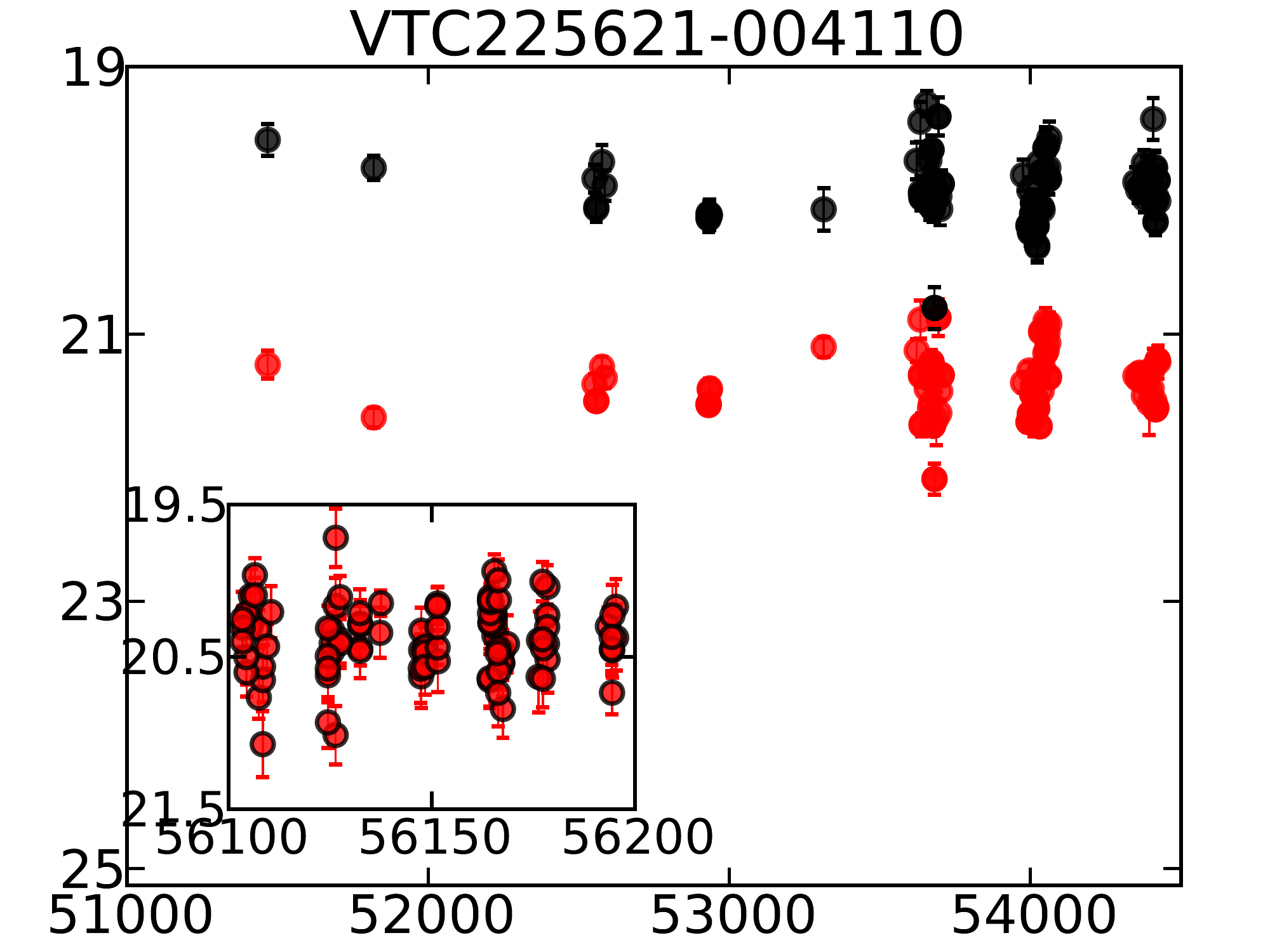}
\includegraphics[width=1.3in]{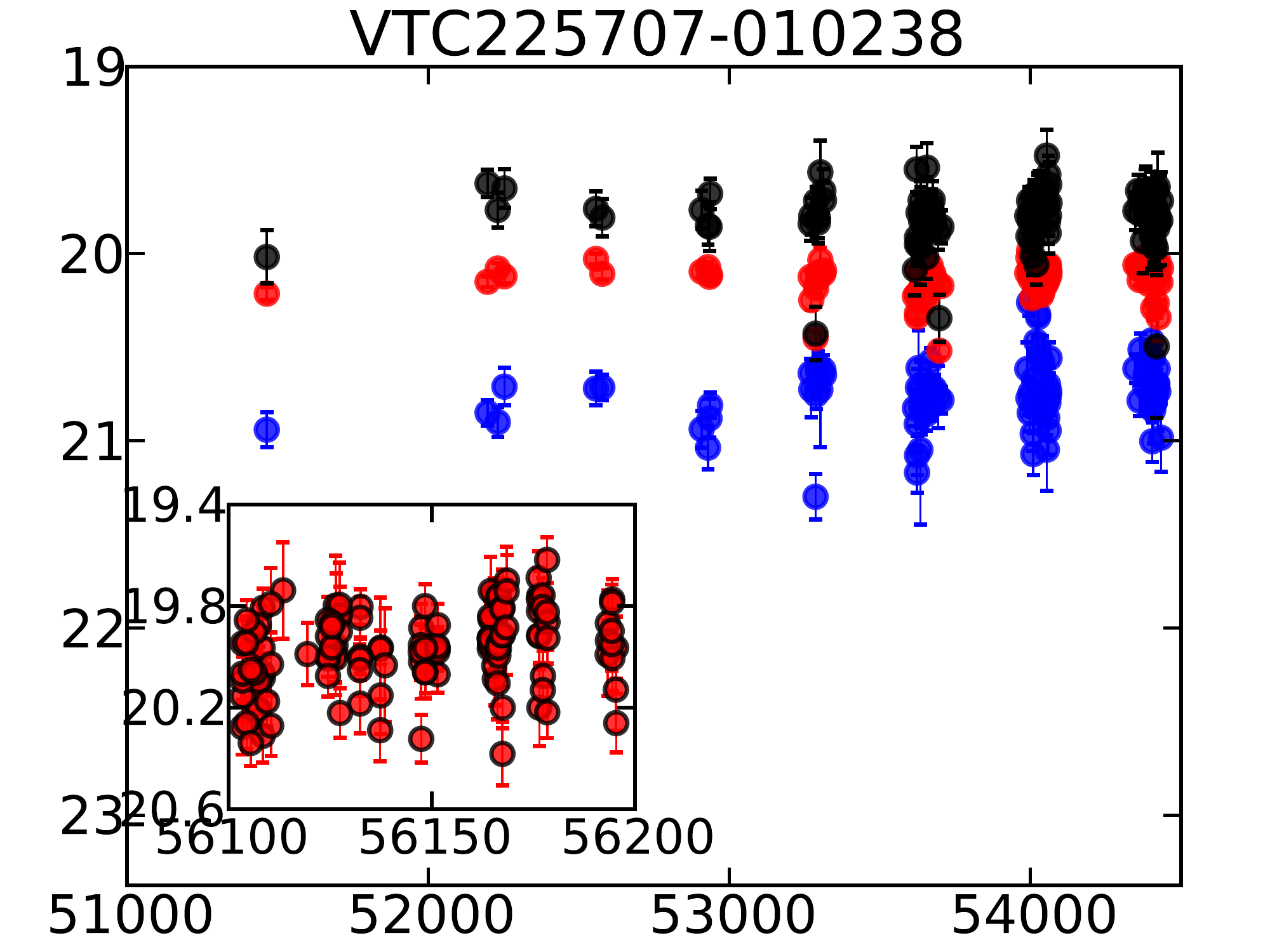}
\includegraphics[width=1.3in]{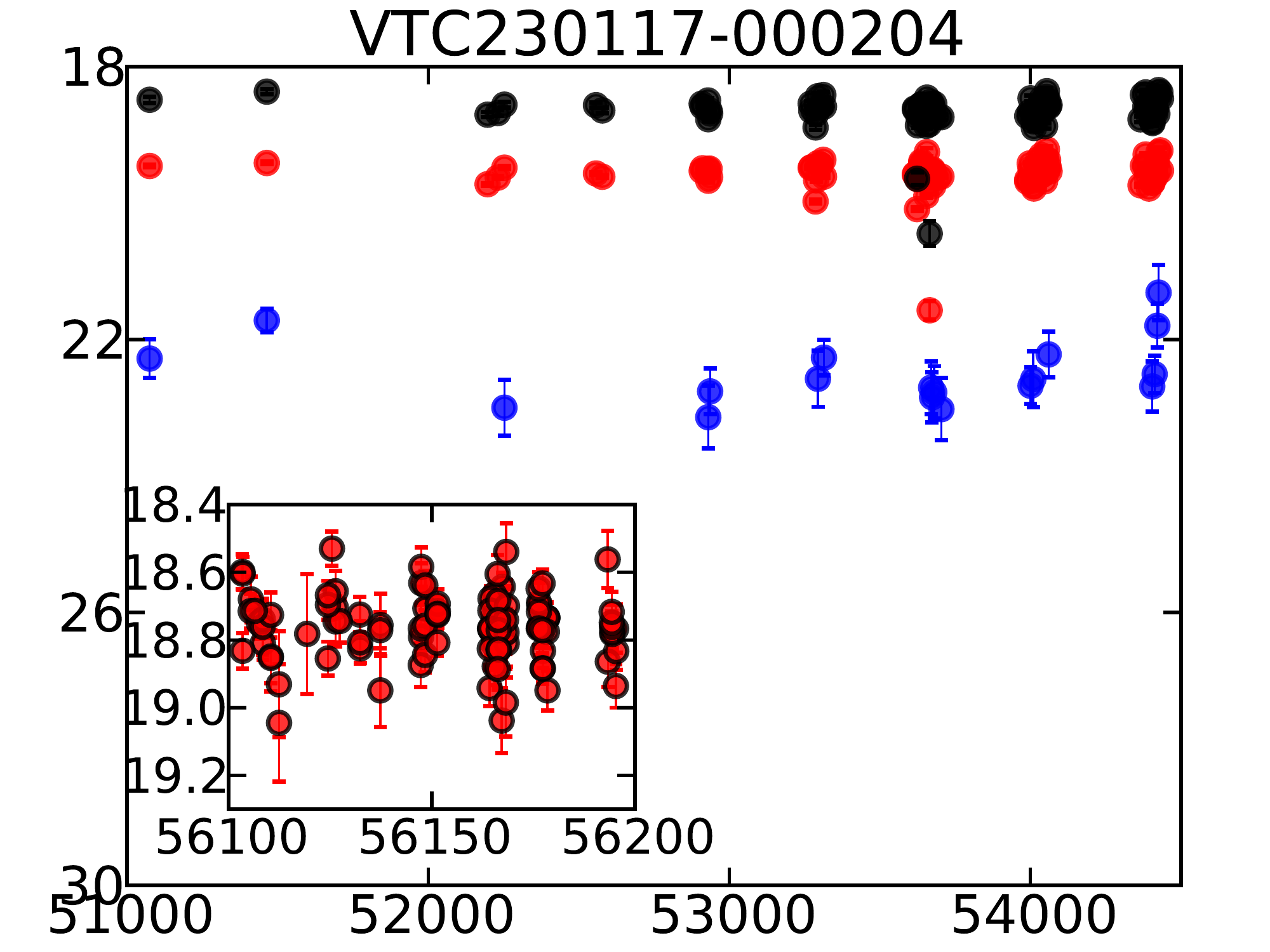}
\includegraphics[width=1.3in]{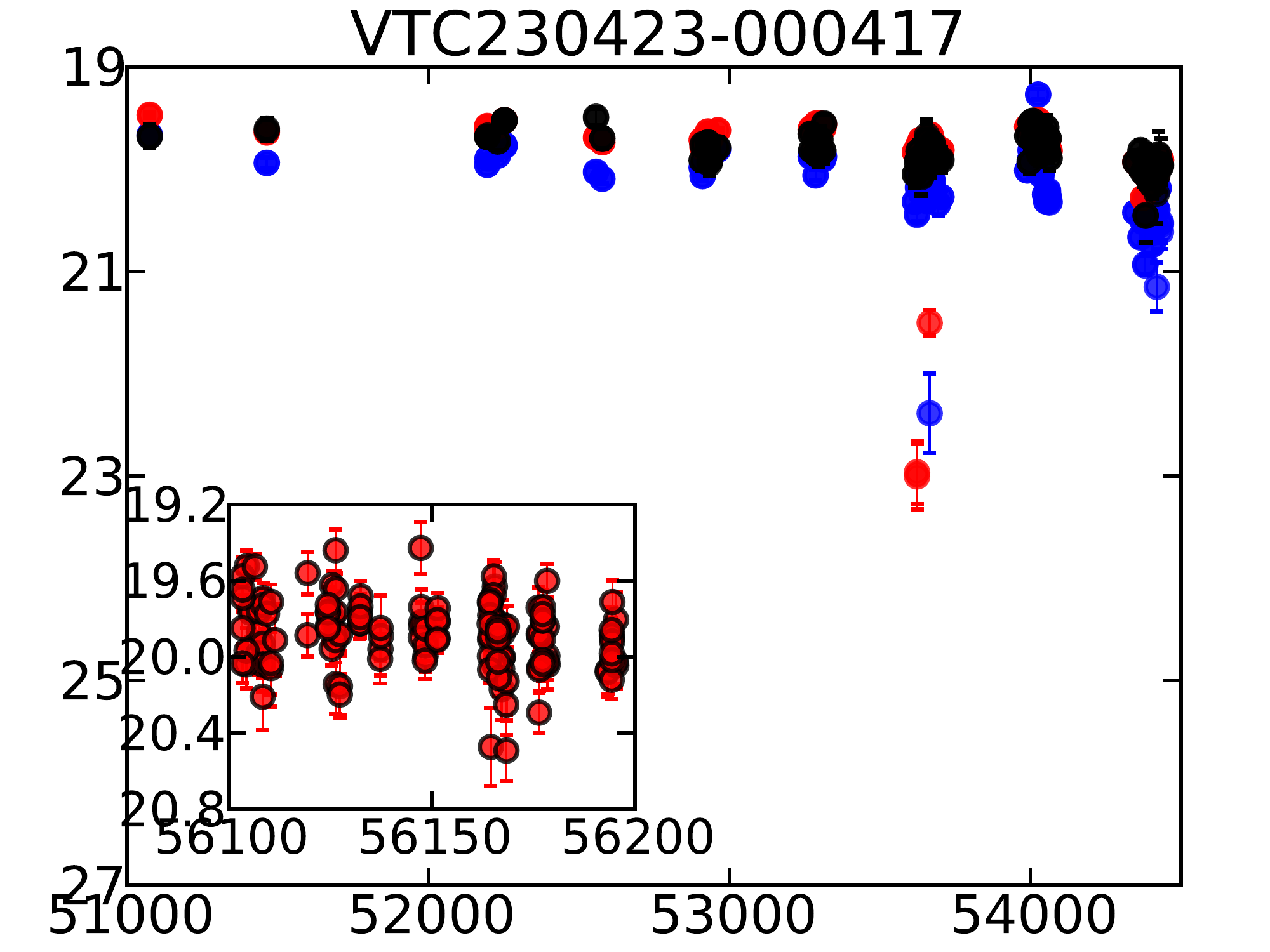}
\includegraphics[width=1.3in]{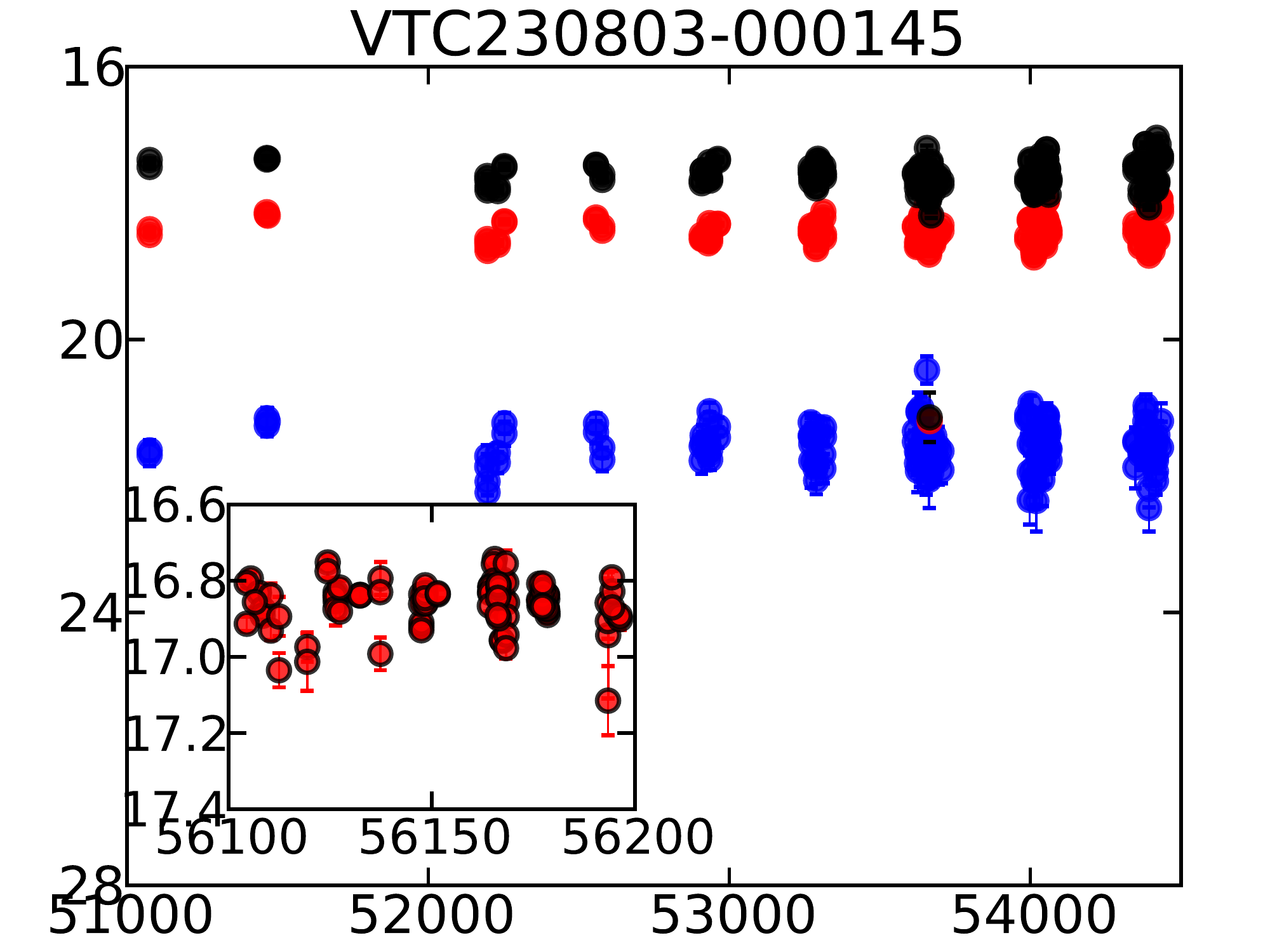}
\includegraphics[width=1.3in]{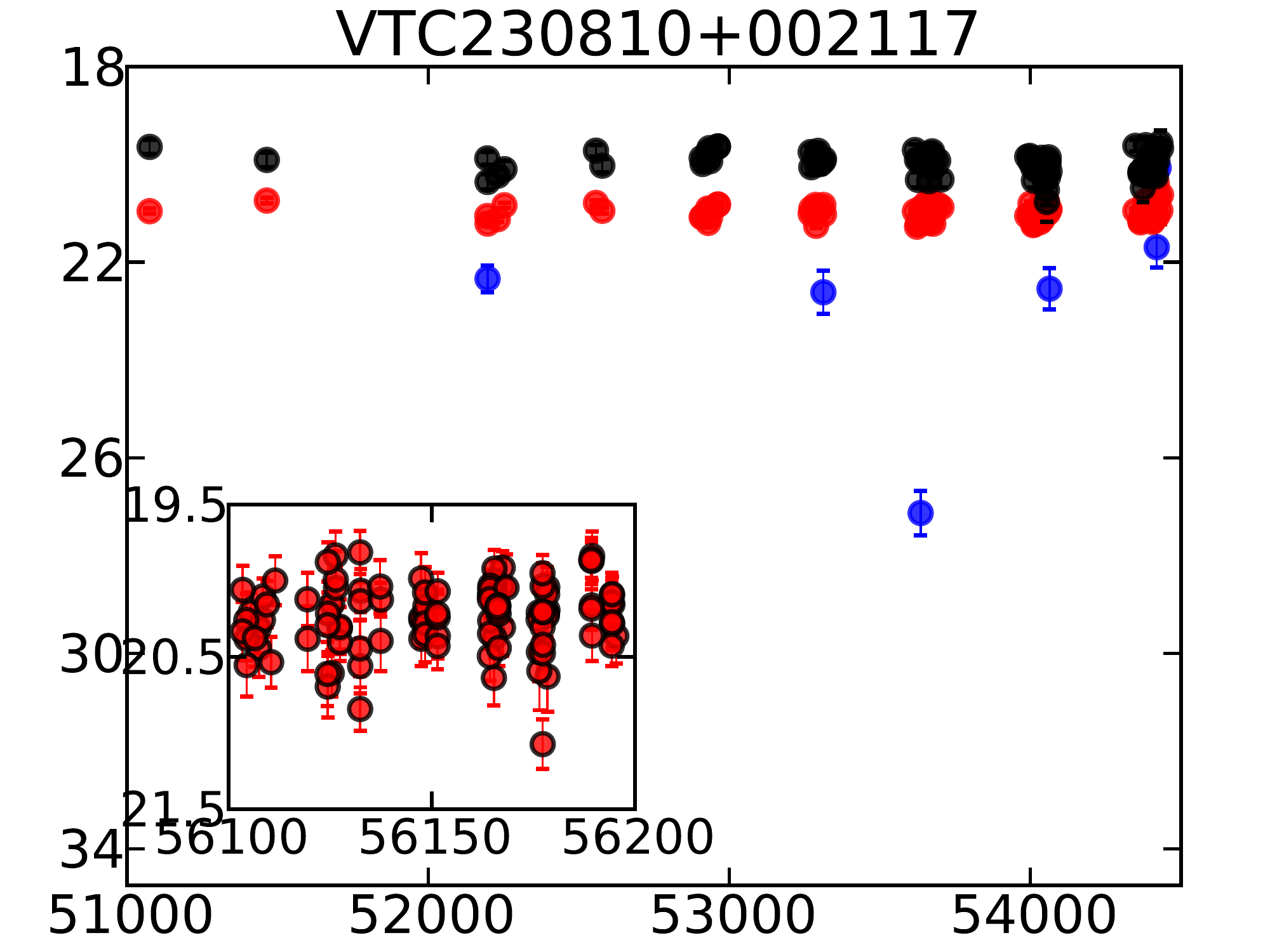}
\includegraphics[width=1.3in]{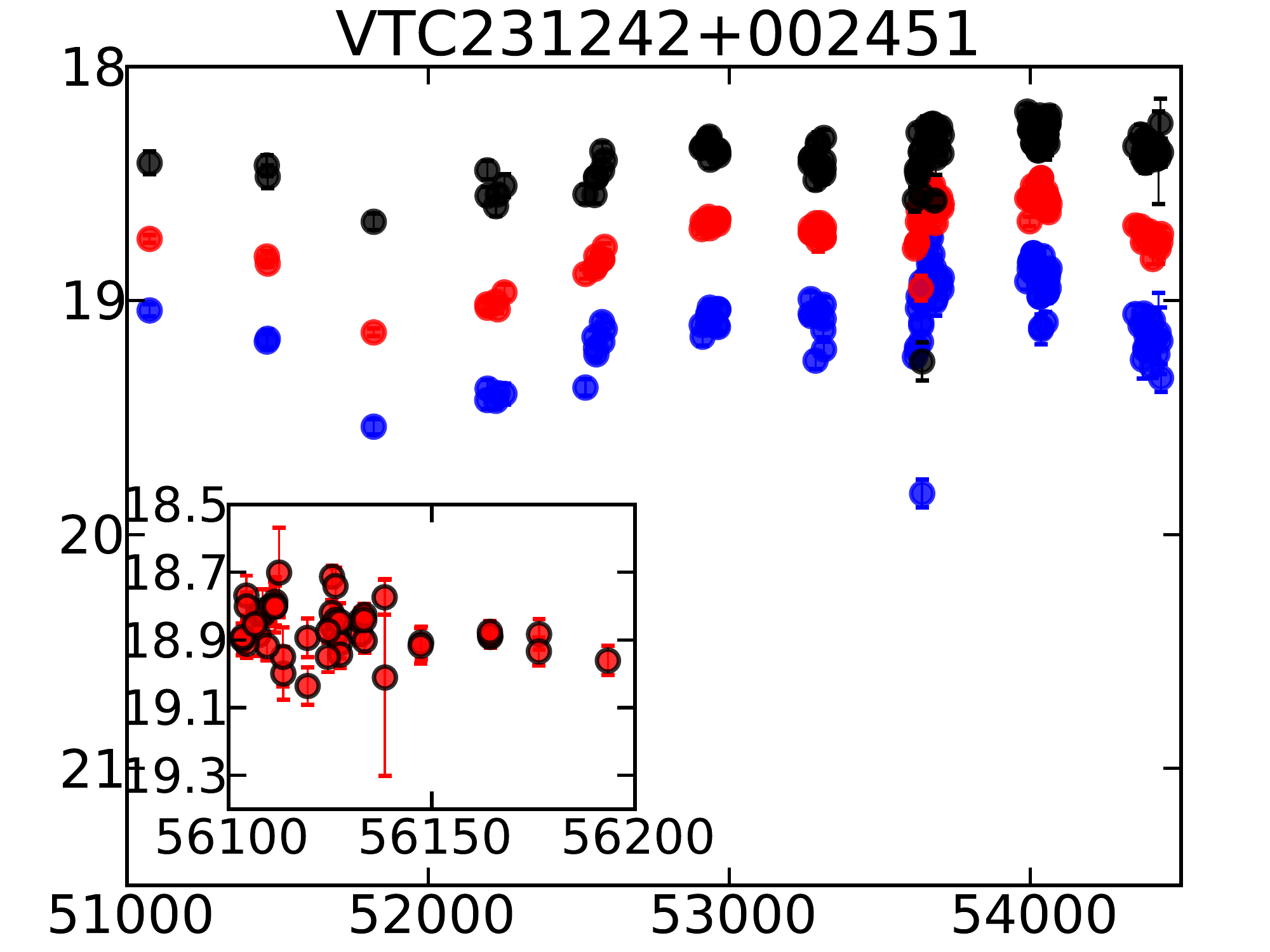}
\includegraphics[width=1.3in]{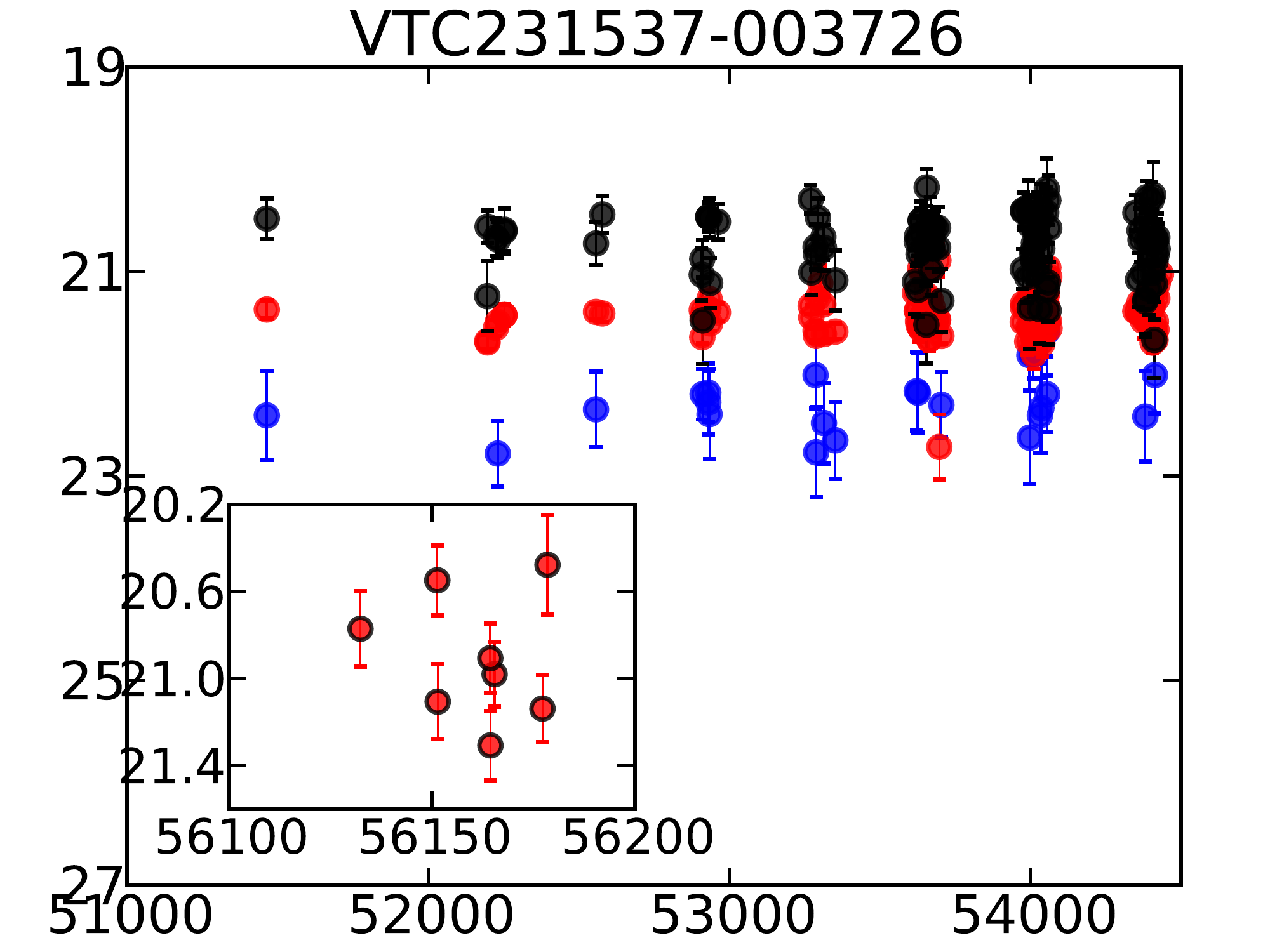}
\includegraphics[width=1.3in]{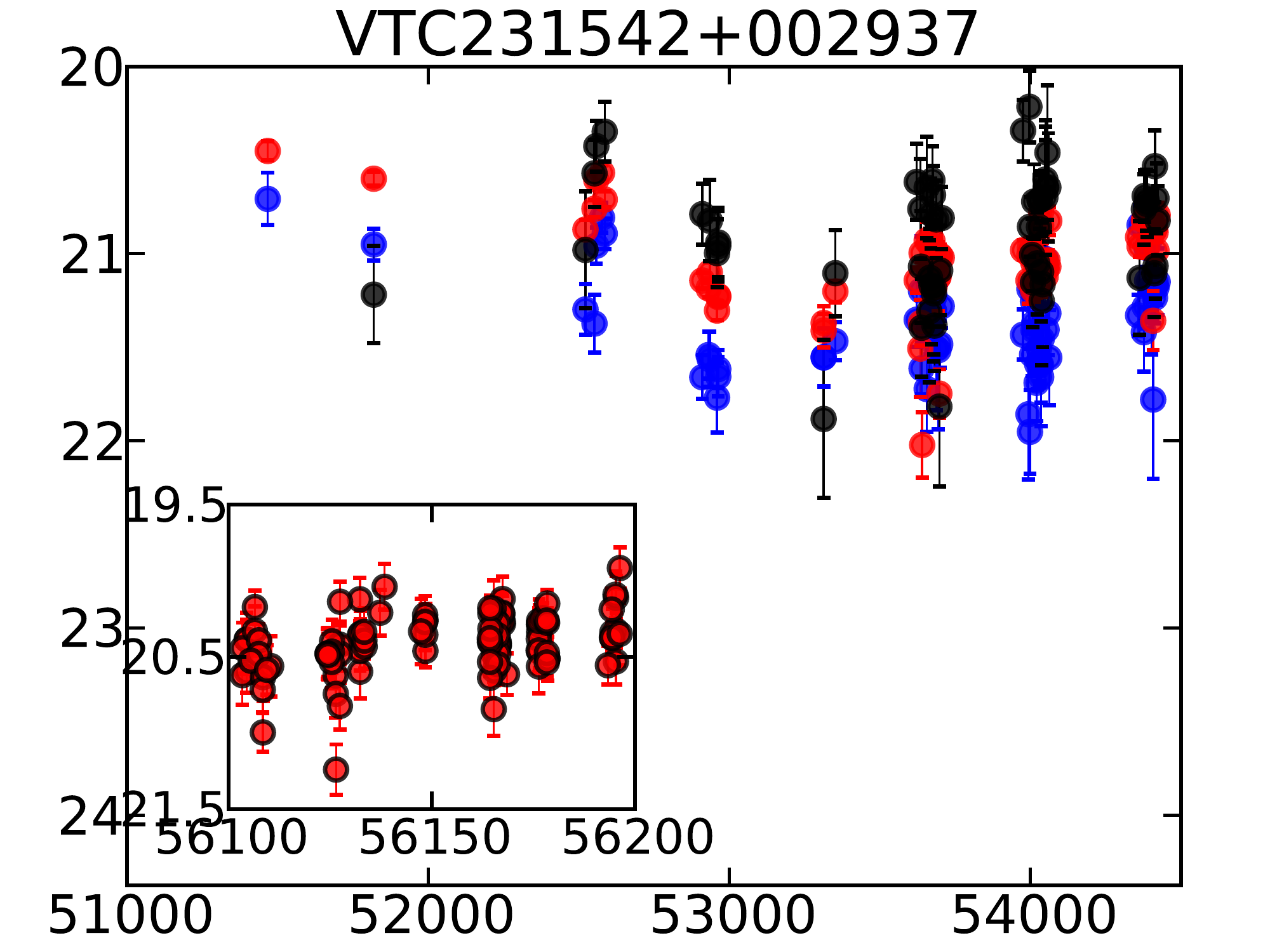}
\includegraphics[width=1.3in]{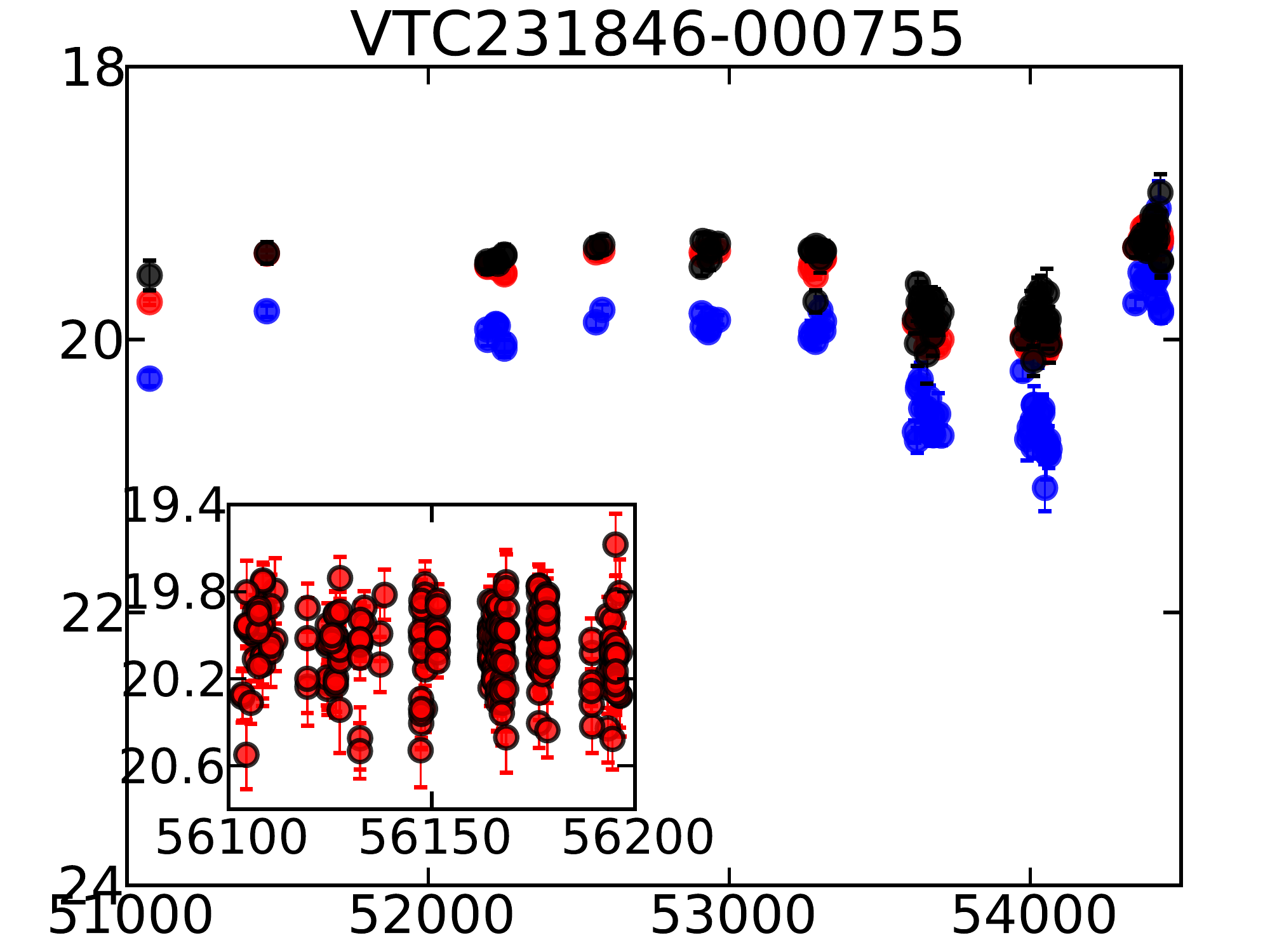}
\includegraphics[width=1.3in]{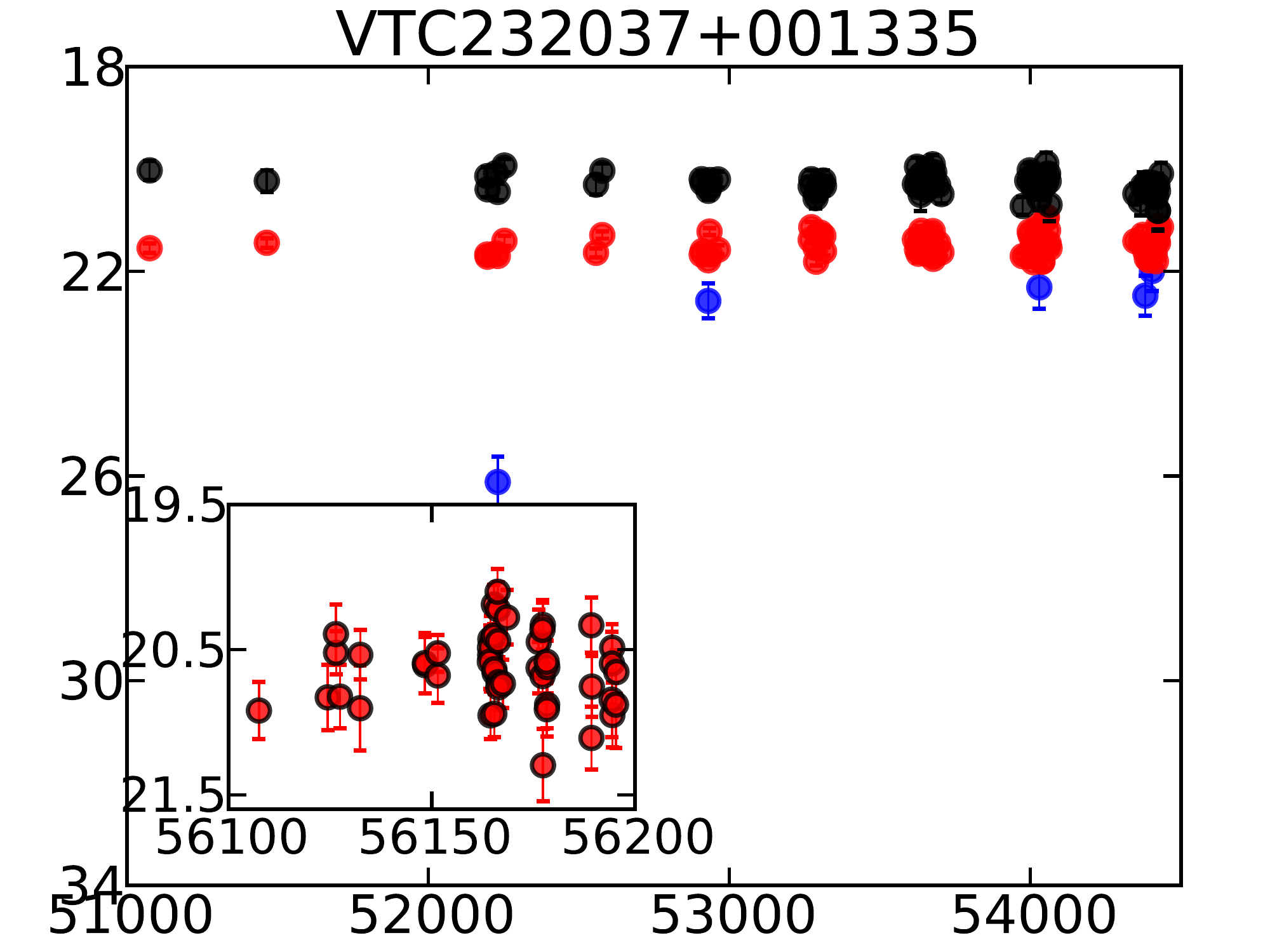}
\includegraphics[width=1.3in]{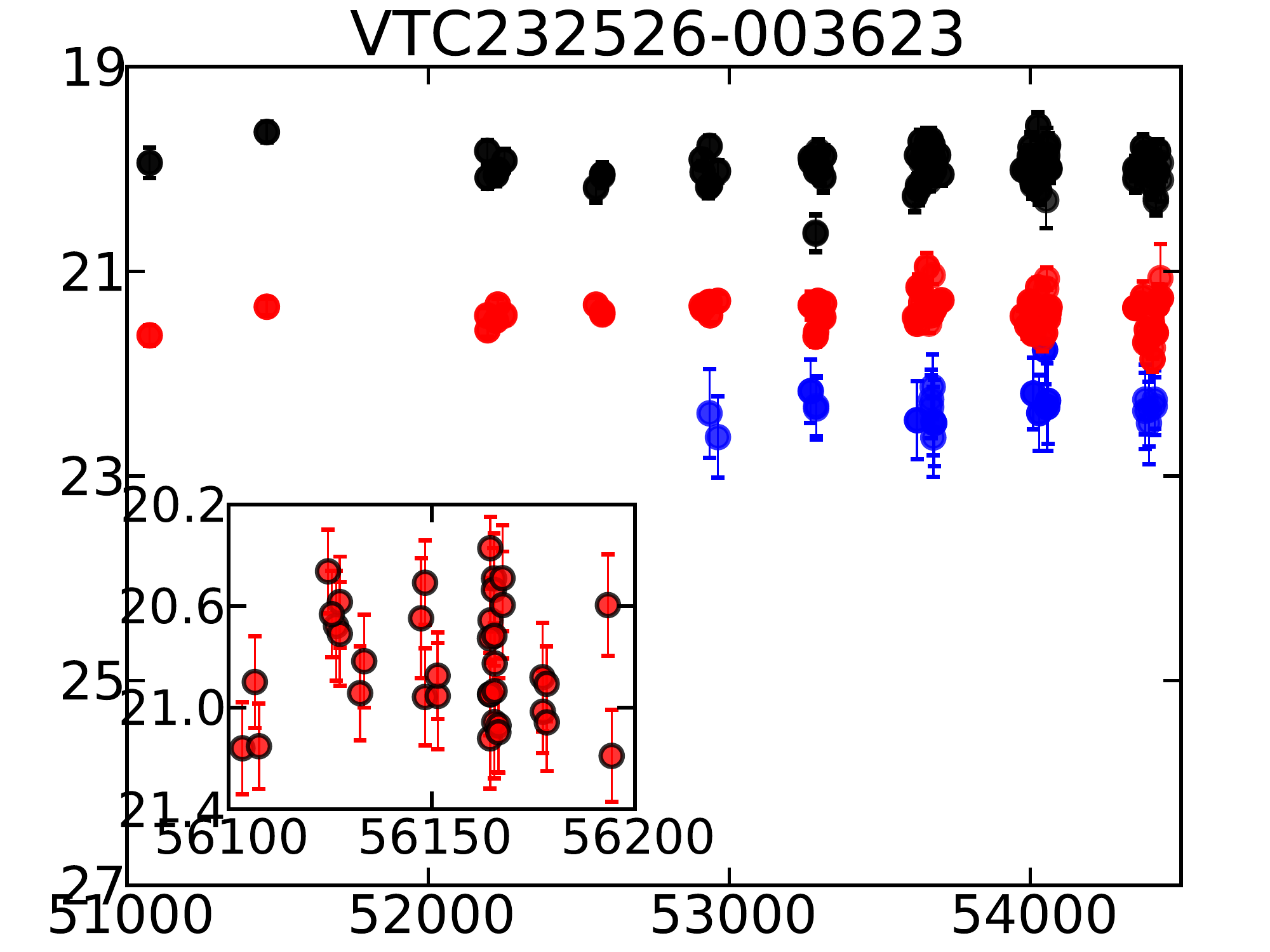}
\includegraphics[width=1.3in]{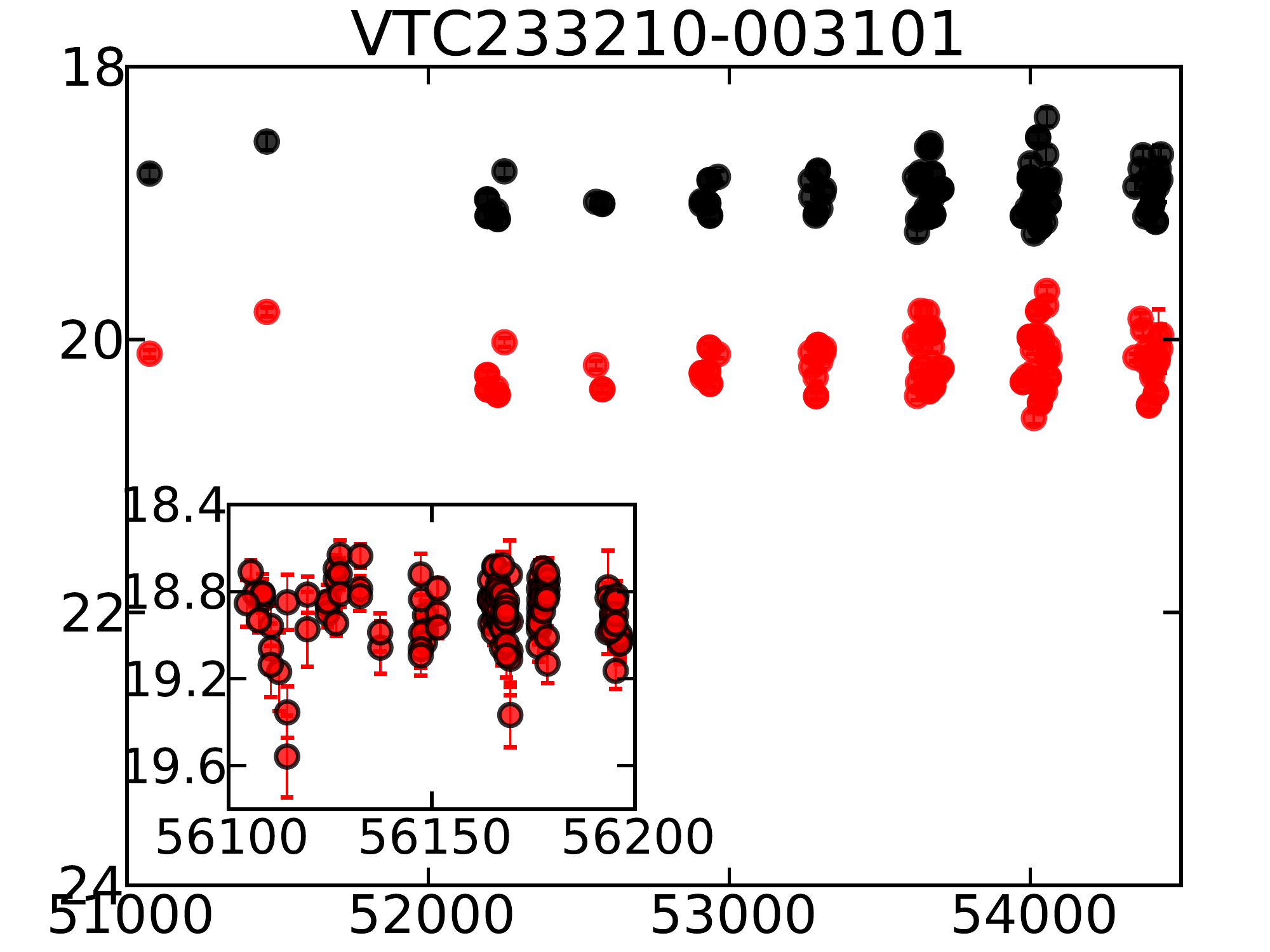}
\includegraphics[width=1.3in]{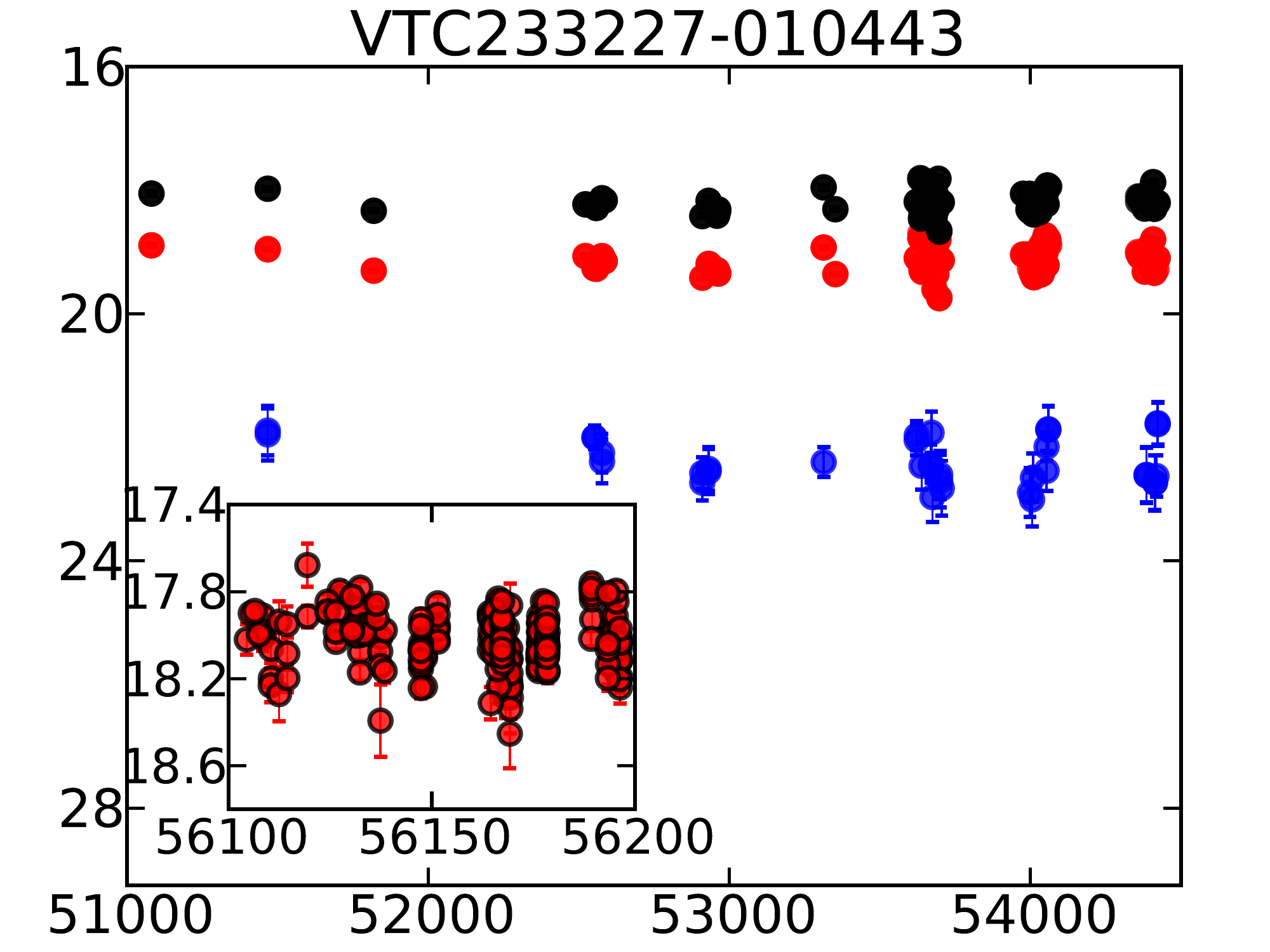}
\caption{Well-sampled optical light curves of radio variable sources from SDSS (showing variability on timescales of years) and PTF (inset; showing variability on timescales of weeks).
Majority of these light curves reveal sub-magnitude or magnitude-level variability on timescales of months to years, indicative of AGN.
For SDSS, the u-, r-, and z-band light curves are shown as blue, red, and black circles with errorbars respectively. 
For PTF, the R band light curve is shown.
The title of the inset describes whether the source is classified as a variable in PTF based on our variability criteria.
See \S\ref{sec:optical_properties} for more details.
The y-axis in each plot is the magnitude and the x-axis is the Modified Julian Date (MJD).
The x-axis runs from 51000 to 54500 in all the panels and from 56100 and 56200 in all insets.
Note that the MJDs 51000, 54000, 56100, and 56200 correspond to Jul 1998, Sep 2006, Jun 2012, and Oct 2012 respectively.}
\label{fig:optical_variability}
\end{figure*}

Here we are less concerned about optical-only transients and will focus instead on the optical variability properties of the radio transients and variables.
A study of optical-only variables and transients from synoptic surveys is better done elsewhere in literature \citep[e.g.][]{drake2009,rau2009}.
We have obtained photometric data from PTF and SDSS after searching for counterparts as described above.
A compilation of the optical light curves of the radio variable sources is given in Figure~\ref{fig:optical_variability}.
Majority of the PTF light curves reveal sub-magnitude variability, while the SDSS light curves show gradual sub-magnitude or magnitude-level variability on timescales of years.
This is characteristic of AGN, where fluctuations in the accretion rate or other causes give rise to optical variability on a wide range of timescales.
Coupled with radio variability information, it seems likely that all of these sources are AGN, excepting the ones associated with known stars.

\section{Summary \& Discussion}\label{sec:summary}

We have carried out a dedicated radio transient survey in a 50 deg$^2$ region of the SDSS Stripe 82. 
This survey is a pilot for the Caltech-NRAO Stripe 82 Survey (CNSS), a multi-epoch survey of the entire $\sim$270 deg$^2$ of Stripe 82, which is currently underway.
The pilot observations were carried out with the Jansky VLA between 2--4 GHz over four epochs spaced logarithmically in time, i.e. having cadences of one week, one month, and 1.5 years.
The median rms noise per epoch is between 50 $\mu$Jy and 90 $\mu$Jy.
With this pilot we have successfully demonstrated the near-real-time calibration, imaging, and transient search capability with the Jansky VLA data.
We have found 142 sources displaying fractional variability beyond 30\%. 
Based on radio follow-up observations, radio and optical variability, radio spectral indices, and multi-wavelength archival photometric data, almost all of these are likely to be due to shocks in AGN jets.
We have also found two bona fide radio transients associated with Galactic objects, a flare from an RS CVn binary and a dKe flare.
Comparison of our pilot survey data with the FIRST and VLA-Stripe 82 surveys has (unexpectedly) revealed a few additional, highly variable and transient, sources on timescales of 5--20 years.
These sources, most of which are either previously-known radio sources or have optical AGN hosts, are likely associated with renewed AGN activity. 
Two other transients, found serendipitously, suggest that there are indeed many more transients in this dataset than the ones which we have found through our stringent selection criteria.
We encourage the astronomical community to take advantage of the data from this pilot radio survey (see \S\ref{sec:radio_processing:final:DR} for the data release website) for finding 
these hidden transients and for other science.

A high-cadence optical survey using PTF was carried out over three months, contemporaneous with the first three epochs of the radio survey.
The motivation behind this joint survey was the selection of radio transients for rapid follow-up, preliminary classification of radio variables and transients based on 
optical light curves, and a direct comparison of the dynamic radio and optical skies.
As noted by previous radio and optical studies (see \S\ref{sec:optical_properties}), 20\% of the persistent radio sources have an optical counterpart above the PTF limiting magnitude of 21, 
while 50\% have counterparts above the SDSS limiting magnitude of 23.5 in the red filter.
We have found primarily slowly-evolving radio transients, and, within the sensitivity limits of our PTF and pilot Jansky VLA surveys, very little overlap between the optical and radio variable skies.
Only 0.05\% of the radio point sources and $<$0.001\% of the optical sources are variable at both frequencies.
The fraction of strong variables per square degree of the radio decimetric sky is at least an order of magnitude smaller than those in the optical sky.
We will now use the radio data to assess the degree of variability and the transient rate of the radio sky, and make recommendations for future surveys.

\subsection{Comparison of variability with previous surveys}\label{sec:summary:variability}

We found only a small fraction ($142/3652 = 3.9^{+0.5}_{-0.9}\%$) of the point sources varying by $>$30\% on week-month-year timescales. 
Majority of these are variable only on a 1.5 year timescale, and as described in \S\ref{sec:radio_transients:final}, this variability fraction is to be taken as the upper limit.
Several previous studies at 1.4 GHz \citep{frail1994,deVries2004,croft2010,bannister2011a,bannister2011b,thyagarajan2011,mooley2013} 
have shown that the fraction of variables on timescales between minutes and years, and flux densities between 0.1 mJy and 100 mJy, is $\sim$1\% or less.
Similar to these studies\footnote{see http://tauceti.caltech.edu/kunal/radio-transient-surveys/index.html for a description of 
past variability studies}, we see that the 3 GHz sky is not highly variable on timescales longer than a few days.
Radio follow-up observations, optical and radio light curves, and multi-wavelength archival data of the radio variable sources suggest that almost all of these are shock-related 
flaring in the jets of AGN.

Our choice of modulation index for measuring variability and selection threshold introduces a bias towards selecting sources with large flux densities.
A source in the PSC would have to have a mean flux density of 1.4 mJy in order to detect a 30\% fractional variability, while at the 7$\sigma$ source detection limit of 
of $\sim$0.5 mJy, a source would have to have $|m|>3/4$ ($f_{\rm var}>$120\%) in order to be identified as a significant variable.
Fifteen strong variables (i.e. sources having $|m|>2/3$, or $f_{\rm var}>$100\%) were identified in our survey, but only 3000 sources in the PSC are bright enough ($>$0.55 mJy) 
to have been identified as a strong variable. 
We can therefore conclude that the fraction of strong variables is less than one percent.

Radio variability appears to be a function of timescale; the variability on hours$\sim$days timescales contrasts with the variability on weeks$\sim$years timescales.
\cite{ofek2011} found that $\sim$30\% of point sources brighter than 1.5 mJy at 5 GHz 
were variable, with majority of the sources varying on timescales $<$10 days.
This variability is primarily small amplitude (modulation indices of less than 0.2).
The structure function of the variable sources constructed by \citeauthor{ofek2011} shows a sharp rise on timescales of $<$1 day, a shallow but steady rise 
between 1--10 days, and a roughly constant value beyond 10 days.
A large fraction of variables displaying small variability amplitudes on short timescales were also found by the MASIV survey \citep{lovell2008} conducted at 5 GHz.
The findings of \citeauthor{ofek2011} and \citeauthor{lovell2008} suggest that the low-amplitude variability on timescales of $<$10 days is mostly extrnisic.
Among the \citeauthor{ofek2011} radio sources, only 0.3\% have modulation indices greater than 0.2 on timescales less than two years.
The variability fraction that we find on week-month timescales in our pilot survey is similar ($\lesssim$1\%), and agrees with the 
variability fraction found in narrow-deep surveys by \cite{frail1994}, \cite{carilli2003}, and \cite{mooley2013}, wide-field surveys such as 
\cite{williams2013}, and other studies between 1--5 GHz on week-month-year timescales.
On 1.5 year timescale we find the variability fraction to be less than 4\%.
Recently, \cite{hodge2013} compared the 1.4 GHz VLA-Stripe 82 and FIRST surveys, and found $\sim$12\% (6 percent per epoch $\times$ 2 epochs) of the sources 
having fractional variablility larger than 0.3 on timescales between 7 years and 22 years, the majority of which were AGN.
While it is possible that, on these longer timescales, the fraction of variable sources in the radio sky is large, some 
of the \cite{hodge2013} variables could be artificial and solely due to angular resolution differences between the VLA-Stripe 82 and FIRST surveys\footnote{However, 
\cite{hodge2013} assert that the fractional variability of 0.30 used in their work is equivalent to a fractional variability of 0.22 in \cite{deVries2004} due to a 
bias that gets introduced from using integrated flux densities.
They also note that the distribution of variability amplitudes seen in their sample is roughly consistent with the extragalactic sample of \citeauthor{deVries2004}.}.
Indeed, \cite{becker2010}, \cite{bannister2011a}, and \cite{croft2010} find that the fraction of variables on similar timescales is a few percent or less.

Radio variability (especially extrinsic) is expected to have a frequency dependence, but this aspect of variability has not been extensively studied.
We can only use the results of past blind surveys to conclude that the fraction of strong variables is less than a few percent between frequencies of 1 GHz and 5 GHz, 
between flux densities of $\sim$0.3 mJy and $\sim$100 mJy, and over a wide gamut of timescales (between one day and several years).



In our pilot survey, we found a single AGN, VTC233002-002736, with flux density $>$3 mJy at 1.4 GHz, that appears to have increased in flux density by more than a factor of ten over the past 15 years.
Similar objects may have been found earlier, by \cite{bannister2011a,bannister2011b}, in the SUMSS survey (e.g. J201524-395949 or J060938-333508).
It is likely that such phenomena are a result of episodes of enhanced accretion leading to increased jet 
activity \citep[see][for discussions of possibly related phenomena]{kunert2006,czerny2009,kunert2010,elitzur2014,keel2014,lamassa2015}.
Assuming a timescale of $\sim$20 years for an enhanced accretion episode, and given that fact that 50 deg$^2$ of the sky has $\sim$2000 AGN with flux density $>$3 mJy \citep{white1997,hopkins2002}, we can estimate 
the period of occurrence of such episodes over the lifetime of an AGN: $\sim$40,000 years.
This is remarkably consistent with previous studies \citep{reynolds1997,czerny2009,kunert2010} suggesting some young radio-loud AGN to have short-lived jets operating on time scales of 10$^4$--10$^5$ years.

\subsection{Transient Rates}\label{sec:summary:rate}

We searched our four-epoch dataset for transients and found only a single source in the PSC (VTC223612+001006; RSCVn) that was present in one epoch and absent in the rest.
Our transient search was carried out over a single-epoch area of $\sim$52 deg$^2$, but the sensitivity is not uniform across this area.
For the first three epochs, our transient search was conducted on single-pointing images out to a radius ($r_{\rm max}$) of $\sim$8\arcmin ~from the pointing center.
Although the fourth epoch has fairly uniform rms noise, the comparison with the first three epochs for transient search diminishes the significance of the added sensitivity for this epoch.
We can approximate the transient rate larger than a threshold flux density, $\kappa(>S)$, using the formulation from Appendix C of \cite{ofek2011}.
We use a Gaussian primary beam response with half-width at half-maximum ($r_{\rm HP}$) of 7.5\arcmin.

\begin{align}
\kappa(>S)&=\kappa_{0}(S/S_{0})^{-3/2} \nonumber\\
\kappa_{0}&= \frac{3 N_b {\rm ln}(2)}{2\pi r^2_{\rm HP}} (1-e^{-3r^2_{\rm max} {\rm ln}(2) / r^2_{\rm HP}} )
\label{Eq:CountFun}
\end{align}

The 2$\sigma$ upper and lower limits for the number of transient events, N$_b$, in 50 deg$^2$ given that we found one event are 5.683 and 2.3$\times$10$^{-2}$ \citep{gehrels1986}.
Since the mean rms noise (flat sky) for the first three epochs of our survey is 70~$\mu$Jy, we use a 7$\sigma$ detection threshold of 0.5 mJy.
For RS CVn variables (active binaries) we calculate a rate of $\kappa$($>$0.5 mJy)$=$0.0081$^{+0.0381}_{-0.0079}$ events deg$^{-2}$. 
For all other types of transients we derive a 95\% confidence upper limit of $\kappa$($>$0.5 mJy)$<$0.024 events deg$^{-2}$. 

\begin{figure*}[htp]
\centering
\includegraphics[width=5.5in]{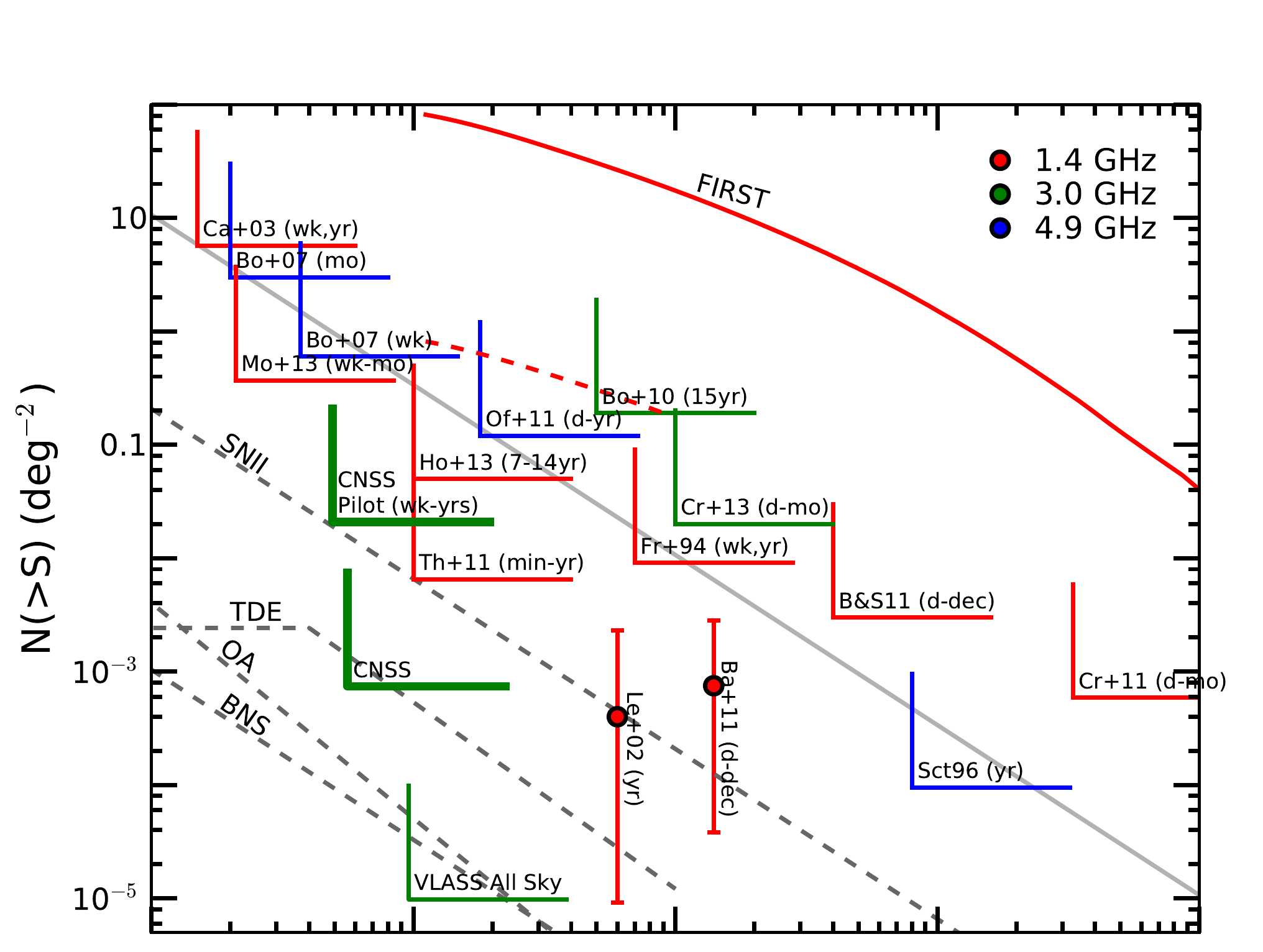}\\
\includegraphics[width=5.5in]{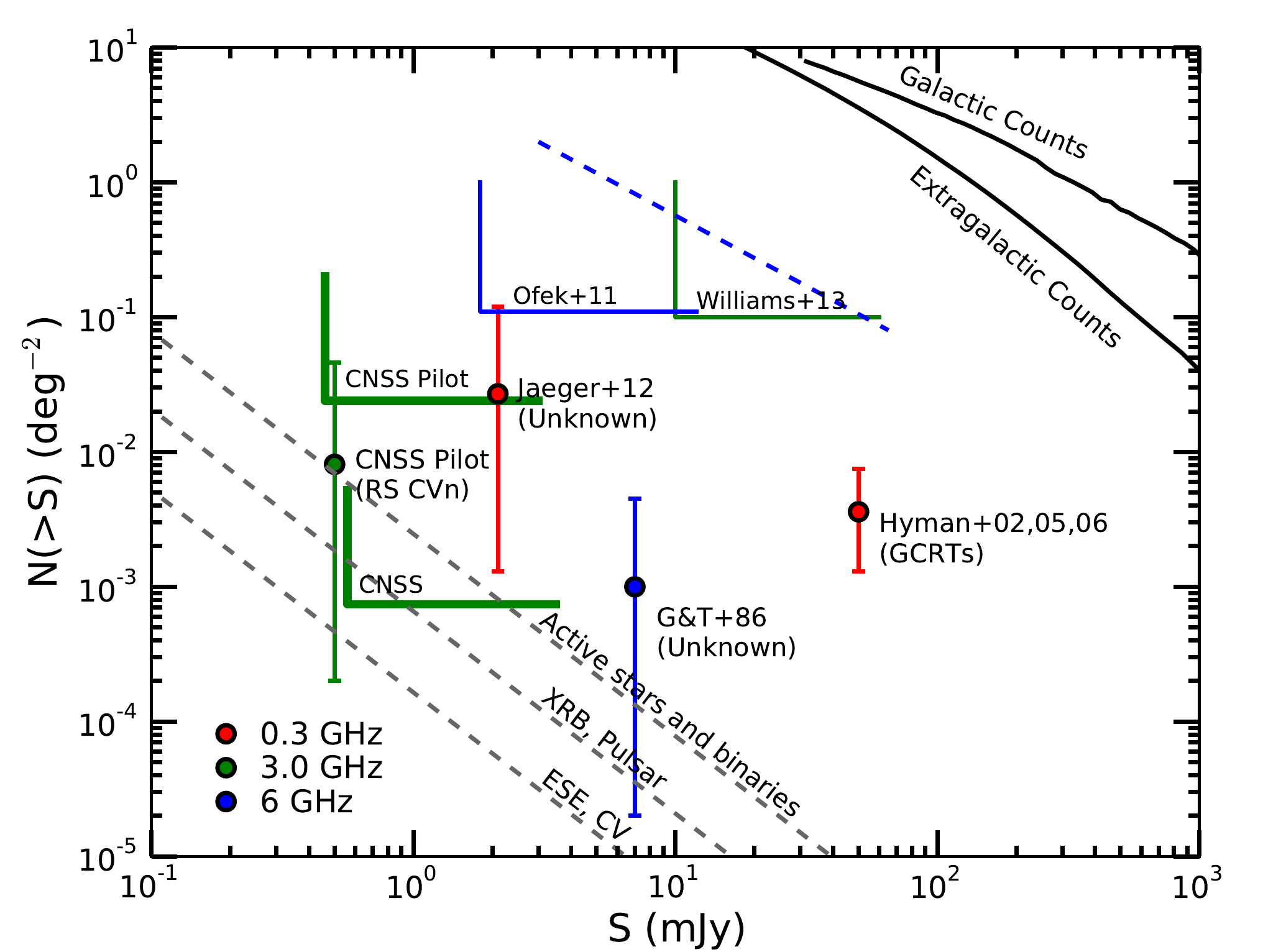}
\caption{{\it Top}: The phase space of slow extragalactic transients. 
The panel shows the upper limits to the transient rates from previous radio surveys (colored wedges; 95\% confidence), 
the rates derived from radio transient detections (2$\sigma$ errorbars), and the expected transient rates. 
The transient detection labeled as 'Le+02' represents a Type II supernova having a peak radio luminosity of $3\times10^{27}$ erg s$^{-1}$ Hz$^{-1}$ and an evolution timescale of $\sim$15 years \citep{levinson2002,gal-yam2006}.
The one labeled 'Ba+11' is a nuclear transient, SUMSS J060938-333508, with a peak radio luminosity of $6\times10^{29}$ erg s$^{-1}$ Hz$^{-1}$ and an evolution timescale of $<$5 years \citep[][K. Bannister, private communication]{bannister2011a,bannister2011b}.
All observed quantities are color-coded according to the observing frequency.
The solid gray line is the rate claimed by \cite{bower2007}, plotted for reference.
The upper limit to the extragalactic transient rate from our pilot survey (this work) and the phase space probed by the full CNSS survey are shown as thick green wedges.
The phase space probed by the VLA Sky Survey all-sky tier (VLASS) is also shown.
The solid red line denotes the source counts from the FIRST survey, and the dashed red line denotes the approximate counts for strong variables at 1.4 GHz (1\% of the persistent sources).
{\it Bottom}: The Galactic transient phase space.
Symbols have similar meanings as in the top panel.
Black solid lines denote the source counts from the FIRST and the MAGPIS 1.4 GHz surveys.
The source counts for variable Galactic sources approximated from \cite{becker2010} are shown as a blue dashed line.
The transient rate for active binaries resulting from our pilot survey is shown by the green errorbar and the upper limit for the rate of all other classes of Galactic transients is denoted by a thick green wedge.
See \S\ref{sec:summary:rate} for more details.}
\label{fig:logNlogS}
\end{figure*}

The logN-logS plots for Galactic and extragalactic slow radio transients are shown in Figure~\ref{fig:logNlogS}.
The top panel shows the upper limits to the transient rates\footnote{see http://www.tauceti.caltech.edu/kunal/radio-transient-surveys/index.html for more details} 
derived from previous radio surveys (colored wedges), the rates derived from radio transient detections (filled circles with errorbars), and the theoretically-expected / empirically-estimated transient 
rates (dashed gray lines; see Table~\ref{tab:transients_summary} for more details). 
It should be noted that the expected transient rates are not sharp lines but are probability surfaces in the logN-logS diagram with the ``most-probable'' rates reported as dashed lines.
The dashed line labeled ``TDE'' represents Swift J1644+57-like \cite{zauderer2011,zauderer2013} events.
Upper limits from a few radio surveys which do not probe any new part of the phase space are not shown in this figure.
All observed quantities are color-coded according to the observing frequency.
The solid gray line is the rate claimed by \cite{bower2007}, plotted for reference.
The upper limit from our pilot survey and the phase space probed by the full CNSS survey are shown as thick green wedges.
The phase space probed by the VLA Sky Survey (VLASS) all-sky tier\footnote{See Table 4 of the VLASS proposal for a summary of the all-sky tier, 
https://safe.nrao.edu/wiki/pub/JVLA/VLASS/VLASS\_final.pdf.} is shown as a thin green wedge.
The source counts from the FIRST survey are represented by the solid red line, and the dashed red line denotes 1\% of these persistent sources, 
representing strong variable sources at 1.4 GHz \citep[e.g.][and references therein]{mooley2013}.
\cite{bannister2011a,bannister2011b} report only a single extragalactic transient, SUMSS J060938-333508, found to be a nuclear source from ATCA follow-up observations (Keith Bannister, private communication).
Hence the transient rate is $7.5\times10^{-4}$ events deg$^{-2}$.
\cite{thyagarajan2011} report 57 transients, but some of these are Galactic and others have indefinite classifications. Hence, we adopt a 95\% confidence level upper limit of 71 transients.
The lower panel of Figure~\ref{fig:logNlogS} shows the Galactic transient phase space.
Symbols have similar meanings as for the extragalactic plot (top panel).
For reference, the source counts from the FIRST and the MAGPIS 1.4 GHz \citep{white2005} surveys are denoted by black solid lines.
The approximate source counts for variable Galactic sources from \cite{becker2010} are denoted by the blue dashed line.
The transient rate for active binaries derived from this work is shown by the green errorbar and the upper limit for all other classes of Galactic transients is denoted by a thick green wedge.
It is evident from these logN-logS diagrams that our pilot survey is not sensitive and wide enough to discover extragalactic explosive transients, but it 
is already in the regime where stellar flares are expected.

Our contemporaneous optical survey with PTF allows us to make a direct comparison between the transient optical and radio skies.
Within the limiting sensitivities of our experiment, the transient skies at these two frequencies appear to be quite distinct.
In the three months of high-cadence PTF search, eight spectroscopically-confirmed supernovae were discovered, none of which were recovered in our radio survey.
Conversely, three stellar flares were detected in the radio, but their optical counterparts are saturated in PTF.
We thus conclude that future joint radio and optical searches, such as MeerKAT and MeerLICHT\footnote{https://www.astro.ru.nl/wiki/research/meerlicht}, 
may only be beneficial if shallow optical surveys are combined with deep radio surveys or vice versa, or if both surveys are deep ($\mu$Jy-level sensitivity).
We find that deep multi-wavelength photometric data and spectroscopy are very important for host and progenitor classification of extragalactic radio transients, and this should inform future slow transient searches.
Given the expensive radio observing time, we have attempted to distinguish between AGN and other transients based on optical light curves from PTF and SDSS.
Although in the present study we have optical light curves only for a small fraction of our radio sources (due to the completeness of optical counterparts), 
we expect to have 100\% completeness in the LSST era, when radio transient classification based on optical light curves may be a feasible approach.

The radio sky at frequencies of a few GHz appears to be quiet, with less than a percent of the persistent sources being strong variables.
From Table~\ref{tab:transients_summary} and Figure~\ref{fig:logNlogS} we find that the rates for a majority of classes of slow radio transients is less than one transient per 10,000 persistent sources.
We also find that the fraction of strong variables and explosive transients among persistent sources in the optical sky is similar to the radio.
However, the large number of persistent optical sources per square degree of the sky (down to the limiting sensitivity of PTF, $R=21$ mag., for example; see 
Figure~\ref{fig:venn}) implies that the number of varying optical sources is much larger than the number of varying 
radio sources (down to the limiting sensitivity of our Jansky VLA survey, $\sim$0.5 mJy).
Accordingly, we conclude that the false positive foreground for future EM-GW searches is lower in the radio than in the optical\footnote{This statement holds 
even for optical and radio surveys that have the same limiting flux density, because the number of optical sources per square degree of the sky is expected to be larger 
than the number of radio sources. Note that the PTF limiting sensitivity of $R=21$ mag. corresponds to 12$\mu$Jy.}.

The rates for extragalactic transients, like core-collapse supernovae and binary neutron star mergers, are significantly low (Table~\ref{tab:transients_summary}), and it is not 
surprising that we found none of these transients in our pilot radio survey.
Surveys over large areas of the sky are thus motivated by the search for such exotica.
Therefore, we recommend wide-field surveys like that of the entire $\sim$270 deg$^2$ of Stripe 82 with the Jansky VLA (the CNSS survey, which is currently underway), all-sky surveys like 
the VLA Sky Survey (VLASS; the all-sky tier has been recently approved), or similar surveys with ASKAP, MeerKAT, and WSRT/Apertif.
Also, shallow radio surveys are preferred to deep surveys due to the added advantage of restricting the transient population to a low redshift space, thus making optical counterpart search feasible.

{\it
Acknowledgemets: The authors extend sincere thanks to Joan Wrobel and other scheduling staff at the NRAO in Socorro for 
extensive help with the scheduling of the VLA observations, and to James Robnett and other computing staff 
for their untiring assistance with the data storage and allocation of computing resources.
The authors also wish to thank Luis Ho, Branimir Sesar, Eran Ofek, Sanjay Bhatnagar, Urvashi Rau, Kumar Golap, Vivek Dhawan, Craig Walker, 
Talvikki Hovatta, Tim Pearson, Anthony Readhead, Chuck Steidel, and Allison Strom for insightful discussions. 
The contribution of PTF collaboration members to optical data processing and optical follow-up observations relevant for this project is acknowledged.
KPM is grateful to NRAO for the Grote Reber Fellowship, and to Yamini Jangir for going over this manuscript and providing useful suggestions.
We thank the anonymous referee for comments which helped in improving certain parts of the manuscript.
The National Radio Astronomy Observatory is a facility of the National 
Science Foundation operated under cooperative agreement by Associated 
Universities, Inc. SRK's research in part is supported by NASA and NSF. 
Some of the data presented herein were obtained at the W.M. Keck Observatory, 
which is operated as a scientific partnership among the California Institute of Technology, 
the University of California and the National Aeronautics and Space Administration. 
The Observatory was made possible by the generous financial support of the W.M. Keck Foundation.
This research has made extensive use of the ADS, CDS (Vizier and SIMBAD), NED, SDSS, and IRSA online databases, 
and the Python, matplotlib, and astropy software packages.}

\begin{appendix}\label{app:vol}

Given a total observing time T, we can either do a wide-shallow or a narrow-deep survey, where wide/narrow 
refers to the area covered, and shallow/deep refers to the sensitivity in terms of limiting flux density.
To find which of these will explore the largest volume, consider two surveys, one which observes a single 
pointing for time T, and the other which observes N pointings with time T/N alloted for each pointing.
We know that the limiting flux density (S$_0$), integration time ($\tau$), maximum distance (d$_0$) and 
the total volume (V$_0$) probed by a survey are related as:
\begin{align}
 S_0 \propto \frac{1}{\sqrt{\tau}} \propto \frac{1}{d_0^2} \Rightarrow V_0 \propto d_0^3 \propto \tau^{3/4}\\
\therefore \frac{V_1}{V_2} = \frac{T^{3/4}}{N \times (T/N)^{3/4}} = \frac{1}{N^{1/4}}
\end{align}
Thus we see that the volume probed by a wide-shallow survey (V$_2$) is larger than that seen by a narrow-deep survey (V$_1$) by 
a factor of the number of pointings to the power of 1/4. However, note that since the distance probed in these two kinds of surveys 
goes as $d_1/d_2 = \sqrt{N}$. This means that if we want to probe transient populations concentrated at 
large distances (redshifts), it is best to do a narrow-deep survey. 
Also see \cite{macquart2014} for survey parameter optimization to maximize the yield of slow transients having well-defined distributions in space.

\end{appendix}


\clearpage
\LongTables
\tabletypesize{\tiny}
\begin{deluxetable*}{lrrrrrrrrrccccl} 
\tablecaption{Summary of the radio transient and variable sources found in this work
\label{tab:transients}}
\tablehead{
\colhead{I}\\
\colhead{I}\\
\colhead{VTC} & \colhead{$\alpha_{\rm J2000}$} & \colhead{$\delta_{\rm J2000}$} & \colhead{$S_{\rm E1}$} & \colhead{$S_{\rm E2}$} & \colhead{$S_{\rm E3}$} & \colhead{$S_{\rm E4}$} & \colhead{$m$} & \colhead{FIRST}  & \colhead{\citeauthor{hodge2011}} & \colhead{$r_{\rm SDSS}$} & \colhead{PTF} & \colhead{$z_{\rm spec}$} & \colhead{log $\overline{L}_\nu$}   & \colhead{Identification} \\
              & \multicolumn{2}{c}{(deg)}                                       &  \multicolumn{4}{c}{(mJy)}                                                                        &               & \colhead{(mJy)}  & \colhead{(mJy)}                  & \colhead{(mag)}          &               &                          & \colhead{(erg cm$^{-2}$ s$^{-1}$)} & \\
}
\colhead{I}\\
\startdata
\hline
\multicolumn{15}{c}{Timescale $<$1 week}\\
\hline
220210-000203 & 330.539900 & -0.034252 &   2.079$\pm$0.109 &   1.064$\pm$0.077 &   1.862$\pm$0.083 &   1.382$\pm$0.059 & -0.56 &         $<$0.50 &        $\ldots$ & $\ldots$ & $\ldots$ & $\ldots$ & $\ldots$  & AGN \\
220609-004654 & 331.536395 & -0.781711 &   3.970$\pm$0.154 &   3.424$\pm$0.106 &   4.637$\pm$0.112 &   3.390$\pm$0.060 & -0.33 &   1.83$\pm$0.14 &   1.47$\pm$0.07 &     20.7 &      nv  & 0.37     &      31.1 & AGN \\
221122+005202 & 332.842184 &  0.867327 &   1.665$\pm$0.096 &   2.037$\pm$0.070 &   1.122$\pm$0.085 &   2.325$\pm$0.041 &  0.55 &   1.25$\pm$0.11 &   0.52$\pm$0.07 &     20.0 &      nv  & 0.31     &      30.7 & AGN \\
221136+011142 & 332.898974 &  1.194886 &   3.420$\pm$0.077 &   3.128$\pm$0.054 &   4.383$\pm$0.064 &   2.368$\pm$0.061 & -0.36 &   1.64$\pm$0.14 &   1.26$\pm$0.16 &     24.1 & $\ldots$ & $\ldots$ & $\ldots$  & AGN \\
221226+004322 & 333.107130 &  0.722873 &   2.408$\pm$0.067 &   2.951$\pm$0.057 &   2.112$\pm$0.058 &   2.391$\pm$0.044 &  0.31 &   2.50$\pm$0.11 &   1.83$\pm$0.06 &     23.1 & $\ldots$ & $\ldots$ & $\ldots$  & AGN \\
221406+002305 & 333.524900 &  0.384618 &   0.809$\pm$0.063 &   0.877$\pm$0.055 &   0.464$\pm$0.049 &   0.604$\pm$0.050 &  0.58 &         $<$0.51 &         $<$0.32 &     18.4 &      nv  & 0.15     &      29.5 & AGN \\
221541+010856 & 333.921049 &  1.148981 &   1.425$\pm$0.081 &   1.041$\pm$0.053 &   1.664$\pm$0.063 &   1.213$\pm$0.059 & -0.48 &         $<$0.69 &   0.68$\pm$0.09 &     19.9 & $\ldots$ & 0.32$^a$ &      30.5 & AGN \\
221713-002216 & 334.306166 & -0.371232 &   1.376$\pm$0.088 &   1.320$\pm$0.068 &   1.783$\pm$0.071 &   1.027$\pm$0.053 & -0.33 &         $<$0.53 &   1.14$\pm$0.07 &     23.8 & $\ldots$ & $\ldots$ & $\ldots$  & AGN \\
221913+011147 & 334.805979 &  1.196250 &   1.015$\pm$0.085 &   1.442$\pm$0.063 &   0.539$\pm$0.074 &   1.007$\pm$0.047 &  0.90 &   1.05$\pm$0.14 &   0.82$\pm$0.07 &     19.1 & $\ldots$ & 0.20$^a$ &      30.0 & AGN \\
222009-010814 & 335.036567 & -1.137346 &   4.386$\pm$0.096 &   3.873$\pm$0.071 &   5.565$\pm$0.081 &   4.895$\pm$0.061 & -0.38 &   5.23$\pm$0.15 &   3.13$\pm$0.10 &     20.6 &      nv  & $\ldots$ & $\ldots$  & QSO \\
222150-000325 & 335.460301 & -0.056997 &   2.239$\pm$0.121 &   2.470$\pm$0.080 &   1.819$\pm$0.086 &   2.106$\pm$0.053 &  0.29 &   2.70$\pm$0.11 &   2.24$\pm$0.07 & $\ldots$ & $\ldots$ & $\ldots$ & $\ldots$  & AGN \\
222232+011104 & 335.634687 &  1.184480 &   0.860$\pm$0.062 &   0.844$\pm$0.048 &   0.348$\pm$0.048 &   0.944$\pm$0.048 &  0.80 &   1.34$\pm$0.14 &   1.50$\pm$0.06 &     19.5 & $\ldots$ & 0.29     &      30.2 & AGN \\
222236+010556 & 335.648844 &  1.098840 &   1.009$\pm$0.078 &   1.054$\pm$0.056 &   0.358$\pm$0.055 &   0.927$\pm$0.050 &  0.96 &         $<$0.72 &   0.53$\pm$0.05 &     18.4 &      nv  & 0.20     &      29.9 & AGN \\
222330+010227 & 335.873148 &  1.040835 &   9.414$\pm$0.134 &   8.921$\pm$0.091 &   5.740$\pm$0.097 &   5.391$\pm$0.044 &  0.40 &   6.11$\pm$0.13 &   6.82$\pm$0.06 &     19.3 &      nv  & $\ldots$ & $\ldots$  & QSO \\
222520+004939 & 336.331883 &  0.827551 &   3.807$\pm$0.115 &   3.500$\pm$0.090 &   5.326$\pm$0.085 &   4.273$\pm$0.054 & -0.44 &   4.69$\pm$0.12 &   3.76$\pm$0.08 &     22.3 & $\ldots$ & 0.80$^a$ &      31.9 & AGN \\
222758+003706 & 336.992096 &  0.618294 &  48.766$\pm$0.240 &  39.161$\pm$0.160 &  51.889$\pm$0.155 &  45.320$\pm$0.095 & -0.30 &  83.75$\pm$0.10 &  64.45$\pm$0.63 &     18.1 &       nv & $\ldots$ & $\ldots$  & QSO \\
223607+005259 & 339.028010 &  0.883015 &   0.951$\pm$0.068 &   0.941$\pm$0.056 &   1.319$\pm$0.055 &   0.940$\pm$0.053 & -0.37 &         $<$0.50 &   0.62$\pm$0.05 & $\ldots$ & $\ldots$ & $\ldots$ & $\ldots$  & AGN \\
223624+004021 & 339.099104 &  0.672614 &   0.886$\pm$0.071 &   0.955$\pm$0.055 &   0.595$\pm$0.060 &   0.828$\pm$0.052 &  0.43 &         $<$0.50 &   0.57$\pm$0.07 &     22.3 & $\ldots$ & 0.59$^a$ &      30.9 & AGN \\
224843-005459 & 342.178945 & -0.916381 &   1.644$\pm$0.085 &   1.429$\pm$0.064 &   1.778$\pm$0.085 &   1.523$\pm$0.057 & -0.27 &         $<$0.72 &   0.69$\pm$0.07 & $\ldots$ & $\ldots$ & $\ldots$ & $\ldots$  & AGN \\
224922+001805 & 342.343018 &  0.301369 &  10.370$\pm$0.161 &  10.029$\pm$0.137 &   7.249$\pm$0.146 &   9.759$\pm$0.059 &  0.27 &   9.46$\pm$0.10 &  11.51$\pm$0.06 & $\ldots$ & $\ldots$ & $\ldots$ & $\ldots$  & AGN \\
225411-010651 & 343.544297 & -1.114119 &   0.644$\pm$0.063 &   0.782$\pm$0.050 &   0.300$\pm$0.056 &   0.300$\pm$0.052 &  0.85 &         $<$0.74 &         $<$0.60 & $\ldots$ & $\ldots$ & $\ldots$ & $\ldots$  & AGN \\
230117-000204 & 345.322041 & -0.034390 &   2.071$\pm$0.079 &   2.600$\pm$0.062 &   1.701$\pm$0.073 &   1.946$\pm$0.054 &  0.37 &   2.67$\pm$0.11 &   2.09$\pm$0.07 &     19.6 &       nv & 0.28$^a$ &     30.6  & AGN \\
230133-002538 & 345.387209 & -0.427294 &   0.676$\pm$0.082 &   0.392$\pm$0.073 &   0.876$\pm$0.079 &   0.524$\pm$0.052 & -0.80 &         $<$0.61 &   0.34$\pm$0.06 & $\ldots$ & $\ldots$ & 0.84$^a$ &     31.0  & AGN \\
230556-001652 & 346.481658 & -0.281193 &   1.071$\pm$0.091 &   0.900$\pm$0.074 &   1.371$\pm$0.079 &   1.031$\pm$0.064 & -0.46 &         $<$0.51 &   0.89$\pm$0.17 & $\ldots$ & $\ldots$ & $\ldots$ & $\ldots$  & AGN \\
230636-002609 & 346.649268 & -0.435896 &   1.741$\pm$0.083 &   1.150$\pm$0.074 &   2.002$\pm$0.084 &   1.607$\pm$0.062 & -0.58 &   1.71$\pm$0.11 &   1.48$\pm$0.09 & $\ldots$ & $\ldots$ & $\ldots$ & $\ldots$  & AGN \\
230918+002642 & 347.324430 &  0.445083 &   0.472$\pm$0.068 &   0.350$\pm$0.055 &   0.716$\pm$0.065 &   0.433$\pm$0.065 & -0.73 &         $<$0.50 &         $<$0.39 & $\ldots$ & $\ldots$ & $\ldots$ & $\ldots$  & AGN \\
231437+003844 & 348.655518 &  0.645515 &   0.774$\pm$0.070 &   0.805$\pm$0.053 &   1.170$\pm$0.060 &   1.026$\pm$0.090 & -0.42 &         $<$0.53 &   0.39$\pm$0.07 &     24.9 & $\ldots$ & $\ldots$ & $\ldots$  & AGN \\
231444+004026 & 348.685137 &  0.673960 &   0.233$\pm$0.060 &   0.606$\pm$0.053 &   0.222$\pm$0.055 &   0.318$\pm$0.081 &  0.89 &         $<$0.53 &   0.34$\pm$0.06 &     23.9 & $\ldots$ & $\ldots$ & $\ldots$  & AGN \\
231559-001205 & 348.994359 & -0.201415 &   4.176$\pm$0.105 &   4.039$\pm$0.095 &   5.282$\pm$0.098 &   4.003$\pm$0.070 & -0.31 &   4.90$\pm$0.10 &   4.21$\pm$0.05 &     23.1 & $\ldots$ & $\ldots$ & $\ldots$  & AGN \\
231746-005355 & 349.441332 & -0.898619 &   1.045$\pm$0.081 &   0.862$\pm$0.074 &   1.284$\pm$0.085 &   1.175$\pm$0.084 & -0.44 &         $<$0.69 &   0.44$\pm$0.05 & $\ldots$ & $\ldots$ & $\ldots$ & $\ldots$  & AGN \\
231942-004547 & 349.925567 & -0.763046 &   3.373$\pm$0.088 &   4.311$\pm$0.076 &   3.093$\pm$0.085 &   3.372$\pm$0.074 &  0.29 &   4.36$\pm$0.14 &   3.89$\pm$0.06 &     23.9 & $\ldots$ & $\ldots$ & $\ldots$  & AGN \\
232217+001252 & 350.569589 &  0.214481 &   0.933$\pm$0.096 &   0.700$\pm$0.078 &   1.226$\pm$0.084 &   1.081$\pm$0.060 & -0.59 &         $<$0.49 &        $\ldots$ & $\ldots$ & $\ldots$ & $\ldots$ & $\ldots$  & AGN \\
232226+010357 & 350.606463 &  1.065718 &   3.082$\pm$0.193 &   2.164$\pm$0.131 &   3.145$\pm$0.168 &   3.577$\pm$0.093 & -0.42 &   4.96$\pm$0.13 &        $\ldots$ &     17.4 & $\ldots$ & 0.12     &      30.0 & AGN \\
232634-010513 & 351.642405 & -1.086815 &   2.481$\pm$0.129 &   1.565$\pm$0.121 &   2.239$\pm$0.128 &   2.241$\pm$0.086 & -0.40 &   1.50$\pm$0.15 &        $\ldots$ &     23.4 & $\ldots$ & 0.79$^a$ &      31.5 & AGN \\
232656-000438 & 351.734689 & -0.077164 &   1.737$\pm$0.060 &   1.477$\pm$0.056 &   2.050$\pm$0.055 &   1.549$\pm$0.065 & -0.37 &   1.09$\pm$0.10 &        $\ldots$ &     24.1 & $\ldots$ & $\ldots$ & $\ldots$  & AGN \\
232723-000507 & 351.843984 & -0.085213 &   1.373$\pm$0.071 &   1.796$\pm$0.062 &   1.316$\pm$0.062 &   1.478$\pm$0.073 &  0.27 &   1.25$\pm$0.10 &        $\ldots$ & $\ldots$ & $\ldots$ & $\ldots$ & $\ldots$  & AGN \\
232933-004002 & 352.386845 & -0.667239 &   8.785$\pm$0.165 &   9.301$\pm$0.143 &   6.812$\pm$0.149 &  12.592$\pm$0.086 &  0.26 &   6.93$\pm$0.14 &        $\ldots$ &     22.4 & $\ldots$ & $\ldots$ & $\ldots$  & AGN \\
233106+002607 & 352.774153 &  0.435175 &   0.841$\pm$0.054 &   0.644$\pm$0.044 &   0.922$\pm$0.047 &   0.648$\pm$0.057 & -0.40 &   1.00$\pm$0.11 &        $\ldots$ &     22.2 & $\ldots$ & 0.74     &      31.1 & AGN \\
\hline
\multicolumn{15}{c}{Timescale $<$1 month}\\
\hline
220127+001402 & 330.360610 &  0.233772 &   1.076$\pm$0.096 &   0.448$\pm$0.071 &   0.480$\pm$0.069 &   0.660$\pm$0.053 &  0.82 &         $<$0.49 &        $\ldots$ &     23.9 & $\ldots$ & $\ldots$ & $\ldots$ & AGN \\
220456-000147 & 331.232587 & -0.029821 &   4.999$\pm$0.169 &   6.578$\pm$0.096 &   5.969$\pm$0.125 &   6.192$\pm$0.076 & -0.27 &   2.29$\pm$0.11 &   2.45$\pm$0.26 & $\ldots$ & $\ldots$ & $\ldots$ & $\ldots$ & AGN \\
220643-003102 & 331.680375 & -0.517180 & 131.189$\pm$0.699 & 194.029$\pm$0.340 & 151.652$\pm$0.298 & 134.492$\pm$0.182 & -0.39 & 122.89$\pm$0.13 &  73.21$\pm$0.76 &     19.5 &       V  & $\ldots$ & $\ldots$ & QSO \\
220910-001601 & 332.293354 & -0.267023 &   6.690$\pm$0.166 &   9.455$\pm$0.120 &   7.796$\pm$0.137 &   8.323$\pm$0.054 & -0.34 &   8.39$\pm$0.10 &   5.78$\pm$0.06 &     19.5 &      nv  &     1.11 &     32.4 & QSO \\
221308-010837 & 333.284558 & -1.143507 &   0.921$\pm$0.075 &   0.376$\pm$0.057 &   0.614$\pm$0.067 &   0.711$\pm$0.059 &  0.84 &         $<$0.72 &   0.95$\pm$0.05 &     23.2 & $\ldots$ & $\ldots$ & $\ldots$ & AGN \\
221350-011130 & 333.460186 & -1.191627 &   1.540$\pm$0.120 &   0.835$\pm$0.086 &   1.083$\pm$0.101 &   1.139$\pm$0.052 &  0.59 &         $<$0.71 &         $<$0.28 &     20.0 & $\ldots$ &     1.96 &     32.0 & QSO \\
221959+011045 & 334.994454 &  1.179204 &   4.827$\pm$0.104 &   3.679$\pm$0.081 &   3.710$\pm$0.107 &   4.981$\pm$0.052 &  0.27 &   4.62$\pm$0.14 &   3.42$\pm$0.07 &     21.0 & $\ldots$ &     0.84 &     31.9 & QSO \\
222123-002509 & 335.345654 & -0.419074 &   3.255$\pm$0.145 &   2.235$\pm$0.107 &   2.724$\pm$0.119 &   2.674$\pm$0.080 &  0.37 &   3.62$\pm$0.10 &   2.70$\pm$0.10 &     23.4 & $\ldots$ & $\ldots$ & $\ldots$ & AGN \\
222942+003556 & 337.423504 &  0.598817 &   3.266$\pm$0.089 &   2.401$\pm$0.070 &   2.772$\pm$0.073 &   2.367$\pm$0.051 &  0.31 &   2.63$\pm$0.11 &   1.97$\pm$0.06 &     24.5 & $\ldots$ & $\ldots$ & $\ldots$ & AGN \\
223317-005009 & 338.321408 & -0.835723 &   1.977$\pm$0.093 &   1.344$\pm$0.068 &   1.448$\pm$0.066 &   1.344$\pm$0.066 &  0.38 &   1.80$\pm$0.14 &   1.27$\pm$0.06 & $\ldots$ & $\ldots$ & $\ldots$ & $\ldots$ & AGN \\
223612+001007 & 339.050505 &  0.168559 &   0.804$\pm$0.086 &   0.271$\pm$0.071 &   0.179$\pm$0.073 &   0.098$\pm$0.057 &  0.99 &         $<$0.52 &         $<$0.39 &      9.7 & $\ldots$ & $\ldots$ & $\ldots$ & RS CVn \\
224036+010852 & 340.150070 &  1.147826 &   2.005$\pm$0.070 &   2.729$\pm$0.054 &   2.411$\pm$0.056 &   1.686$\pm$0.057 & -0.31 &   2.13$\pm$0.14 &   1.41$\pm$0.06 &     21.0 & $\ldots$ &     2.99 &     32.7 & QSO \\
224448-000619 & 341.200443 & -0.105397 &   8.637$\pm$0.086 &   6.604$\pm$0.074 &   5.485$\pm$0.079 &   6.716$\pm$0.064 &  0.27 &   6.90$\pm$0.10 &   7.21$\pm$0.07 &     18.9 &      nv  & $\ldots$ & $\ldots$ & QSO \\
224657+005240 & 341.737102 &  0.877754 &   2.641$\pm$0.063 &   1.928$\pm$0.057 &   1.702$\pm$0.071 &   1.982$\pm$0.054 &  0.31 &   1.40$\pm$0.13 &   0.75$\pm$0.06 & $\ldots$ & $\ldots$ & $\ldots$ & $\ldots$ & AGN \\
225649-005401 & 344.205944 & -0.900221 &   1.935$\pm$0.086 &   1.397$\pm$0.074 &   1.364$\pm$0.077 &   1.480$\pm$0.059 &  0.32 &   1.63$\pm$0.14 &   1.74$\pm$0.06 & $\ldots$ & $\ldots$ & $\ldots$ & $\ldots$ & AGN \\
225707-010238 & 344.280373 & -1.043805 &   0.602$\pm$0.073 &   1.218$\pm$0.069 &   0.914$\pm$0.065 &   1.192$\pm$0.056 & -0.68 &         $<$0.74 &         $<$0.39 &     20.1 &      nv  &     1.56 &     31.8 & QSO \\
230236+005739 & 345.647977 &  0.960878 &   4.146$\pm$0.142 &   3.165$\pm$0.114 &   3.344$\pm$0.144 &   4.485$\pm$0.054 &  0.27 &   3.43$\pm$0.11 &   4.92$\pm$0.06 & $\ldots$ & $\ldots$ & $\ldots$ & $\ldots$ & AGN \\
230803-000145 & 347.012572 & -0.029111 &   0.376$\pm$0.058 &   0.698$\pm$0.053 &   0.453$\pm$0.056 &   0.351$\pm$0.062 & -0.60 &         $<$0.52 &         $<$0.29 &     18.5 &      nv  &     0.15 &     29.5 & AGN \\
230810+002117 & 347.039702 &  0.354644 &   1.200$\pm$0.063 &   1.771$\pm$0.055 &   1.322$\pm$0.061 &   0.903$\pm$0.078 & -0.38 &   1.33$\pm$0.11 &   1.03$\pm$0.06 &     21.1 &      nv  & 0.40$^a$ &     30.7 & AGN \\
231242+002451 & 348.174063 &  0.414044 &   4.347$\pm$0.095 &   3.198$\pm$0.079 &   3.767$\pm$0.083 &   2.386$\pm$0.062 &  0.30 &   1.19$\pm$0.10 &   0.86$\pm$0.06 &     18.7 &      nv  &     1.90 &     32.5 & QSO \\
231334-001645 & 348.390098 & -0.279113 &   2.343$\pm$0.061 &   1.440$\pm$0.054 &   1.353$\pm$0.061 &   1.657$\pm$0.064 &  0.48 &   1.87$\pm$0.10 &   2.20$\pm$0.05 &     22.6 & $\ldots$ &     0.74 &     31.4 & AGN \\
231455+002456 & 348.728453 &  0.415556 &   1.177$\pm$0.071 &   0.757$\pm$0.064 &   0.873$\pm$0.064 &   0.855$\pm$0.075 &  0.43 &   0.96$\pm$0.10 &   0.81$\pm$0.06 &     24.8 & $\ldots$ & $\ldots$ & $\ldots$ & AGN \\
232125-004845 & 350.352737 & -0.812522 &   2.527$\pm$0.072 &   1.829$\pm$0.065 &   1.822$\pm$0.068 &   1.905$\pm$0.071 &  0.32 &   1.94$\pm$0.14 &        $\ldots$ &     21.8 & $\ldots$ & 0.58$^a$ &     31.2 & AGN \\
232526-003623 & 351.359745 & -0.606393 &   2.039$\pm$0.092 &   2.672$\pm$0.075 &   2.098$\pm$0.078 &   2.535$\pm$0.076 & -0.27 &   3.26$\pm$0.13 &        $\ldots$ &     21.4 &      nv  &     0.63 &     31.4 & AGN \\
232548-011134 & 351.451958 & -1.192766 &   5.404$\pm$0.148 &   7.212$\pm$0.145 &   5.776$\pm$0.140 &   5.925$\pm$0.083 & -0.29 &   6.21$\pm$0.14 &        $\ldots$ &     20.7 & $\ldots$ &     0.46 &     31.5 & AGN \\
\hline
\multicolumn{15}{c}{Timescale $<$1.5 years}\\
\hline
215701+005124 & 329.252431 &  0.856537 &   0.923$\pm$0.074 &   1.142$\pm$0.052 &   0.811$\pm$0.059 &   0.764$\pm$0.043 &  0.40 &   1.06$\pm$0.10 &        $\ldots$ &     22.9 & $\ldots$ & 0.57$^a$ &     30.9 & AGN \\
215732+010748 & 329.384654 &  1.130011 &   1.219$\pm$0.099 &   1.195$\pm$0.060 &   1.048$\pm$0.065 &   0.806$\pm$0.053 &  0.39 &         $<$0.78 &        $\ldots$ &     19.1 &      nv  &     0.30 &     30.4 & AGN \\
215929+004723 & 329.872536 &  0.789814 &   1.777$\pm$0.118 &   1.586$\pm$0.095 &   1.500$\pm$0.087 &   2.091$\pm$0.043 & -0.27 &   1.52$\pm$0.11 &        $\ldots$ & $\ldots$ & $\ldots$ & $\ldots$ & $\ldots$ & AGN \\
215951+010041 & 329.963762 &  1.011278 &   1.405$\pm$0.102 &   1.375$\pm$0.075 &   1.300$\pm$0.074 &   2.005$\pm$0.044 & -0.37 &   2.03$\pm$0.12 &        $\ldots$ & $\ldots$ & $\ldots$ & $\ldots$ & $\ldots$ & AGN \\
220005+002309 & 330.022192 &  0.385698 &   1.357$\pm$0.068 &   1.316$\pm$0.054 &   1.217$\pm$0.066 &   0.962$\pm$0.056 &  0.31 &         $<$0.51 &        $\ldots$ & $\ldots$ & $\ldots$ & $\ldots$ & $\ldots$ & AGN \\
220017-000134 & 330.072365 & -0.026093 &   5.912$\pm$0.100 &   5.995$\pm$0.080 &   4.965$\pm$0.087 &   4.612$\pm$0.067 &  0.26 &   7.06$\pm$0.09 &        $\ldots$ &     19.8 &      nv  &     0.61 &     31.7 & QSO \\
220109+010124 & 330.288600 &  1.023416 &  12.343$\pm$0.147 &  11.386$\pm$0.107 &  11.627$\pm$0.096 &  15.358$\pm$0.049 & -0.30 &  37.55$\pm$0.14 &        $\ldots$ & $\ldots$ & $\ldots$ & $\ldots$ & $\ldots$ & AGN \\
220110+002547 & 330.293140 &  0.429823 &   2.504$\pm$0.077 &   2.715$\pm$0.058 &   2.335$\pm$0.056 &   2.049$\pm$0.056 &  0.28 &   3.76$\pm$0.11 &        $\ldots$ & $\ldots$ & $\ldots$ & $\ldots$ & $\ldots$ & AGN \\
220221+001114 & 330.587782 &  0.187335 &   1.438$\pm$0.078 &   1.623$\pm$0.065 &   1.616$\pm$0.060 &   1.212$\pm$0.045 &  0.29 &   2.00$\pm$0.11 &        $\ldots$ &     24.2 & $\ldots$ & $\ldots$ & $\ldots$ & AGN \\
220445+005129 & 331.187680 &  0.857985 &   3.041$\pm$0.089 &   2.913$\pm$0.066 &   2.641$\pm$0.069 &   3.801$\pm$0.050 & -0.26 &   4.76$\pm$0.13 &   3.10$\pm$0.15 & $\ldots$ & $\ldots$ & $\ldots$ & $\ldots$ & AGN \\
220708+010126 & 331.784654 &  1.023799 &   7.029$\pm$0.173 &   6.286$\pm$0.122 &   6.481$\pm$0.131 &   8.313$\pm$0.053 & -0.28 &  15.68$\pm$0.13 &  10.72$\pm$0.06 &     18.8 &      nv  &     2.91 &     33.2 & QSO \\
220804+000556 & 332.018501 &  0.098983 &   6.253$\pm$0.163 &   5.422$\pm$0.118 &   5.585$\pm$0.128 &   7.053$\pm$0.058 & -0.26 &  10.06$\pm$0.11 &   7.99$\pm$0.07 &     23.2 & $\ldots$ & $\ldots$ & $\ldots$ & AGN \\
220904+004607 & 332.266020 &  0.768647 &   1.317$\pm$0.160 &   1.712$\pm$0.133 &   1.042$\pm$0.143 &   1.008$\pm$0.043 &  0.52 &         $<$0.51 &   1.00$\pm$0.05 & $\ldots$ & $\ldots$ & $\ldots$ & $\ldots$ & AGN \\
221160-003139 & 332.998688 & -0.527367 &   1.362$\pm$0.077 &   1.073$\pm$0.057 &   1.174$\pm$0.061 &   1.414$\pm$0.051 & -0.27 &   2.10$\pm$0.10 &   2.12$\pm$0.05 &     23.9 & $\ldots$ & $\ldots$ & $\ldots$ & AGN \\
221257-005711 & 333.235700 & -0.953020 &   1.897$\pm$0.160 &   1.319$\pm$0.125 &   1.807$\pm$0.136 &   2.194$\pm$0.055 & -0.50 &   3.51$\pm$0.15 &   2.83$\pm$0.06 & $\ldots$ & $\ldots$ & $\ldots$ & $\ldots$ & AGN \\
221548-001031 & 333.948344 & -0.175409 &   0.882$\pm$0.102 &   0.755$\pm$0.083 &   1.083$\pm$0.081 &   1.283$\pm$0.051 & -0.52 &   1.30$\pm$0.10 &   1.36$\pm$0.06 &     22.6 & $\ldots$ & 0.54$^a$ &     30.9 & AGN \\
221609-005708 & 334.037121 & -0.952234 &   9.000$\pm$0.158 &   7.481$\pm$0.149 &   7.805$\pm$0.149 &  11.971$\pm$0.064 & -0.46 &   6.56$\pm$0.15 &   5.35$\pm$0.05 &     17.5 &      nv  &     2.40 &     33.2 & QSO \\
221642-004904 & 334.174200 & -0.817899 &   1.734$\pm$0.069 &   1.828$\pm$0.055 &   2.074$\pm$0.062 &   2.383$\pm$0.054 & -0.26 &   1.24$\pm$0.15 &   2.46$\pm$0.06 &     22.8 & $\ldots$ & 0.69$^a$ &     31.4 & AGN \\
222038-001209 & 335.158511 & -0.202623 &   1.916$\pm$0.205 &   1.596$\pm$0.157 &   1.922$\pm$0.158 &   2.574$\pm$0.047 & -0.47 &  10.26$\pm$0.11 &   7.29$\pm$0.07 & $\ldots$ & $\ldots$ & $\ldots$ & $\ldots$ & AGN \\
222109-001940 & 335.286275 & -0.327789 &   0.959$\pm$0.125 &   0.907$\pm$0.095 &   0.892$\pm$0.105 &   1.563$\pm$0.061 & -0.53 &   1.59$\pm$0.10 &   1.20$\pm$0.08 &     22.4 & $\ldots$ &     2.36 &     32.2 & QSO \\
222127-001530 & 335.362235 & -0.258239 &   1.779$\pm$0.121 &   1.720$\pm$0.085 &   1.558$\pm$0.085 &   2.316$\pm$0.055 & -0.30 &   2.06$\pm$0.11 &   1.67$\pm$0.10 &     23.2 & $\ldots$ & 0.46$^a$ &     31.0 & AGN \\
222201-005008 & 335.505891 & -0.835608 &   8.209$\pm$0.141 &   7.158$\pm$0.113 &   8.512$\pm$0.109 &  12.122$\pm$0.070 & -0.51 &   5.56$\pm$0.15 &   4.02$\pm$0.07 &     20.1 &      nv  & 0.33$^a$ &     31.4 & AGN \\
222359+011148 & 335.997306 &  1.196745 &   1.044$\pm$0.116 &   1.006$\pm$0.087 &   1.118$\pm$0.084 &   1.478$\pm$0.054 & -0.38 &   1.70$\pm$0.14 &   0.75$\pm$0.15 &     22.4 & $\ldots$ & $\ldots$ & $\ldots$ & AGN \\
222443+002858 & 336.180822 &  0.482741 &   1.718$\pm$0.109 &   1.832$\pm$0.075 &   1.917$\pm$0.084 &   2.447$\pm$0.044 & -0.29 &   2.12$\pm$0.11 &   1.52$\pm$0.06 &     19.5 &      nv  & 0.26$^a$ &     30.5 & AGN \\
222524-001837 & 336.348824 & -0.310389 &   0.934$\pm$0.080 &   1.058$\pm$0.056 &   0.862$\pm$0.060 &   0.704$\pm$0.055 &  0.40 &         $<$0.55 &   1.11$\pm$0.06 & $\ldots$ & $\ldots$ & $\ldots$ & $\ldots$ & AGN \\
222546+004038 & 336.439976 &  0.677156 &   1.572$\pm$0.069 &   1.607$\pm$0.060 &   1.667$\pm$0.060 &   1.146$\pm$0.060 &  0.34 &   2.18$\pm$0.10 &   2.92$\pm$0.07 & $\ldots$ & $\ldots$ & $\ldots$ & $\ldots$ & AGN \\
222605-010441 & 336.521610 & -1.078109 &   4.218$\pm$0.075 &   4.657$\pm$0.053 &   4.261$\pm$0.054 &   3.465$\pm$0.055 &  0.29 &   5.22$\pm$0.15 &   4.06$\pm$0.06 &     22.9 & $\ldots$ & 0.42$^a$ &     31.3 & AGN \\
222630-001248 & 336.625294 & -0.213342 &   4.690$\pm$0.134 &   4.418$\pm$0.104 &   4.561$\pm$0.121 &   5.855$\pm$0.052 & -0.28 &  10.75$\pm$0.11 &   7.66$\pm$0.07 & $\ldots$ & $\ldots$ & $\ldots$ & $\ldots$ & AGN \\
222647+005211 & 336.694041 &  0.869751 & 359.852$\pm$0.879 & 324.569$\pm$0.639 & 362.536$\pm$0.610 & 433.361$\pm$0.246 & -0.29 & 617.48$\pm$0.10 & 285.28$\pm$5.57 &     21.3 &      nv  &     2.26 &     34.7 & QSO \\
222704+011055 & 336.766725 &  1.182021 &   1.665$\pm$0.150 &   1.488$\pm$0.098 &   1.302$\pm$0.090 &   2.291$\pm$0.065 & -0.42 &   1.11$\pm$0.14 &         $<$3.06 &     17.4 & $\ldots$ & 0.06$^a$ &     29.1 & AGN \\
222907+001908 & 337.277758 &  0.318936 &   1.107$\pm$0.098 &   0.918$\pm$0.086 &   1.031$\pm$0.079 &   1.423$\pm$0.048 & -0.43 &   1.51$\pm$0.11 &   0.91$\pm$0.06 & $\ldots$ &      nv  &     1.80 &     32.0 & QSO \\
222930-000845 & 337.373252 & -0.145791 &   3.183$\pm$0.141 &   3.171$\pm$0.106 &   3.020$\pm$0.100 &   4.163$\pm$0.059 & -0.27 &   4.40$\pm$0.11 &   2.66$\pm$0.06 &     22.9 & $\ldots$ & 0.56$^a$ &     31.5 & AGN \\
223024+004334 & 337.598552 &  0.726077 &   1.417$\pm$0.082 &   1.180$\pm$0.069 &   1.210$\pm$0.066 &   1.586$\pm$0.047 & -0.29 &   1.44$\pm$0.10 &   0.67$\pm$0.08 &     18.3 &      nv  &     0.13 &     29.7 & AGN \\
223047+003160 & 337.694749 &  0.533246 &   2.088$\pm$0.153 &   1.877$\pm$0.122 &   1.881$\pm$0.126 &   2.860$\pm$0.047 & -0.41 &   3.93$\pm$0.11 &   3.70$\pm$0.06 &     17.4 &      nv  &     0.09 &     29.7 & AGN \\
223140+002305 & 337.917233 &  0.384585 &   0.775$\pm$0.092 &   0.641$\pm$0.071 &   0.616$\pm$0.068 &   1.125$\pm$0.057 & -0.55 &   1.83$\pm$0.10 &   1.36$\pm$0.06 & $\ldots$ & $\ldots$ & $\ldots$ & $\ldots$ & AGN \\
223225+003431 & 338.103745 &  0.575289 &   1.785$\pm$0.086 &   1.871$\pm$0.071 &   1.385$\pm$0.067 &   1.370$\pm$0.048 &  0.31 &   2.19$\pm$0.11 &   2.13$\pm$0.06 & $\ldots$ & $\ldots$ & $\ldots$ & $\ldots$ & AGN \\
223409+010618 & 338.537649 &  1.105054 &  14.140$\pm$0.123 &  16.928$\pm$0.104 &  16.556$\pm$0.110 &  22.451$\pm$0.069 & -0.28 &  27.15$\pm$0.13 &  24.37$\pm$0.24 &     22.0 & $\ldots$ & $\ldots$ & $\ldots$ & AGN \\
223516-005849 & 338.817742 & -0.980367 &   3.230$\pm$0.144 &   3.480$\pm$0.110 &   3.026$\pm$0.114 &   4.715$\pm$0.059 & -0.30 &   2.91$\pm$0.15 &   4.54$\pm$0.06 &     19.1 &      nv  &     1.18 &     32.2 & QSO \\
223908+003232 & 339.784225 &  0.542353 &   4.063$\pm$0.249 &   4.083$\pm$0.167 &   4.267$\pm$0.184 &   5.843$\pm$0.058 & -0.35 &   3.83$\pm$0.10 &   4.01$\pm$0.08 &     19.3 &      nv  & 0.24$^a$ &     30.9 & AGN \\
224023-003555 & 340.096938 & -0.598702 &   2.583$\pm$0.100 &   2.467$\pm$0.079 &   2.953$\pm$0.093 &   3.737$\pm$0.069 & -0.41 &   5.67$\pm$0.13 &   3.17$\pm$0.08 &     18.9 &      nv  &     1.16 &     32.0 & QSO \\
224628-001214 & 341.615430 & -0.203797 &  56.423$\pm$0.343 &  56.155$\pm$0.291 &  55.536$\pm$0.307 &  73.059$\pm$0.112 & -0.26 &  70.42$\pm$0.10 & 100.88$\pm$0.65 &     21.8 &      nv  & 0.55$^a$ &     32.7 & AGN \\
224730+000006 & 341.875813 &  0.001783 & 281.818$\pm$1.200 & 272.963$\pm$0.769 & 260.412$\pm$1.024 & 545.119$\pm$0.403 & -0.67 & 322.29$\pm$0.10 & 397.62$\pm$2.48 &     18.1 &       V  &     0.97 &     34.1 & QSO \\
224733+010817 & 341.885647 &  1.138013 &   1.415$\pm$0.061 &   1.364$\pm$0.053 &   1.202$\pm$0.065 &   0.973$\pm$0.054 &  0.34 &   1.78$\pm$0.14 &   0.86$\pm$0.09 & $\ldots$ & $\ldots$ & $\ldots$ & $\ldots$ & AGN \\
224803+003959 & 342.013923 &  0.666301 &   7.771$\pm$0.131 &   6.594$\pm$0.099 &   6.750$\pm$0.111 &   9.238$\pm$0.068 & -0.33 &  11.93$\pm$0.11 &   8.39$\pm$0.06 &     22.3 & $\ldots$ & 0.82$^a$ &     32.2 & AGN \\
225103+000156 & 342.760572 &  0.032354 &   2.359$\pm$0.126 &   2.126$\pm$0.114 &   2.042$\pm$0.119 &   3.040$\pm$0.060 & -0.35 &   5.33$\pm$0.10 &   3.64$\pm$0.06 &     23.3 & $\ldots$ & $\ldots$ & $\ldots$ & AGN \\
225307-010950 & 343.277710 & -1.163845 &   2.320$\pm$0.079 &   2.124$\pm$0.060 &   2.011$\pm$0.071 &   1.464$\pm$0.053 &  0.37 &   1.39$\pm$0.14 &   1.73$\pm$0.07 &     20.5 &      nv  &     0.34 &     30.7 & AGN \\
225438-001641 & 343.658007 & -0.277990 &   3.564$\pm$0.080 &   3.759$\pm$0.072 &   3.375$\pm$0.077 &   2.759$\pm$0.074 &  0.31 &   2.92$\pm$0.11 &   3.52$\pm$0.09 &     22.0 & $\ldots$ & 0.38$^a$ &     31.1 & AGN \\
225510+002526 & 343.791849 &  0.423809 &   0.548$\pm$0.067 &   0.762$\pm$0.056 &   0.495$\pm$0.060 &   0.401$\pm$0.046 &  0.62 &         $<$0.53 &   0.42$\pm$0.07 & $\ldots$ & $\ldots$ & $\ldots$ & $\ldots$ & AGN \\
225525-000956 & 343.854026 & -0.165458 &   4.053$\pm$0.128 &   3.776$\pm$0.104 &   3.613$\pm$0.112 &   5.175$\pm$0.081 & -0.31 &   6.19$\pm$0.10 &   4.77$\pm$0.12 &     22.6 & $\ldots$ & 0.71$^a$ &     31.8 & AGN \\
225621-004110 & 344.085666 & -0.686069 &   0.982$\pm$0.107 &   1.318$\pm$0.091 &   0.960$\pm$0.098 &   0.866$\pm$0.050 &  0.41 &   1.22$\pm$0.15 &   0.94$\pm$0.07 &     21.5 &      nv  &     0.56 &     30.9 & AGN \\
225934+010821 & 344.891408 &  1.139151 &   0.742$\pm$0.060 &   0.966$\pm$0.052 &   0.984$\pm$0.052 &   0.610$\pm$0.046 &  0.45 &         $<$0.69 &         $<$0.34 &     23.7 & $\ldots$ & $\ldots$ & $\ldots$ & AGN \\
225936-003356 & 344.898924 & -0.565593 &   2.109$\pm$0.128 &   1.989$\pm$0.110 &   1.844$\pm$0.124 &   2.683$\pm$0.066 & -0.30 &   5.78$\pm$0.12 &   4.94$\pm$0.06 & $\ldots$ & $\ldots$ & $\ldots$ & $\ldots$ & AGN \\
230112-002112 & 345.299432 & -0.353462 &   1.611$\pm$0.108 &   1.523$\pm$0.084 &   1.484$\pm$0.087 &   2.122$\pm$0.070 & -0.33 &   5.52$\pm$0.11 &   2.97$\pm$0.06 & $\ldots$ & $\ldots$ & $\ldots$ & $\ldots$ & AGN \\
230132-010319 & 345.382529 & -1.055362 &   2.802$\pm$0.106 &   2.772$\pm$0.091 &   2.890$\pm$0.106 &   4.338$\pm$0.056 & -0.44 &  11.78$\pm$0.14 &   7.11$\pm$0.07 & $\ldots$ & $\ldots$ & $\ldots$ & $\ldots$ & AGN \\
230158+000352 & 345.490838 &  0.064493 &   7.164$\pm$0.143 &   5.936$\pm$0.123 &   6.574$\pm$0.154 &  12.024$\pm$0.059 & -0.68 &   5.38$\pm$0.11 &   5.02$\pm$0.07 &     23.1 & $\ldots$ & $\ldots$ & $\ldots$ & AGN \\
230218-005817 & 345.576721 & -0.971495 &   2.119$\pm$0.152 &   2.245$\pm$0.122 &   2.255$\pm$0.126 &   2.936$\pm$0.063 & -0.27 &  14.39$\pm$0.15 &  10.93$\pm$0.09 & $\ldots$ & $\ldots$ & 0.91$^a$ &     31.8 & AGN \\
230334-004006 & 345.890039 & -0.668195 &   3.817$\pm$0.155 &   3.400$\pm$0.120 &   3.223$\pm$0.140 &   4.479$\pm$0.070 & -0.27 &   7.41$\pm$0.14 &   5.79$\pm$0.07 &     22.0 & $\ldots$ & 0.70$^a$ &     31.7 & AGN \\
230423-000417 & 346.096113 & -0.071504 &   4.115$\pm$0.066 &   3.415$\pm$0.065 &   3.576$\pm$0.066 &   2.608$\pm$0.064 &  0.27 &   2.44$\pm$0.10 &   3.84$\pm$0.08 &     19.7 &      nv  &     1.05 &     32.0 & QSO \\
230748+002213 & 346.950170 &  0.370314 &   4.502$\pm$0.083 &   4.657$\pm$0.069 &   4.229$\pm$0.077 &   2.940$\pm$0.103 &  0.45 &   7.68$\pm$0.10 &   6.26$\pm$0.06 &     22.4 & $\ldots$ & 0.71$^a$ &     31.8 & AGN \\
230847+010904 & 347.196936 &  1.151171 &   1.371$\pm$0.063 &   1.303$\pm$0.054 &   1.295$\pm$0.059 &   0.872$\pm$0.064 &  0.40 &   2.44$\pm$0.14 &   1.52$\pm$0.14 & $\ldots$ & $\ldots$ & $\ldots$ & $\ldots$ & AGN \\
231014+002531 & 347.559929 &  0.425187 &  17.643$\pm$0.098 &  18.021$\pm$0.074 &  16.694$\pm$0.080 &  13.110$\pm$0.117 &  0.32 &  65.78$\pm$0.11 &  37.53$\pm$0.43 & $\ldots$ & $\ldots$ & $\ldots$ & $\ldots$ & AGN \\
231210-003135 & 348.042870 & -0.526485 &   2.677$\pm$0.073 &   2.661$\pm$0.070 &   2.580$\pm$0.079 &   1.596$\pm$0.098 &  0.50 &   5.86$\pm$0.10 &   5.50$\pm$0.07 &     23.3 & $\ldots$ & $\ldots$ & $\ldots$ & AGN \\
231517+002630 & 348.822686 &  0.441566 &   3.696$\pm$0.080 &   3.709$\pm$0.065 &   3.481$\pm$0.065 &   2.853$\pm$0.100 &  0.26 &   8.88$\pm$0.10 &   9.35$\pm$0.06 & $\ldots$ & $\ldots$ & $\ldots$ & $\ldots$ & AGN \\
231537-003726 & 348.902796 & -0.623911 &   1.678$\pm$0.134 &   1.286$\pm$0.126 &   1.564$\pm$0.127 &   3.021$\pm$0.091 & -0.81 &   2.05$\pm$0.14 &   1.20$\pm$0.06 & $\ldots$ &      nv  & $\ldots$ & $\ldots$ & AGN \\
231542+002937 & 348.923546 &  0.493578 &  12.632$\pm$0.170 &  13.768$\pm$0.137 &  12.830$\pm$0.135 &   9.258$\pm$0.120 &  0.39 &  17.92$\pm$0.11 &  17.03$\pm$0.06 &     21.0 &      nv  &     1.35 &     32.8 & QSO \\
231557+005001 & 348.986515 &  0.833474 &   1.945$\pm$0.095 &   2.160$\pm$0.082 &   2.012$\pm$0.089 &   1.516$\pm$0.069 &  0.35 &         $<$0.57 &   3.41$\pm$0.07 &     19.7 & $\ldots$ &     2.52 &     32.5 & QSO \\
231713+000256 & 349.305229 &  0.048771 &  13.192$\pm$0.181 &  12.635$\pm$0.161 &  11.598$\pm$0.160 &  16.734$\pm$0.078 & -0.28 &  31.86$\pm$0.11 &  27.90$\pm$0.20 & $\ldots$ & $\ldots$ & $\ldots$ & $\ldots$ & AGN \\
231846-000755 & 349.690889 & -0.131884 &   4.550$\pm$0.086 &   3.684$\pm$0.072 &   3.976$\pm$0.080 &   5.134$\pm$0.074 & -0.33 &   4.18$\pm$0.11 &   3.14$\pm$0.06 &     19.5 &      nv  &     0.86 &     31.9 & QSO \\
232025+002744 & 350.106120 &  0.462184 &  30.309$\pm$0.157 &  25.698$\pm$0.134 &  26.054$\pm$0.122 &  35.256$\pm$0.095 & -0.31 &  35.98$\pm$0.10 &  21.68$\pm$0.27 & $\ldots$ & $\ldots$ &     2.89 &     33.8 & QSO \\
232037+001335 & 350.153115 &  0.226407 &   0.771$\pm$0.081 &   0.957$\pm$0.062 &   0.788$\pm$0.070 &   0.542$\pm$0.066 &  0.55 &         $<$0.51 &         $<$0.39 & $\ldots$ &      nv  & 0.42$^a$ &     30.5 & AGN \\
232236-000712 & 350.650440 & -0.119883 &  11.177$\pm$0.163 &  10.612$\pm$0.144 &  10.984$\pm$0.125 &  14.273$\pm$0.087 & -0.29 &  24.53$\pm$0.10 &        $\ldots$ &     25.3 & $\ldots$ & $\ldots$ & $\ldots$ & AGN \\
232311-003122 & 350.794080 & -0.522647 &   1.702$\pm$0.082 &   1.786$\pm$0.077 &   1.369$\pm$0.076 &   1.289$\pm$0.066 &  0.32 &   1.80$\pm$0.13 &        $\ldots$ & $\ldots$ & $\ldots$ & $\ldots$ & $\ldots$ & AGN \\
232324+003328 & 350.849824 &  0.557710 &  14.938$\pm$0.124 &  13.590$\pm$0.099 &  12.629$\pm$0.109 &   9.361$\pm$0.063 &  0.37 &  14.55$\pm$0.11 &        $\ldots$ & $\ldots$ & $\ldots$ & $\ldots$ & $\ldots$ & AGN \\
232656+000303 & 351.732481 &  0.050827 &   2.510$\pm$0.131 &   1.996$\pm$0.118 &   2.536$\pm$0.119 &   2.968$\pm$0.054 & -0.39 &   0.93$\pm$0.10 &        $\ldots$ &     23.6 & $\ldots$ & $\ldots$ & $\ldots$ & AGN \\
232804+001904 & 352.017137 &  0.317754 &   7.877$\pm$0.090 &   6.876$\pm$0.079 &   6.344$\pm$0.100 &   4.273$\pm$0.067 &  0.47 &   6.15$\pm$0.10 &        $\ldots$ &     21.6 & $\ldots$ & $\ldots$ & $\ldots$ & AGN \\
233210-003101 & 353.043262 & -0.516945 &   1.307$\pm$0.079 &   1.573$\pm$0.075 &   1.229$\pm$0.076 &   1.102$\pm$0.064 &  0.35 &   2.50$\pm$0.13 &        $\ldots$ &     20.3 &      nv  &     0.45 &     30.8 & AGN \\
233227-010443 & 353.111298 & -1.078566 &   2.130$\pm$0.101 &   2.225$\pm$0.093 &   1.983$\pm$0.096 &   1.543$\pm$0.075 &  0.36 &   3.24$\pm$0.15 &        $\ldots$ & $\ldots$ &      nv  &     0.26 &     30.5 & AGN \\
233260-005129 & 353.249596 & -0.857978 &  15.958$\pm$0.120 &  14.168$\pm$0.102 &  14.493$\pm$0.101 &  21.498$\pm$0.098 & -0.41 &  45.33$\pm$0.13 &        $\ldots$ & $\ldots$ & $\ldots$ & $\ldots$ & $\ldots$ & AGN \\
233301-004501 & 353.255643 & -0.750179 &   2.952$\pm$0.134 &   2.697$\pm$0.122 &   3.097$\pm$0.132 &   4.850$\pm$0.076 & -0.57 &   6.75$\pm$0.14 &        $\ldots$ & $\ldots$ & $\ldots$ & $\ldots$ & $\ldots$ & AGN \\
\hline
\multicolumn{15}{c}{Timescale $<$20 years}\\
\hline
221650+005429 & 334.210170 &  0.908094 &   1.673$\pm$0.067 &   1.765$\pm$0.059 &   1.598$\pm$0.064 &   1.925$\pm$0.050 & -0.09 &         $<$0.57 &         $<$0.31 &     21.4 &       nv & 0.55$^a$ &     31.1 & AGN \\
221711+011038 & 334.294275 &  1.177200 &  14.097$\pm$0.108 &  15.139$\pm$0.081 &  13.095$\pm$0.077 &  18.508$\pm$0.057 & -0.20 &   3.84$\pm$0.15 &   7.51$\pm$0.07 &     22.2 & $\ldots$ & 0.49$^a$ &     32.0 & AGN \\
221813-010344 & 334.554017 & -1.062315 &   8.724$\pm$0.082 &   8.955$\pm$0.066 &   7.971$\pm$0.072 &   8.745$\pm$0.057 &  0.02 &         $<$0.72 &   0.79$\pm$0.10 & $\ldots$ & $\ldots$ & $\ldots$ & $\ldots$ & AGN \\
223041-001644 & 337.672656 & -0.278969 &   3.805$\pm$0.084 &   3.508$\pm$0.068 &   3.763$\pm$0.070 &   3.240$\pm$0.057 &  0.08 &         $<$0.53 &   0.49$\pm$0.06 & $\ldots$ & $\ldots$ & $\ldots$ & $\ldots$ & AGN \\
223514-001425 & 338.806545 & -0.240294 &   3.858$\pm$0.161 &   4.042$\pm$0.138 &   4.265$\pm$0.169 &   4.090$\pm$0.075 & -0.01 &  10.45$\pm$0.11 &   8.36$\pm$0.10 & $\ldots$ & $\ldots$ &     0.14 &     30.3 & AGN \\
230113-002941 & 345.306025 & -0.494588 &   9.942$\pm$0.094 &   9.409$\pm$0.084 &   9.257$\pm$0.092 &  10.574$\pm$0.066 & -0.12 &   2.54$\pm$0.13 &   2.51$\pm$0.06 & $\ldots$ & $\ldots$ & $\ldots$ & $\ldots$ & AGN \\
233002-002736 & 352.507328 & -0.460065 &   5.492$\pm$0.157 &   5.342$\pm$0.143 &   5.742$\pm$0.147 &   5.510$\pm$0.073 & -0.03 &         $<$0.52 &        $\ldots$ &     21.4 & $\ldots$ &     1.65 &     32.5 & QSO \\
\hline
\multicolumn{15}{c}{Serendipitous}\\
\hline
221515-005028 & 333.811588 & -0.841078 &   2.569$\pm$0.086 &   1.989$\pm$0.062 &   1.778$\pm$0.071 &   1.787$\pm$0.058 &  0.11 &         $<$0.68 &   0.94$\pm$0.05 &     21.4 &       nv & 0.44$^a$ &     31.0 & AGN \\
223634-003352 & 339.141345 & -0.564383 &   0.114$\pm$0.150 &   0.897$\pm$0.123 &   0.373$\pm$0.122 &   0.394$\pm$0.069 &  0.78 &         $<$0.66 &         $<$0.63 &      8.9 & $\ldots$ & $\ldots$ & $\ldots$ & RS CVn \\
230241+003450 & 345.672648 &  0.580639 &   0.090$\pm$0.078 &   0.422$\pm$0.069 &   0.046$\pm$0.076 &   0.127$\pm$0.065 &  1.61 &         $<$0.51 &         $<$0.28 &     10.9 & $\ldots$ & $\ldots$ & $\ldots$ & dKe \\
\colhead{I}\\
\enddata
\tablecomments{(1) The PTF column lists the variability properties of the optical counterparts of the radio variable sources; V=variable, nv=not variable. 
See S\ref{sec:optical_properties} for the definition of variability used for PTF sources in this work.
(2) The $z_{\rm spec}$ column gives the spectroscopic redshift from SDSS. Values possessing a superscript 'a' represent photometric redshift estimates from SDSS.
(3) For sources variable on multiple timescales, we place it in the smallest timescale section where the source shows significant variability.
For example, VTC220456-000147 is picked up as a variable on timescales of $<$1 month and $<$20 years based on our selection criteria. 
It is placed in the $<$1 month timescale section in this table.
(4) All flux density upper limits are 3$\sigma$.}
\end{deluxetable*}

\begin{table}
\centering
\tiny
\caption{Summary of the radio follow-up observations of variable and transient sources reported in this work}
\label{tab:radio_followup}
\begin{tabular}{lrrr}
\hline\hline 
Obs. Date   & Freq.& S     & $\sigma_S$\\
            & (GHz)& (mJy) & (mJy)\\
\hline
\multicolumn{4}{c}{VTC225411-010651}\\
\hline
01 Sep 2012 &  2.4 & 0.504 & 0.039\\
01 Sep 2012 &  3.2 & 0.604 & 0.028\\
01 Sep 2012 &  3.8 & 0.569 & 0.033\\
01 Sep 2012 &  4.5 & 0.596 & 0.024\\
01 Sep 2012 &  5.1 & 0.623 & 0.023\\
01 Sep 2012 &  7.1 & 0.641 & 0.021\\
01 Sep 2012 &  7.7 & 0.646 & 0.021\\
01 Sep 2012 & 13.2 & 0.584 & 0.035\\
01 Sep 2012 & 13.8 & 0.672 & 0.032\\
01 Sep 2012 & 14.2 & 0.639 & 0.035\\
17 Sep 2012 &  1.2 & 1.130 & 0.079\\
17 Sep 2012 &  1.8 & 1.363 & 0.077\\
17 Sep 2012 &  2.4 & 1.109 & 0.051\\
17 Sep 2012 &  3.1 & 0.841 & 0.040\\
17 Sep 2012 &  3.8 & 0.671 & 0.068\\
17 Sep 2012 &  4.5 & 0.596 & 0.049\\
17 Sep 2012 &  5.1 & 0.623 & 0.047\\
17 Sep 2012 &  7.1 & 0.641 & 0.044\\
17 Sep 2012 &  7.7 & 0.646 & 0.045\\
\hline
\multicolumn{4}{c}{VTC232939-004755}\\
\hline
01 Sep 2012 & 2.4  & 0.739 & 0.065\\
01 Sep 2012 & 3.0  & 0.701 & 0.048\\
01 Sep 2012 & 3.4  & 0.740 & 0.054\\
01 Sep 2012 & 4.8  & 0.778 & 0.027\\
01 Sep 2012 & 7.4  & 0.741 & 0.021\\
01 Sep 2012 & 13.5 & 0.598 & 0.026\\
01 Sep 2012 & 14.5 & 0.568 & 0.027\\
\hline
\multicolumn{4}{c}{VTC233002-002736}\\
\hline
01 Sep 2012 &  2.4 & 6.846 & 0.095\\
01 Sep 2012 &  3.2 & 9.294 & 0.071\\
01 Sep 2012 &  3.8 & 9.641 & 0.108\\
01 Sep 2012 &  4.5 & 10.519 & 0.073\\
01 Sep 2012 &  5.1 & 10.558 & 0.071\\
01 Sep 2012 &  7.1 & 9.612 & 0.070\\
01 Sep 2012 &  7.7 & 9.381 & 0.073\\
01 Sep 2012 & 13.2 & 7.269 & 0.102\\
01 Sep 2012 & 13.8 & 6.836 & 0.097\\
01 Sep 2012 & 14.2 & 6.487 & 0.103\\
01 Sep 2012 & 14.8 & 6.217 & 0.098\\
17 Sep 2012 &  1.2 & 1.829 & 0.189\\
17 Sep 2012 &  1.8 & 3.791 & 0.179\\
17 Sep 2012 &  2.4 & 7.437 & 0.082\\
17 Sep 2012 &  3.1 & 9.327 & 0.067\\
17 Sep 2012 &  3.8 & 10.380 & 0.123\\
29 May 2014 &  1.3 & 2.396 & 0.179\\
29 May 2014 &  1.9 & 4.253 & 0.240\\
29 May 2014 &  2.4 & 7.095 & 0.188\\
29 May 2014 &  3.0 & 7.511 & 0.109\\
29 May 2014 &  3.4 & 7.272 & 0.094\\
29 May 2014 &  8.4 & 4.984 & 0.051\\
29 May 2014 &  9.4 & 4.562 & 0.053\\
29 May 2014 & 10.4 & 3.955 & 0.055\\
29 May 2014 & 11.4 & 3.474 & 0.081\\
29 May 2014 & 13.2 & 3.295 & 0.074\\
29 May 2014 & 13.8 & 3.505 & 0.070\\
29 May 2014 & 15.7 & 2.937 & 0.076\\
29 May 2014 & 16.3 & 2.949 & 0.076\\
\hline
\multicolumn{4}{c}{VTC221515-005028}\\
\hline
29 May 2014 &  1.3 & 2.759 & 0.137\\
29 May 2014 &  1.9 & 2.549 & 0.175\\
29 May 2014 &  2.4 & 2.654 & 0.080\\
29 May 2014 &  3.0 & 2.373 & 0.069\\
29 May 2014 &  3.4 & 1.935 & 0.067\\
29 May 2014 &  4.5 & 2.155 & 0.061\\
29 May 2014 &  5.1 & 2.058 & 0.055\\
29 May 2014 &  7.1 & 1.337 & 0.047\\
29 May 2014 &  7.7 & 1.198 & 0.048\\
29 May 2014 & 13.2 & 0.487 & 0.059\\
29 May 2014 & 13.8 & 0.710 & 0.054\\
29 May 2014 & 15.7 & 0.554 & 0.060\\
29 May 2014 & 16.3 & 0.464 & 0.059\\
\hline
\end{tabular}
\end{table}

\end{document}